\documentclass{aa}  

\usepackage{graphicx}
\usepackage{txfonts}
\usepackage{newtxtext,newtxmath}
\usepackage{graphicx}
\usepackage{xspace}	    
\usepackage{xcolor}
\usepackage{url}

\usepackage{longtable}

\usepackage{hyperref}
\hypersetup{
    colorlinks=true,
    linkcolor=blue,
    filecolor=black, 
    citecolor=gray,
    urlcolor=blue,
    }
\newcommand{\simgt}{\,\rlap{\lower 3.5 pt \hbox{$\mathchar \sim$}} \raise 1pt \hbox {$>$}\,}
\newcommand{\simlt}{\,\rlap{\lower 3.5 pt \hbox{$\mathchar \sim$}} \raise 1pt \hbox {$<$}\,}

\newcommand{\BE}{\begin{equation}}
\newcommand{\EE}{\end{equation}}
\newcommand{\BEA}{\begin{eqnarray}}
\newcommand{\EEA}{\end{eqnarray}}

\newcommand{\DV}{\ifmmode{\Delta v}\else $\Delta v$\xspace\fi}

\newcommand{\HI}{\ifmmode{\textsc{hi}}\else H\textsc{i}\fi\xspace}
\newcommand{\HII}{\ifmmode{\textsc{hii}}\else H\textsc{ii}\fi\xspace}

\newcommand{\Msun}{\ifmmode{M_\odot}\else $M_\odot$\xspace\fi}
\newcommand{\MUV}{\ifmmode{M_\textsc{uv}}\else $M_\textsc{uv}$\xspace\fi}
\newcommand{\fesc}{\ifmmode{f_\textrm{esc}}\else $f_\textrm{esc}$\xspace\fi}
\newcommand{\lya}{\ifmmode{\mathrm{Ly}\alpha}\else Ly$\alpha$\xspace\fi}

\newcommand{\nh}[1][]{\ifmmode{\overline{n}_\textsc{h}^{#1}}\else $\overline{n}_\textsc{h}$\xspace\fi}

\newcommand{\xHI}{\ifmmode{x_\HI}\else $x_\HI$\xspace\fi}
\newcommand{\NHI}{\ifmmode{N_\HI}\else $N_\HI$\xspace\fi}

\newcommand{\xHImean}{\ifmmode{\overline{x}_\HI}\else $\overline{x}_\HI$\xspace\fi}
\newcommand{\xHIImean}{\ifmmode{\overline{x}_\HII}\else $\overline{x}_\HII$\xspace\fi}
\newcommand{\trec}{\ifmmode{t_\textrm{rec}}\else $t_\textrm{rec}$\xspace\fi}
\newcommand{\clump}[1][]{\ifmmode{C_\HII^{#1}}\else $C_\HII$\xspace\fi}

\newcommand{\Nion}{\ifmmode{\dot{N}_{\mathrm{ion}}}\else $\dot{N}_\mathrm{ion}$\xspace\fi}
\newcommand{\Rion}[1][]{\ifmmode{R_\mathrm{ion}^{#1}} \else $R_\mathrm{ion}$\xspace\fi}

\newcommand{\kp}{\ifmmode{k_\textrm{peak}}\else $k_\textrm{peak}$\xspace\fi}
\newcommand{\hp}{\ifmmode{h_\textrm{peak}}\else $h_\textrm{peak}$\xspace\fi}
\newcommand{\hMpc}{\ifmmode{\,h^{-1}\textrm{Mpc}}\else \,$h^{-1}$Mpc\xspace\fi}

\newcommand{\Tb}{\ifmmode{T_{21}}\else $T_{21}$\xspace\fi}
\newcommand{\aesc}{\ifmmode{\alpha_\mathrm{esc}}\else $\alpha_\mathrm{esc}$\xspace\fi}
\newcommand{\fescII}{\ifmmode{f_\mathrm{esc,10}^\textsc{ii}}\else $f_\mathrm{esc,10}^\textsc{ii}$\xspace\fi}
\newcommand{\fescIII}{\ifmmode{f_\mathrm{esc,7}^\textsc{ii}}\else $f_\mathrm{esc,7}^\textsc{iii}$\xspace\fi}
\newcommand{\astarII}{\ifmmode{\alpha_\star^\textsc{ii}}\else $\alpha_\star^\textsc{ii}$\xspace\fi}
\newcommand{\astarIII}{\ifmmode{\alpha_\star^\textsc{iii}}\else $\alpha_\star^\textsc{iii}$\xspace\fi}
\newcommand{\fstarII}{\ifmmode{f_{\star,10}^\textsc{ii}}\else $f_{\star,10}^\textsc{ii}$\xspace\fi}
\newcommand{\fstarIII}{\ifmmode{f_{\star,7}^\textsc{iii}}\else $f_{\star,7}^\textsc{iii}$\xspace\fi}
\newcommand{\tstar}{\ifmmode{t_\star}\else $t_\star$\xspace\fi}
\newcommand{\Mturn}{\ifmmode{M_\mathrm{turn}}\else $M_\mathrm{turn}$\xspace\fi}
\newcommand{\LX}{\ifmmode{L_X/{\dot{M}_\star}}\else $L_X/{\dot{M}_\star}$\xspace\fi}
\newcommand{\nuX}{\ifmmode{E_0}\else $E_0$\xspace\fi}
\newcommand{\AVCB}{\ifmmode{A_\mathrm{VCB}}\else $A_\mathrm{VCB}$\xspace\fi}
\newcommand{\ALW}{\ifmmode{A_\mathrm{LW}}\else $A_\mathrm{LW}$\xspace\fi}
\newcommand{\Mpcinv}{\ifmmode{\,\mathrm{Mpc}^{-1}}\else \,Mpc$^{-1}$\xspace\fi} 

\newcommand{\kms}{\,\ifmmode{\mathrm{km}\,\mathrm{s}^{-1}}\else km\,s${}^{-1}$\fi\xspace}
\newcommand{\cm}{\,\ifmmode{\mathrm{cm}}\else cm\fi\xspace}

\newcommand{\nsample}{99}
\newcommand{\nhighz}{12}

\newcommand{\nsamplefitNHI}{83}
\newcommand{\nsamplefitxHI}{14}
\newcommand{\nhighzfit}{6}

\begin{document} 

\titlerunning{JWST damping wings}
\authorrunning{Mason et al.}

   \title{Constraints on the $z\sim6-13$ intergalactic medium from JWST spectroscopy of Lyman-alpha damping wings in galaxies}

   \author{
          Charlotte A. Mason \inst{1, 2}
          \and
          Zuyi Chen \inst{3}
          \and
          Daniel P. Stark \inst{4}
          \and
          Ting-Yi Lu\inst{1, 2}
            \and
          Michael Topping \inst{3}
            \and
          Mengtao Tang \inst{3}
          }

   \institute{Cosmic Dawn Center (DAWN)
        \and
            Niels Bohr Institute, University of Copenhagen, Jagtvej 128, 2200 Copenhagen N, Denmark
        \and
        Steward Observatory, University of Arizona, 933 N Cherry Ave, Tucson, AZ 85721, USA
        \and
        Department of Astronomy, University of California, Berkeley, CA 94720 USA
             }
   \date{Received January 20, 2025; accepted November 11, 2025}

 
  \abstract
   {
   JWST provides a unique dataset for studying the earliest stages of reionization at $z>9$, promising insights into the first galaxies. 
   Many JWST/NIRSpec prism spectra of $z>5$ galaxies reveal smooth Lyman-alpha breaks, implying damping wing scattering by neutral hydrogen. 
   }
   {We investigate what current prism spectra imply about the intergalactic medium (IGM) at $z>6$, and how best to use NIRSpec spectra to recover IGM properties. We use a sample of \nsample\ $z\sim5.5-13$ galaxies with high S/N prism spectra in the public archive, including \nhighz\ at $z>10$. 
   }
   {We analyse these spectra using damping wing sightlines from inhomogeneous reionizing IGM simulations, mapping between the distance of a source from the neutral IGM and the average IGM neutral fraction. We marginalise over absorption by local neutral hydrogen around the galaxies, and Lyman-alpha emission.
   }
   {We observe a decline in the median and variance of flux around the \lya break with increasing redshift, consistent with an increasingly neutral IGM, as ionized regions become smaller and rarer. 
   At $z\simgt9$ the spectra become consistent with an almost fully neutral IGM.
   We find S/N$>$15 per pixel is required to robustly estimate IGM properties from prism spectra.
   We fit a sub-sample of high S/N spectra and infer mean IGM neutral fractions $\xHImean=0.33^{+0.18}_{-0.27}, 0.64^{+0.17}_{-0.23}$ ($>0.70$ excluding GNz11) at $z \approx 6.5, 9.3$.
   We also investigate local HI absorption, finding a median column density of $\log_{10}\NHI\approx10^{20.8}$\,cm$^{-2}$, comparable to $z\sim3$ Lyman-break galaxies, with no significant redshift evolution $z\simgt5.5$. We find galaxies showing the highest column density absorption are more likely to be in close associations of sources ($\simlt 500$\,pkpc), implying absorption is enhanced in massive dark matter halos.
    Future deep prism and grating spectroscopy of $z>9$ sources will provide tighter constraints on the earliest stages of reionization, key for understanding the onset of star formation.}
   {}

               
   \maketitle
%

\section{Introduction}  \label{sec:intro}

Understanding the reionization of intergalactic hydrogen in the early universe has long been a frontier in astronomy.
In the past two decades significant progress has been made in constraining the end stages of reionization, with multiple independent observations demonstrating reionization was complete by $z\sim5.3-6$ and on-going at $z\sim7-8$ \citep[e.g.,][]{Stark2010,Ouchi2017,Planck2018,Mason2018,Davies2018b,Qin2024}. 
However, until the launch of JWST, we had no observational constraints on the earliest stages of reionization at $z\simgt8$. 
Constraints on the early IGM promise crucial information about the onset of star formation and the higher-than-expected UV luminosity density detected by JWST at $z>9$ \citep[e.g.,][]{Castellano2022,Naidu2022,Adams2022,Donnan2022,Harikane2022,Finkelstein2022} as the collective ionizing output of galaxies below even JWST's detection limits will be felt in the IGM.

JWST finally provides the ability to chart the earliest stages of reionization through measurements of the Lyman-alpha (\lya) damping wing, due to scattering by neutral hydrogen in the IGM, in $z>8$ galaxies.
The damping wing feature results in smooth absorption up to several thousand km/s redward of \lya in the spectra of high redshift sources \citep[e.g.,][]{Miralda-Escude1998b}. The strength of absorption depends on the density and spatial distribution of neutral hydrogen along the line of sight, and thus can be used to constrain the properties of the high-redshift IGM. 
Before the launch of JWST, \lya damping wings had been observed in just four bright quasars at $z\sim7-7.5$ \citep{Mortlock2011,Banados2018,Wang2020,Yang2020_quasar}. In galaxies, fainter but orders of magnitude more numerous than quasars, the integrated impact of the damping wing had been detected as a decrease in the equivalent width distribution of galaxies' \lya emission \citep[e.g.,][]{Stark2010,Pentericci2014,Mason2019b,jung_texas_2020,Bolan2022} and the decline in \lya-emitter luminosity functions \citep[e.g.,][]{Ouchi2017,Hu2019,Morales2021,Umeda2024} at $z\simgt6$.

The spectral sensitivity of JWST/NIRSpec \citep{Jakobsen202} has enabled the first detection of the UV continuum for typical star-forming galaxies at $z>5$, and thus direct observations of the IGM damping wing. 
Excitingly, early JWST results have revealed many $z>9$ galaxies show strong damping wing features in their spectra \citep[e.g.,][]{Curtis-Lake2022,Umeda2023,Heintz2023b}, and a continued decline in the \lya equivalent width distribution at $z\simgt8$ \citep{Nakane2024,Tang2024c,Jones2025,Kageura2025}, implying we are finally detecting galaxies in an almost fully neutral IGM.
However, JWST spectroscopy has also provided hints of early ionized bubbles via surprising detections of \lya from galaxies at $z\approx11$ and $z\approx13$ \citep{Bunker2023,Witstok2024b}. 
Placing these detections in context requires a large census of $z\simgt9$ spectra: 
current and future spectroscopic surveys with JWST provide the potential to precisely chart the earliest stages of reionization via both the decline in \lya emission and the impact of IGM damping on the UV continuum redward of \lya. 

However, JWST spectra also present unique challenges for inferring properties of the IGM from the UV continuum. The most efficient spectroscopic mode is the NIRSpec prism. The low spectral resolution of the prism around $1\,\mu$m ($R\sim40$) means the damping wing appears in only $\sim 5$ pixels. Moderate Lyman-alpha emission (\lya, EW$\simlt50$\,\AA) can be spread across these pixels and confused with a high continuum flux \citep{Keating2023a,Chen2024_fesc,Jones2024_Lya,Park2024}, in addition to NV P-Cygni stellar wind lines which may be present in sources with $\simlt 10$\,Myr massive stars \citep[e.g.,][]{Chisholm2019} and interstellar absorption lines.
Furthermore, galaxies at all redshifts are commonly observed with absorption around \lya due to dense neutral hydrogen in the ISM and CGM, and proximate absorbers along the line of sight, \citep[e.g.,][]{Shapley2003,Reddy2016,Hu2023,Heintz2024}. These features also change the shape of the UV continuum, though, as we show below, with a different wavelength dependence than the neutral IGM, but may be hard to distinguish from the IGM with low resolution, low S/N spectra.

\citet{Umeda2023} presented the most comprehensive study of galaxy damping wings to-date, fitting the spectra of 27 spectroscopically confirmed $z>7$ galaxies, including the impact of \lya emission and neutral hydrogen (HI) in the host galaxies in the spectra, to infer the IGM neutral fraction \xHImean at $z\sim7-13$, finding evidence for an increasing neutral fraction with redshift. However, to fit the IGM damping wing from galaxy spectra, this, and most previous works with JWST, have assumed a simple analytical model for the \lya transmission, which approximates the IGM as ionized within the galaxies' host bubble and uniform beyond the bubble with neutral fraction \xHImean \citep{Miralda-Escude1998b}. While this model reproduces the median IGM transmission in realistic IGM simulations at fixed \xHImean \citep{Keating2023a}, using it to fit individual sources can bias inferred \xHImean as it overestimates the contribution of neutral gas at large distances \citep{Mesinger2008}. The damping wing optical depth most strongly depends on the distance of a galaxy to the first neutral patch, thus accurate \xHImean inferences requires a realistic mapping between the ionized bubble size distribution as a function of redshift and \xHImean.
In this work, we present an analysis of galaxy damping wings using sightlines from realistic, inhomogeneous IGM simulations which can capture this mapping.

In this paper we seek to understand what current NIRSpec prism spectra imply about the IGM at $z\simgt6$ and how best to use NIRSpec galaxy continuum spectra to robustly recover IGM properties. We use a sample of \nsample\ $z>5.5$ galaxies, including \nhighz\ at $z>10$, to explore the redshift evolution of the \lya break. We find a decrease in both the flux and variance of the strength of the break which we interpret as most likely due to an increasingly neutral IGM, as large ionized regions become smaller and rarer.
We describe an approach for fitting the UV continuum using damping wing sightlines from realistic IGM simulations to forward-model galaxy spectra, accounting for the inhomogeneous nature of the reionizing IGM, marginalising over galaxies' \lya emission and local absorption systems, to infer constraints on galaxies' distances from neutral gas and the mean neutral fraction \xHImean. 

This paper is structured as follows: in Section~\ref{sec:dw} we present our method for modelling the \lya damping wing optical depth, due to both neutral IGM and local absorbers. We describe our observational sample, obtained from public JWST Cycle 1 and 2 NIRSpec spectra, in Section~\ref{sec:obs}, and our spectral fitting approach in Section~\ref{sec:fitting}. We present the evolution of the spectra and our fits to these spectra in Section~\ref{sec:results}. We discuss our results in the context of the reionization process and local absorption systems in Section~\ref{sec:disc}, and present our conclusions in Section~\ref{sec:conclusions}.

We fix use the best fit cosmological parameters from Planck 2018 data \citep[TT,TE,EE+lowE+lensing+BAO from][]{Planck2018}, and all distances are comoving unless specified otherwise.

\section{Modelling the Lyman-alpha damping wing}  \label{sec:dw}

Following e.g. \citet{Mesinger2015} and \citet{Mason2018} we model the contribution of diffuse neutral gas in the IGM (Section~\ref{sec:dw_igm}), and the dense HI in the surroundings of galaxies (Section~\ref{sec:dw_DLA}) separately, i.e. $\tau_\alpha = \tau_\textsc{igm} + \tau_\textsc{dla}$, as we describe below.

\subsection{Optical depth through the reionizing IGM} \label{sec:dw_igm}

\begin{figure*}
    \centering
\includegraphics[width=0.65\textwidth]{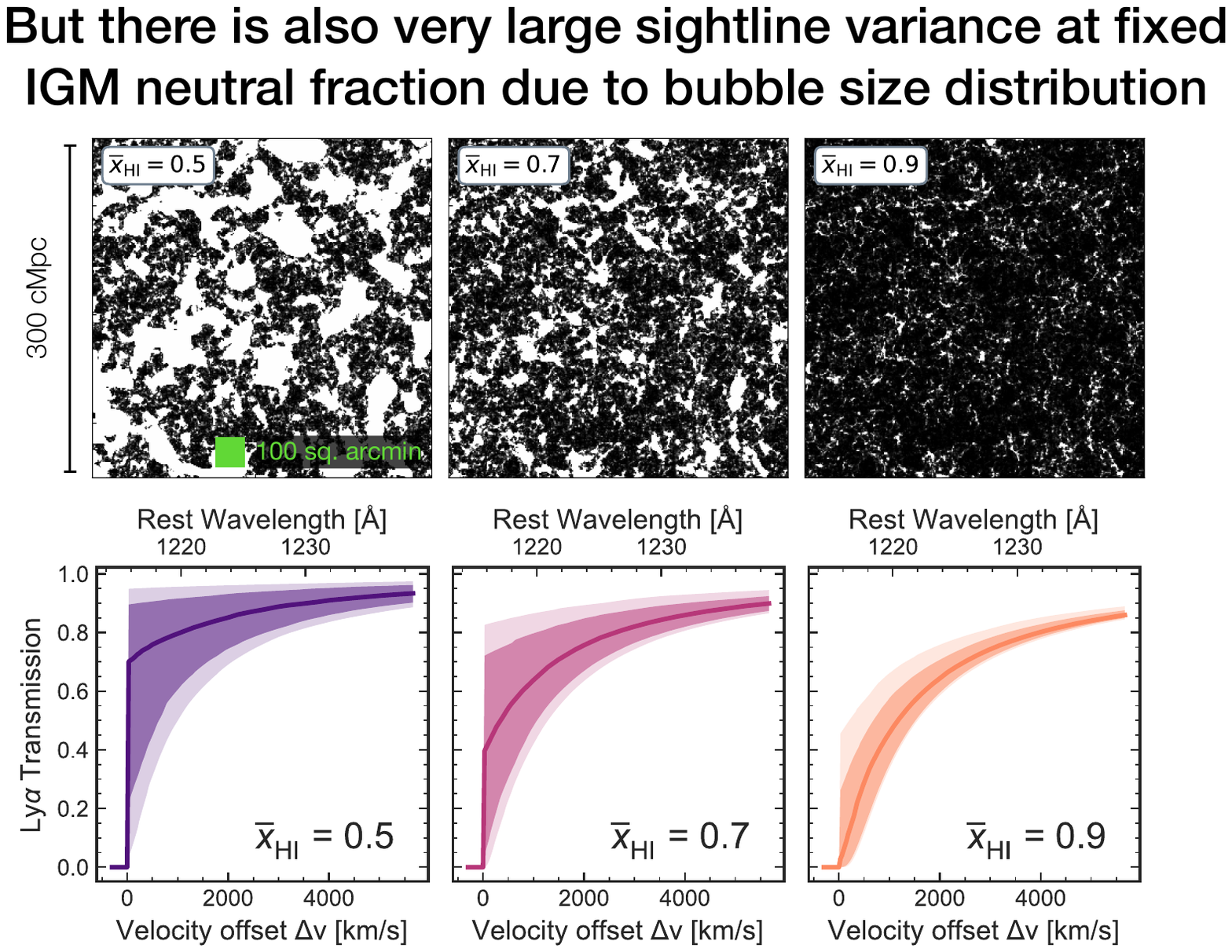}
    \caption{\textbf{Top panels:} Example 300 cMpc $\times$ 300 cMpc slices of the ionization field (white patches show ionized gas, black neutral gas) in our (1.6\,cGpc)$^3$ simulations at $\xHImean=[0.5,0.7,0.9]$ at $z=8$, as described in Section~\ref{sec:dw_sims}. We show a 100 sq. arcmin field as a green square (similar to e.g., the CEERS and JADES survey coverage) for comparison. \textbf{Bottom panels:} \lya transmission profiles, due to the optical depth from the IGM, to galaxies in the corresponding simulations. We truncate the transmission blueward of \lya to account for residual neutral gas inside ionized regions \citep{Mason2020}. We show the median (solid line), 68\%, and 95\% range (shaded regions) of transmission profiles from sightlines to $\sim2000$ galaxies in the simulations with $\MUV\sim-19$. When $\xHImean \simlt 0.7$ the sightline variance in the IGM is significant, meaning a large number of sightlines are required to accurately estimate \xHImean(see Section~\ref{sec:disc}).}
    \label{fig:T_igm_sightline}
\end{figure*}

We first describe the general properties of the \lya optical depth due to the inhomogeneous distribution of neutral hydrogen in the IGM along the line-of-sight expected during reionization. In Section~\ref{sec:dw_sims} we then describe the simulations we use to model the distribution of neutral hydrogen in the reionizing IGM.

For each sightline to a galaxy, the optical depth due to diffuse neutral hydrogen in the IGM can be approximated by the integral over the damping wing component of the optical depth in every neutral patch along the sightline:

\BE \label{eqn:tau_damp}
\tau_\textsc{igm}(\lambda_\mathrm{obs}) = \int_{D_{b}}^{D_\mathrm{max}} d\tau_\textsc{igm}
\EE
where $D_{\mathrm{b}}$ is the distance of the galaxy from the edge of its host ionized bubble along the line-of-sight. We set $D_\mathrm{max}=1.6$\,cGpc, wrapping around our simulation cubes (see Section~\ref{sec:dw_sims}) assuming periodic boundaries \citep[the optical depth converges after $\sim200$\,cMpc, e.g.,][]{Mesinger2008}. The contribution to the optical depth from each neutral patch $i$ along the line-of-sight is given by \citep[e.g.,][]{Miralda-Escude1998b}:
\BEA \label{eqn:tau_damp_int}
d\tau_{\textsc{igm},i}(z) &=& 6.43\times10^{-9} \xHI_{,i} \tau_\mathrm{GP}(z) \nonumber \\
&\times& \left[ I\left(\frac{1+z_{\mathrm{begin},i}}{1+z}\right) - I\left(\frac{1+z_{\mathrm{end},i}}{1+z}\right) \right] \nonumber \\
\EEA
where $\xHI_{,i}$ is the neutral fraction in each patch (we assume $\xHI_{,i}=1$), $\tau_\mathrm{GP}(z) = \pi e^2 f_\alpha \overline{n}_H(z)/[m_e c H(z)]$ is the \citet{Gunn1965b} optical depth, where $e$ is the electron/proton charge, $f_\alpha=0.416$ is the \lya oscillator strength, $m_e$ is the electron mass and $\overline{n}_H(z)$ is the mean hydrogen number density. We define $1 + z = (1+z_g)\lambda_\mathrm{emit}/\lambda_{\mathrm{Ly}\alpha}$, where $z_g$ is the redshift of the galaxy and $\lambda_{\mathrm{Ly}\alpha}$ is the rest-frame wavelength of \lya (1216\,\AA). $z_{\mathrm{begin},i} \approx z_g - c D_{b,i}/H(z)$ is the redshift of the beginning of a neutral patch a comoving distance $D_{b,i}$ from the galaxy and $z_{\mathrm{end},i}$ is the redshift of the end of the neutral patch, and finally,
\BE \label{eqn:tau_damp_ME}
I(x) = \frac{x^{9/2}}{1-x} + \frac{9}{7}x^{7/2} + \frac{9}{5}x^{5/2} + 3x^{3/2} + 9x^{1/2} - \frac{9}{2} \ln{ \left| \frac{1+x^{1/2}}{1-x^{1/2}} \right|}.
\EE
We assume gas inside ionized regions is optically thick to \lya photons at resonance (i.e. $\tau(\lambda_\mathrm{emit} \leq \lambda_{\mathrm{Ly}\alpha}) \rightarrow \infty$), truncating the blue side of \lya \citep[][]{Mason2020}, as expected given the opacity in the \lya forest at $z\simgt6$ \citep{Bosman2022}. Gravitational infall of the IGM will shift this truncation redward of the \lya resonant wavelength \citep[e.g.,][]{Santos2004a,Dijkstra2007}, which we also include (see Section~\ref{sec:fitting}). 

We gain two important insights by considering the limit $z_\mathrm{end} \ll z_\mathrm{begin}$ (i.e. a single ionized patch out to $D_b$ from the source, followed by neutral gas up to a distance $D_\mathrm{max} \gg D_b$): (1) the IGM optical depth is sensitive to neutral gas within $\sim100$\,cMpc, i.e. very large distances, (2) because $I(x)$ increases very steeply with $x$, neutral gas closest to the galaxy has the highest contribution to the damping wing. (2) has important consequences for reionization inferences.

\begin{figure}
    \centering
    \includegraphics[width=\columnwidth]{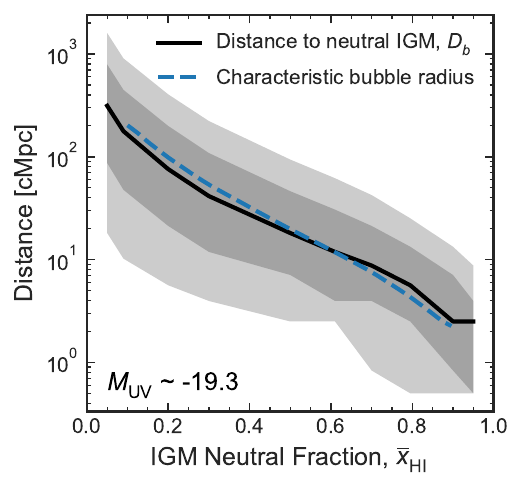}
    \caption{Median, 68\%, 95\% range of distances of $\MUV \sim -19.3$ galaxies (the median in our sample) to the neutral IGM, $D_b$ as a function of \xHImean from the simulations used in this work \citep[see Section~\ref{sec:dw_sims}][]{Lu2024}. The median $D_b$ closely tracks the characteristic (mean) bubble radius in the simulations (blue dashed line), and shows a very broad distribution as the size distribution of ionized bubbles is broad (see Figure~\ref{fig:T_igm_sightline}) and galaxies can sit in a range of locations inside bubbles, not always in the centre.}
    \label{fig:Db_dist}
\end{figure}

In particular, it implies that individual galaxy spectra mostly tell us about distance of a galaxy to the neutral IGM, $D_b$, and are much less sensitive to the global neutral fraction, \xHImean \citep[see Appendix~\ref{app:ME98};][]{Mesinger2008,Chen2024_DW,Keating2023b}. Simulations predict a wide distribution of bubble sizes at fixed \xHImean, and galaxies sit in a variety of positions within bubbles, resulting in significant variance in $D_b$ and thus damping wing profiles \citep[see Figure~\ref{fig:T_igm_sightline},][]{Lu2024}. In Figure~\ref{fig:Db_dist} we plot the distribution of $D_b$ as a function of \xHImean from the simulations we use in this work (see Section~\ref{sec:dw_sims} below), clearly demonstrating the broad distribution. The median $D_b$ tracks the mean bubble radius\footnote{Assuming a spherically symmetric bubble of radius $R_b$ and galaxies uniformly distributed inside the bubble $\overline{D}_b = 3/4 R_b$.} \citep[as estimated using the mean free path method,][]{Lu2024}, but the large variance in $D_b$, in addition to measurement uncertainties on $D_b$ from fitting damping wings (e.g. $\sigma(\log_{10}{D_b})\simgt0.5-0.7$\,dex when fitting NIRSpec prism spectra, see discussion in Section~\ref{sec:disc_future}), means a single galaxy cannot precisely constrain the mean bubble size and thus \xHImean.

Accurately inferring \xHImean from damping wing observations therefore requires observations of tens of galaxies at a given redshift (see discussion in Section~\ref{sec:disc_future}) to recover a distribution of $D_b$ (or damping wing optical depths), and a mapping between these distribution and \xHImean. 
Such a mapping is possible using damping wing sightlines or bubble size distributions from simulations run on grids of fixed \xHImean, which allow us to statistically link the observed distribution of \lya transmission properties to \xHImean. For single sources, even with very high S/N spectroscopy, the resulting uncertainty in \xHImean due to the broad distribution of $D_b$ at fixed \xHImean can be $\sim0.3$ (see e.g., \citealt{Greig2017,Davies2018b,Wang2020,Yang2020_quasar}, for damping wing analyses of $z>7$ quasars, and \citealt{Kist2025a} for an extensive discussion of this sightline variance). Combining many sources thus provides tighter constraints on \xHImean \citep[e.g., considering the evolution of the \lya EW distribution in $z>6$ galaxies, see][]{Mesinger2015,Mason2018,Bolan2022,Tang2024c}.

Most analyses of \lya damping wings in galaxies observed by JWST to-date have assumed a uniform IGM with a volume-averaged neutral fraction \xHImean, and, but not always, with a single ionized bubble around each galaxy \citep[following e.g.,][]{Miralda-Escude1998b}. 
However, because the damping wing is more sensitive to $D_b$ than to \xHImean, not only does this neglect the sightline variance, this approach can lead to significant biases in the inferred \xHImean \citep[discussed in detail by][]{Mesinger2008}. 
In particular, the uniform IGM approximation can underestimate the damping wing for galaxies in small bubbles during the late stages of reionization and overestimate it during earlier stages (see Appendix~\ref{app:ME98}). Relatively accurate estimates of $D_b$ can be obtained assuming a fully neutral IGM beyond the first bubble \citep[e.g.,][]{Mason2020,Hayes2023}, but still require a mapping from the inferred distribution of $D_b$ to \xHImean \citep[e.g.,][]{Tang2024c}.
In this work we analyse JWST spectra using sightlines from realistic, inhomogeneous IGM simulations from \citet{Lu2024}, described in Section~\ref{sec:dw_sims}, allowing us to map between inferred damping wings parameterised by $D_b$, and the IGM neutral fraction \xHImean.
Here, we consider a single reionization morphology model, but we do not expect this to significantly impact our \xHImean inference -- as demonstrated by \citet{Sobacchi2015,Greig2017,Greig2019,Mason2018b}, the \lya damping wing transmission is fairly insensitive to the reionization morphology when considering $>L^\star$ galaxies (as we will for our study due to their higher SNR spectra) which are likely to be in the largest ionized regions in all reionization models \citep{Lu2024}, and when considering redshift bins broader than the typical scales of bubbles ($\Delta z \simgt 0.1$, as will be the case with our study) which smooths out information about the reionization morphology \citep{Sobacchi2015}. In an upcoming work we will discuss cases where differences in the reionization morphology could be distinguished (i.e. for UV-faint galaxies observed in narrower redshift windows).

\begin{figure*}
    \centering
    \includegraphics[width=0.7\textwidth]{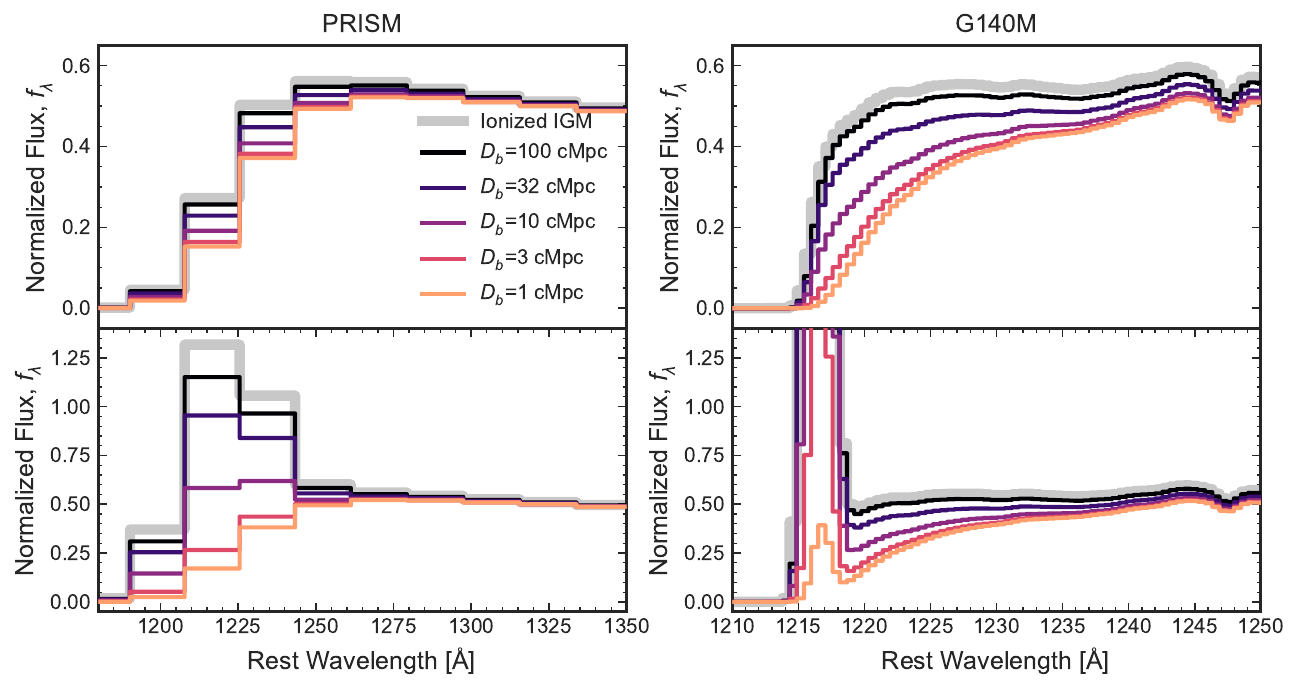}
    \caption{Mock spectra at $z=10$ demonstrating the impact of the distance from neutral IGM, $D_b$. Left (right) panels show the spectrum convolved to the resolution of the NIRSpec prism (G140M grating). The grey line shows the spectrum in an almost fully ionized IGM (i.e. \lya is attenuated only blueward of resonance by the \citet{Gunn1965b} optical depth). Coloured lines show the spectrum if the galaxy is a distance $D_b=1-100$\,cMpc from the neutral IGM. The top panels show a case with no \lya emission, the bottom panels show the same spectrum including \lya emission with pre-IGM $EW=100$\,\AA, FWHM=200\,km s$^{-1}$ and $\DV$=200\,km\,s$^{-1}$, where weak \lya due to the smallest $D_b$ can only be clearly identified in the G140M spectrum.}
    \label{fig:example_damped_Db}
\end{figure*}

In Figure~\ref{fig:example_damped_Db} we show mock spectra, convolved to the resolution of the NIRSpec prism and G140M gratings, showing the impact of the neutral IGM. Here we have taken a template high resolution spectrum at $z\sim10$ (see Section~\ref{sec:fitting}), adding a Gaussian \lya emission line with $EW=100$\,\AA, FWHM=200\,km s$^{-1}$ and velocity offset from systemic $\DV$=200\,km s$^{-1}$. We apply IGM damping wings using Equation~\ref{eqn:tau_damp}, assuming a single ionized region in a fully neutral IGM.
Figure~\ref{fig:example_damped_Db} shows large ($>1$\,dex) changes in the distance to neutral IGM, $D_b$, can be clearly distinguished. However, distinguishing $D_b \simlt 3$\,cMpc is challenging in the prism, whereas these can be distinguished with G140M, especially if \lya emission is present. This is because the gradient of the damping wing is steepest closest to line center, making deep constraints on \lya emission most important for measuring $D_b$ in the early stages of reionization when bubbles are expected to have $R\simlt 10$\,cMpc \citep{Lu2024}. We discuss prospects for constraining the damping wing signal with grating spectra in Section~\ref{sec:disc_future}. Overall, we see the NIRSpec prism provides an efficient, though relatively blunt, tool for constraining IGM properties.

\begin{figure}
    \centering
     \includegraphics[width=\columnwidth]{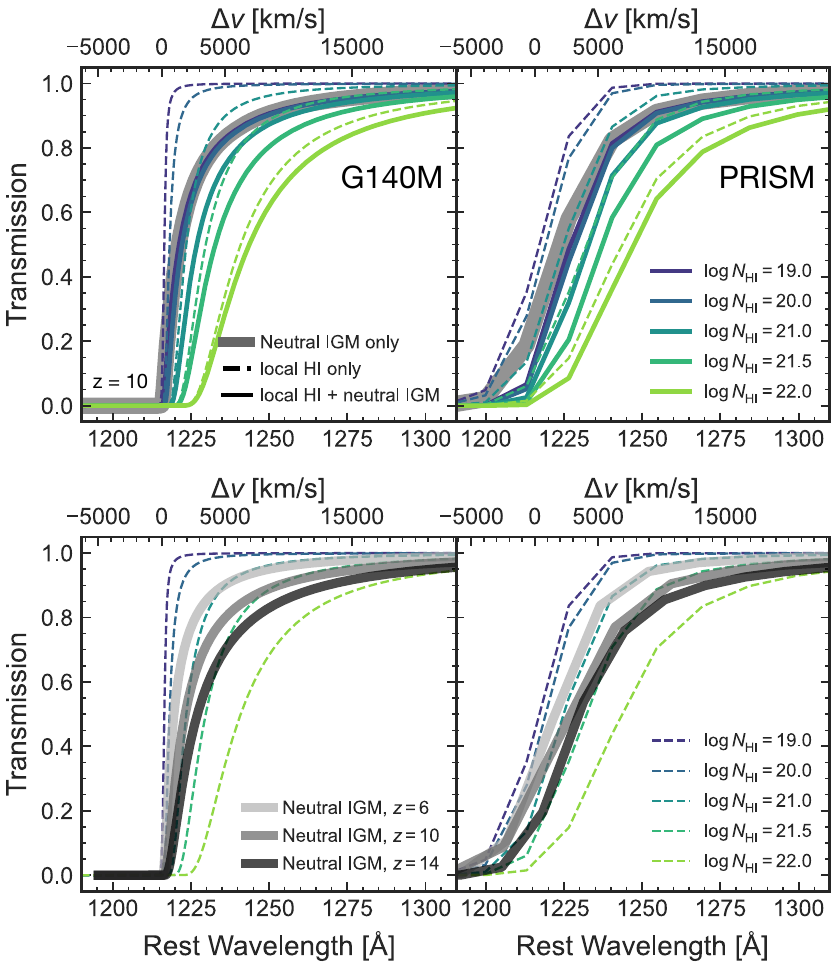}
    \caption{Transmission ($e^{-\tau}$) as a function of wavelength around \lya due to the neutral IGM and local absorbers for high resolution ($R\simgt1000$, left panels) and convolved with the resolution of the prism (right panels). \textbf{Top panels:} For a source at $z=10$. The thick grey line shows the transmission expected in the fully neutral IGM (Section~\ref{sec:dw_sims}). Dashed coloured lines show the absorption profiles expected for local absorbers in an ionized IGM (Section~\ref{sec:dw_DLA}). Solid coloured lines show the profiles for the combinations of both local absorbers and neutral IGM. For $\NHI \simlt 10^{20.5}$\,cm$^{-2}$ the neutral IGM dominates the damping wing profile. For higher column densities $\NHI \simgt 10^{22}$\,cm$^{-2}$, the shape becomes dominated by the local absorption, though the neutral IGM causes more absorption at redder wavelengths than a local absorber alone. \textbf{Bottom panels:} The neutral IGM damping wing at $z=6,10,14$ (grey solid lines) compared to only local absorption (dashed coloured lines, same as top panel). By $z\sim14$ the IGM damping wing becomes similar in strength to a $\NHI \simgt 10^{21.5}$\,cm$^{-2}$ local absorber.
    }
    \label{fig:example_damped}
\end{figure}

\subsubsection{Reionization simulations}  \label{sec:dw_sims}

To obtain realistic IGM damping wings we use semi-numerical reionization simulations by \citet{Lu2024}, which are optimised for comparison to JWST observations, and refer the reader there for full details.
The simulations are created using the semi-numerical code \texttt{21cmFAST-v2} \citep{Mesinger2007,Sobacchi2014a,Mesinger2016}. \texttt{21cmFAST-v2} generates IGM properties from a 3D density field, flagging cells as ionized when the rate of ionizations exceeds the rate of recombinations. The ionization rate is set by the collapsed matter fraction in a cell multiplied by an ionization parameter.

We create a grid of simulation cubes, using the same initial conditions, at fixed redshifts $z\sim6-16$, with $\Delta z = 1$, which are each (1.6\,cGpc)$^3$ volume -- sufficient to sample 100s of $\MUV \sim -22$ galaxies, with $\sim1$\,cMpc resolution in the IGM. 
For each cube, we use a halo-filtering approach in Extended Press-Schechter theory \citep{Sheth2001} to generate a halo catalog from the associated density field (see \citet{Mesinger2007} for a full description of the method) and assign UV luminosities to halos based on the \citet{Mason2015} luminosity function model, which successfully reproduces observations over $z\sim0-10$ \citep[see][for a comparison of the simulated LFs and observations]{Lu2024}, enabling us to account for the expected large-scale environment of the observed galaxies. We note, while this model doesn't match $z>10$ JWST LFs well \citep{Mason2023a}, as shown by \citet{Whitler2020,Lu2024} the exact environment of galaxies as a function of UV luminosity is less important that the average neutral fraction in determining \lya visibility, so this should not significantly impact our results. We include 0.5\,mag scatter in the halo mass -- UV luminosity mapping to include the impact of stochastic star formation \citep[e.g.,][]{Ren2019,Mason2023a,Gelli2024}, but note this has only a small impact on galaxies' \lya transmission \citep{Whitler2020}.
We vary the ionization parameter to produce IGM cubes from the density field at neutral fractions $\xHImean =0-1$, with spacing $\Delta \xHImean \approx 0.05$, from which we sample IGM damping wings to every halo to generate a catalogue of damping wings as a function of redshift, \xHImean, $D_b$ and \MUV.

In Figure~\ref{fig:T_igm_sightline} we show example slices from our IGM cubes at $z=8$, showing $\xHImean=0.5,0.7,0.9$, and the corresponding median and 68\% and 95\% ranges of IGM damping wings. 
Figure~\ref{fig:T_igm_sightline} demonstrates there is large sightline variance in \lya transmission due to the broad bubble size distributions, especially during the mid-stages of reionization \citep[see also, e.g.][]{Mesinger2008,Mason2018b,Keating2023a}, and thus the importance of using simulations to map from damping wing observations to \xHImean estimates.
In Section~\ref{sec:disc_future}, we demonstrate we require $\sim20$ sightlines, i.e. galaxies, per redshift bin to accurately recover \xHImean. 

\subsection{Optical depth from local absorbers}  \label{sec:dw_DLA}

In addition to the \lya damping wing from the IGM, sources can also experience \lya damping absorption from dense HI gas on local ($<1$\,pMpc) scales. Spectroscopic studies at $z\sim0-4$ have shown that roughly half of Lyman-break galaxies show absorption around \lya, often in addition to \lya emission \citep{Shapley2003,Heckman2011,Rivera-Thorsen2015,Reddy2016,Pahl2019,Hu2023,Begley2024}, which has recently been extended to $z>5$ with JWST \citep[e.g.,][]{Chen2024_fesc,Heintz2023b,Heintz2024,Hainline2024}. These results imply neutral gas in the ISM and/or CGM with column densities $\NHI \simgt 10^{20}$\,cm$^{-2}$, i.e. damped \lya absorbers (DLAs), though with a non-uniform covering fraction \citep[e.g.,][]{Heckman2011,Reddy2016}. Proximate absorbers along the line of sight may also provide additional opacity \citep[e.g.,][]{Davies2023}.

A key question is to what extent this local absorption affects our ability to estimate the impact of the IGM at $z\simgt6$. \citet{McQuinn2008} and \citet{Lidz2021} have discussed this in the context of measuring IGM damping wings in gamma ray burst (GRB) spectra and demonstrated the absorption profiles due to the IGM and local gas are significantly different. The optical depth from local HI gas can be approximated by:
\begin{equation} \label{eqn:tau_DLA}
    \tau_\mathrm{DLA}(\Delta \lambda) = \NHI \sigma_\alpha(\Delta\lambda, T)
\end{equation}
where we use the approximation for the \lya optical depth $\sigma_\alpha$ given by \citet{Tasitsiomi2006}. As the damping wings are set by natural line broadening, the temperature, $T$, of the absorbing gas has negligible impact on the optical depth, so we set $T=10^4$\,K.

The Lorentzian wing of the \lya optical depth \citep[e.g., see][for a review]{Dijkstra2014} implies $\tau_\mathrm{DLA} \sim 1/(\Delta \lambda)^2$, while the IGM damping wing, being an integral over a much longer path length, follows a shallower profile, $\tau_\mathrm{IGM} \sim 1/(\Delta \lambda)$. This means the impact of the IGM and DLAs can be distinguished in the UV continuum.

We demonstrate this in Figure~\ref{fig:example_damped} where we show \lya transmission profiles for a fully neutral IGM versus local absorption through various column densities \NHI, and the combination of both local absorption and the neutral IGM for both grating and prism resolution. The steep local HI absorption profile relative to the IGM damping wing is clearly seen for $\NHI \simlt 10^{21}$\,cm$^{-2}$ where more flux is reduced at linecenter. At fixed \NHI, the addition of neutral IGM suppresses flux at longer wavelengths. Only for extremely high column densities, $\NHI \simgt 10^{22}$\,cm$^{-2}$ does the DLA damping wing start to dominate over the IGM damping wing. Recent JWST observations have presented evidence for DLAs reaching $\NHI \approx 10^{22.0-22.5}$ around some sources \citep{Heintz2023b,Chen2024_fesc,DEugenio2023}, but as we will show in Section~\ref{sec:disc_DLA} this is likely a tail of the distribution and the majority of $z\simgt5$ sources have lower inferred column densities. In Appendix~\ref{app:DLAs} we show mock spectra for both the prism and G140M resolution grating, finding that, given sufficient S/N, the IGM can be distinguished from local absorption. In Figure~\ref{fig:example_damped} we also show how the damping wing for a fully neutral IGM evolves with redshift, becoming similar in strength to a $\NHI \simgt 10^{21-21.5}$\,cm$^{-2}$ DLA at $z\sim6-14$.

In our fiducial models, we fix the absorber to the redshift of the source, assuming most absorption happens in the ISM/CGM, and assume a uniform covering fraction of local neutral gas, $f_\mathrm{cov}=1$.
We also consider models with non-uniform covering fraction $f_\mathrm{cov}<1$,
and with proximate absorbers, though in the majority of cases these do not provide better fits. 
In Appendix~\ref{app:DLAs} we describe the transmission profile including the covering fraction and discuss the impact of other variables on the transmission profiles.
We also show in Appendix~\ref{app:DLAs} that even without a precise spectroscopic redshift from emission lines it should still be possible to get information about the IGM damping relative to DLAs. 

In addition to DLAs, an increase in lower column density systems (Lyman-limit systems and sub-DLAs, $\NHI \sim 10^{17-20}$\,cm$^{-2}$) is expected in the ionized IGM as the UV background drops during reionization \citep[e.g.,][]{Bolton2013}. We will discuss this further in Section~\ref{sec:disc_DLA} but do not expect this to strongly affect our results as the damping wing shape is barely changed at the resolution of the prism in the presence of sub-DLAs (see Figures~\ref{fig:example_damped} and \ref{fig:example_spec_DLAs}).

\begin{figure}
    \centering
\includegraphics[width=\columnwidth]{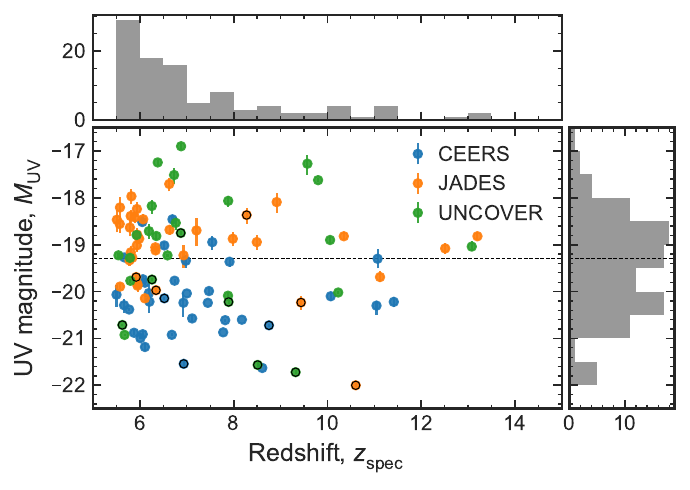}
    \caption{UV magnitude versus spectroscopic redshift for our sample. We show sources from CEERS, JADES and UNCOVER in blue, orange and green respectively, and highlight sources with sufficient S/N($>15$) for robust IGM fitting (see Section~\ref{sec:fitting}) with black outlines.}
    \label{fig:Muv_z}
\end{figure}

\section{Data and sample selection}  \label{sec:obs}

We select our sample from public JWST NIRSpec data from CEERS \citep[GO-1345, DDT-2750,][]{Finkelstein2022,ArrabalHaro2023}, UNCOVER \citep[GO-2561,][]{Bezanson2022} and JADES GOODS-S \citep[GTO-1210, GO-3215,][]{Eisenstein2023a,Eisenstein2023b}.
The NIRSpec spectra are reduced and inspected in the same way as described by \citet{Tang2023,Tang2024c,Chen2024_fesc} using the JWST data reduction pipeline\footnote{\url{https://jwst- pipeline.readthedocs.io/en/latest/}} and we refer the reader there for more details. 
We extract 1D spectra using boxcar extraction with an aperture matching the continuum or emission line profile in the spatial direction, with a typical width of $\sim0.5$\,arcsec.
We applied slit-loss corrections assuming a point source, given that the majority of sources in our sample are compact.

We note work at lower redshifts has shown \lya profiles can depend on spectroscopic aperture, whereby if the aperture, $R_\mathrm{ap}$, is smaller than galaxies' half-light radius, $R_\mathrm{eff}$ (i.e. if $R_\mathrm{eff}/R_\mathrm{ap} > 1$) some of the emission from the extended \lya halo can be missed or `vignetted', resulting in \lya absorption profiles \citep[e.g.,][]{Hayes2013,Scarlata2015,Henry2015a,Hu2023,Huberty2025}. 
We test the expected impact of this in our sample. At $z>5$, typical UV sizes are smaller 
\citep[$R_\mathrm{eff}\simlt400\,$pc, $\simlt0.08$\,arcsec, e.g.,][]{Yang2022,Morishita2024_size} than a single MSA shutter ($0.20\arcsec \times 0.46\arcsec$). Assuming the \MUV-size relation, including intrinsic scatter, derived by \citet{Morishita2024_size} for the median $\MUV \approx - 19.3$ for our sample (see below), we estimate an effective source area $<2(<0.5)$\,pkpc$^2$ at $z\sim5(14)$ (upper 84\%). The MSA shutter area corresponds to $\approx 4(1)$\,pkpc$^2$ at $z\sim5(14)$. Thus throughout the redshift range of this study, the MSA shutter area is more than double the effective source area in physical units, and the ratio between these areas does not significantly evolve with redshift.
Thus, while sources may not be perfectly centred in MSA shutters, we do not expect aperture vignetting to systematically increase the prevalence of observed \lya absorption with redshift in our sample.

We select all sources with $z_\mathrm{spec} \geq 5.5$, requiring the detection of multiple emission lines (usually the [OIII] doublet). For $z>10$ sources, we also include spectra with spectroscopic confirmation from only the \lya break. 
To establish a sample with sufficient S/N for fitting the damping wing we perform a S/N cut on the continuum.
We find the noise produced by the pipeline underestimates variance in the spectra, particularly in the rest-frame UV. Thus we rescale the error spectra to match the standard deviation of the flux over the range $2200-2400$\,\AA\ (to avoid strong UV emission lines) for each source. This results in rescalings of $\sim 1-3 \times$ the pipeline error spectra \citep[see also, e.g.,][for a similar rescaling of CEERS spectra]{ArrabalHaro2023}. Ensuring the S/N of flux blue-ward of the Lyman-limit is normally distributed (as the flux should be zero due to IGM absorption), results in comparable rescaling factors for every spectrum.

We select sources where the median S/N over $1300-1500$\,\AA\ (after rescaling) is $\geq3$ per pixel.
This results in \nsample\ sources, spanning $\MUV \approx [-17,-22]$ (median $-19.3$) and $z=5.5 - 13.2$, including \nhighz\ sources at $z>10$. The median S/N per pixel $\approx 8$, and \nsamplefitxHI\ sources have S/N$>$15 per pixel\footnote{At the time of writing there are 11 additional sources in the public archive from Cycle 1+2 with S/N$>$15, but all are $z<7.7$. As our focus is the earliest stages of reionization at $z\simgt9$ we leave analysis of these lower redshift sources to future work.}, sufficient to robustly recover ionized bubble sizes (see Appendix~\ref{app:fitting}). Figure~\ref{fig:Muv_z} shows the UV magnitude -- redshift distribution of our sample. 

For each source, in addition to spectroscopic redshift, we measure \MUV and [OIII]+H$\beta$ EW. \MUV is derived from NIRCam photometry using the filters nearest to the rest-frame 1500\,\AA, as done by \citet{Tang2023}. [OIII]+H$\beta$ EW is derived with prism when the optical continuum has good SNR, or from BEAGLE modelling to NIRCam photometry otherwise \citep[following the approach by][]{Chen2024_fesc}. We used the following sources for NIRCam photometric catalogs: the CEERS catalog from \citet{Endsley2022}, JADES DR2 \citep{Rieke2023,Eisenstein2023b} and UNCOVER DR2 \citep{Weaver2024}, using the lensing map by \citet{Furtak2023} to correct for magnification.

\section{Spectral fitting}  \label{sec:fitting}

Here we describe our approach for fitting the prism continuum spectra to recover IGM and local HI properties. We first describe how we forward-model each galaxy's spectrum after transmission through local HI and the IGM. We then describe our likelihood function which accounts for the covariance in prism spectra and discuss the S/N requirements for recovering robust IGM constraints. We describe the setup for our Bayesian inference and priors in more detail in Appendix~\ref{app:bayes}.

We perform the following steps to forward-model prism spectra for each observed source:
\begin{enumerate}
    \item \textbf{Create an intrinsic continuum model for $<1500$\AA} by fitting the observed spectrum at $>$1500\AA\ using the photoionization modelling code BEAGLE \citep{Chevallard2016}. By fitting the spectrum including all nebular emission lines, BEAGLE predicts the nebular continuum at $<1500$\AA.
    The BEAGLE fits are performed using a constant star formation history, a \citet{Chabrier2003} IMF (upper mass cut 100\,\Msun), \citet{Pei1992} SMC extinction curve (uniform prior on the V-band optical depth from 0 to 6), uniform prior on $\log {U}$ from -4 to -1, and a uniform prior on the ionizing photon escape fraction $f_\mathrm{esc}$ from $0-1$. Non-zero escape fractions are required to fit very blue, $\beta < -2.6$, UV slopes, as seen in a small fraction of $z>5$ spectra \citep{Topping2024}. 
    
    To test the accuracy of the continuum models we compare the predicted and observed spectra at $1400-1500$\AA. This is blueward of the range we used to fit the spectrum with BEAGLE but redward of where the IGM and local absorbers can significantly change the continuum. We calculate the residual spectrum over $1400-1500$\AA\ (observed - predicted/observed). We find: 1) the distribution of mean (over $1400-1500$\AA) residuals is peaked at zero, indicating no systematic bias above or below the observed continuum, and 2) the distribution of standard deviations of residuals (equivalent to the fractional error on the continuum models) across our sample has a median at 10\% (6-22\%, 16-84\% range, with the uncertainty decreasing with increasing S/N).
    A range of dust attenuation laws may also impact the shape of the UV continuum, though we note the majority of our sources are fit with negligible dust attenuation. Deep, high resolution grating spectra of the UV continuum will provide further tests of photoionization models, which will be important future work.
    Our recovered uncertainties are comparable to the uncertainty on fits to quasar continua at $z<7.5$ \citep[$\sim5-10\%$,][]{Greig2024a,Hennawi2024}. 
    These recovered errors are however $\approx5\times$ higher than the uncertainty of the continuum models output from BEAGLE, so we rescale all output continuum models uncertainties by a factor of five.
    \item \textbf{Add attenuation by local absorbers}, $\tau_\mathrm{DLA}(\NHI_\mathrm{,eff})$, with effective column density $\NHI_\mathrm{,eff}$ at the redshift of the source\footnote{Given the resolution of the prism at $1\,\mu$m ($\Delta z \simgt 0.05, \simgt 100$\,pkpc) this can be due to an unresolved ensemble of absorbers, thus we use the subscript \textit{eff} to denote it is an effective optical depth, and add a single absorber at the redshift of the source.}. 
    As described in Section~\ref{sec:dw_DLA} we also consider fits with non-uniform covering fraction, $f_\mathrm{cov}<1$, or with a proximate absorption system along the line-of-sight. 
    In the majority of sources these do not provide significantly better fits.
    \item \textbf{Add emergent \lya emission} using a single Gaussian emission line with equivalent width EW$_\lya$, FWHM and velocity offset from systemic, \DV. This is the emission before transmission through the IGM, and must be included even in sources without apparent \lya emission to account for the contribution of weak \lya which is unresolved by the prism \citep{Jones2024_Lya,Chen2024_fesc,Keating2023a}.
    Lower redshift studies have shown a large fraction of the variance in \lya emission can be predicted by galaxy properties linked to the production and escape of \lya photons \citep[see e.g.,][]{Trainor2019a,Hayes2023b}
    Thus, we use conditional empirical priors for the above \lya properties as a function of each galaxy's emission properties (\MUV, OIII+H$\beta$ EW), based on $z\sim5-6$ observations \citep[][]{Tang2024}. We describe these priors in more detail in Appendix~\ref{app:bayes}. 
    \item \textbf{Add resonant scattering attenuation due to gas infalling to the halo}. By $z\simgt5$ dense residual HI in the ionized IGM resonantly scatters photons emitted blue-ward of \lya linecenter. Gravitational infall of gas around halos can shift this attenuation red-ward, to approximately the circular velocity of the halo, as \lya photons appear blue-shifted in the frame of infalling gas \citep[e.g.,][]{Santos2004a,Dijkstra2007}. Following \citet{Mason2018} we cut transmission blueward of the circular velocity of the halo, and add a random scatter of 10\%, motivated by hydrodynamical simulations by \citet{Park2021} demonstrating moderate sightline variance.
    \item \textbf{Add attenuation by the neutral IGM} at a distance $D_b$ from the source: using IGM damping wing optical depths $\tau_\mathrm{IGM}(D_b, \xHImean)$ drawn from the simulated sightlines described in Section~\ref{sec:dw_sims}.
    For each galaxy we draw sightlines from halos in the simulation with UV magnitudes $\MUV$ within 0.2 mag of the observed magnitude.
    \item \textbf{Convolve the model spectrum with the resolution of the prism.}\footnote{We calculate the line spread function using observations from CAL-1125 of a planetary nebula IRAS-05248-7007 in the LMC. We find $R\sim45$ around $\sim1\,\mu$m.}
\end{enumerate}

\noindent Thus the final model spectrum given galaxy parameters $\theta_\mathrm{gal}$ is:
\BEA \label{eqn:f_model}
f_\mathrm{mod}(\lambda, \xHI, D_b, \theta_\mathrm{gal}) = f_\mathrm{emit}(\lambda, \theta_{\mathrm{Ly}\alpha}) \times \mathcal{T}(\lambda, \xHImean, D_b, \theta_\textsc{dla})
\EEA
where $f_\mathrm{emit}$ is the continuum model (step 1) plus intrinsic \lya emission (step 2) multiplied by the transmission curve:
\BE \label{eqn:T_tot}
\mathcal{T}(\lambda, \xHImean, D_b, \theta_\textsc{dla}) = e^{-\tau_\textsc{dla}(\lambda, \theta_\textsc{dla}) - \tau_\textsc{igm}(\lambda, D_b, \xHImean)}
\EE
where $\theta_{\mathrm{Ly}\alpha}=(\mathrm{EW}_\lya, \DV, FWHM)$ are the \lya emission parameters (step 2), $\theta_\textsc{dla} = (\NHI_\mathrm{,eff}, f_\mathrm{cov}, z_\mathrm{DLA})$ are the DLA parameters (step 3) and $\tau_\textsc{igm}(D_b, \xHImean)$ is the IGM optical depth (step 5).

We fit the spectra using Bayesian inference. Because resampling of prism spectra introduces covariance between adjacent pixels \citep{Jakobsen202}, we use the following likelihood for each source:
\BE \label{eqn:likelihood}
\ln p(\mathbf{f}_\mathrm{obs} \,|\, \mathbf{\theta}) = -\frac{1}{2}\mathbf{r}^\intercal K^{-1} \mathbf{r} - \frac{1}{2}\ln \left[{2\pi \, \mathrm{det}(K)}\right]
\EE
where $\mathbf{r} = \mathbf{f}_\mathrm{obs} -\mathbf{f}_\mathrm{mod}$ is the residual vector, $K$ is the $N\times N$ covariance matrix\footnote{Where $N$ is the number of spectral pixels we fit over.}. 

Based on estimates of the NIRSpec prism covariance matrix from multiple exposures in the GTO surveys (P. Jakobsen, priv. comm.), we assume the covariance matrix can be approximated as a near-diagonal matrix:
\BE \label{eqn:cov}
K_{ij} = \sigma^2_i \delta_{ij} + k(\lambda_i, \lambda_j)
\EE
where $\sigma_i$ is the observational noise 
in spectral pixel $\lambda_i$, $\delta_{ij}$ is the Kronecker delta and $k$ is a covariance function between adjacent spectral pixels $\lambda_i, \lambda_j$:

\BE \label{eqn:cov_k}
k(\lambda_i, \lambda_j) = \frac{1}{2}\sigma_i\sigma_j
\EE
i.e. $k=0$ if $j\neq i\pm 1$.

We use Bayesian inference to infer the parameters $\xHImean$, $D_b$ and $\theta_\mathrm{gal}$ ($=\theta_\mathrm{\lya}, \theta_\mathrm{DLA}$) for each galaxy.
We note that, since the damping wing fits are most sensitive to $D_b$ (see Appendix~\ref{app:ME98}), using sightlines labelled with both $D_b$ and \xHImean makes the inference of \xHImean for each galaxy explicitly conditional on $D_b$, thereby propagating uncertainties in the inferred $D_b$ values into the uncertainties on \xHImean.
For $z>10$ sources without spectroscopic redshifts from emission lines we also fit for $z_\mathrm{spec}$ as a free parameter, using a Gaussian prior for the redshift based on an initial fit to the \lya break.
We describe the setup for the inference and priors in Appendix~\ref{app:bayes}.

To understand the S/N requirements to obtain robust inferences we perform fits to mock spectra. As described in Section~\ref{sec:dw_igm}, to accurately recover information about \xHImean requires being able to infer the distribution of distances of each galaxy to the neutral IGM, $D_b$. We find that to robustly recover bubble sizes $D_b \sim 5$\,cMpc (typical of the early stages of reionization, see Figure~\ref{fig:Db_dist}) with prism spectra requires S/N$\geq$15 per pixel, while $\log_{10}\NHI\simgt20$ can be recovered with S/N$\geq$5 per pixel. Because of the low resolution of the prism we can obtain only upper limits on smaller bubble sizes and column densities. These constraints are vastly improved with higher resolution data, as we will discuss in Section~\ref{sec:disc_future}. In Section~\ref{sec:disc_future} we also discuss prospects for reducing uncertainties in \xHImean with larger samples.
We describe the mock tests and validation of our model-fitting in more detail in Appendix~\ref{app:fitting}.

We plot all individual spectra, their best-fit BEAGLE models, and damping wing fits in Appendix~\ref{app:spec}.

\section{Results}  \label{sec:results}

We first present empirical results from our sample in Section~\ref{sec:res_obs}, finding the redshift evolution around the \lya break shows strong evidence for an increasingly neutral IGM. In Section~\ref{sec:res_fit} we then present our fits to the individual spectra and the inferred evolution of IGM properties.

\subsection{Redshift evolution of $z>6$ spectra}  \label{sec:res_obs}

\begin{figure}
    \centering
    \includegraphics[width=\columnwidth]{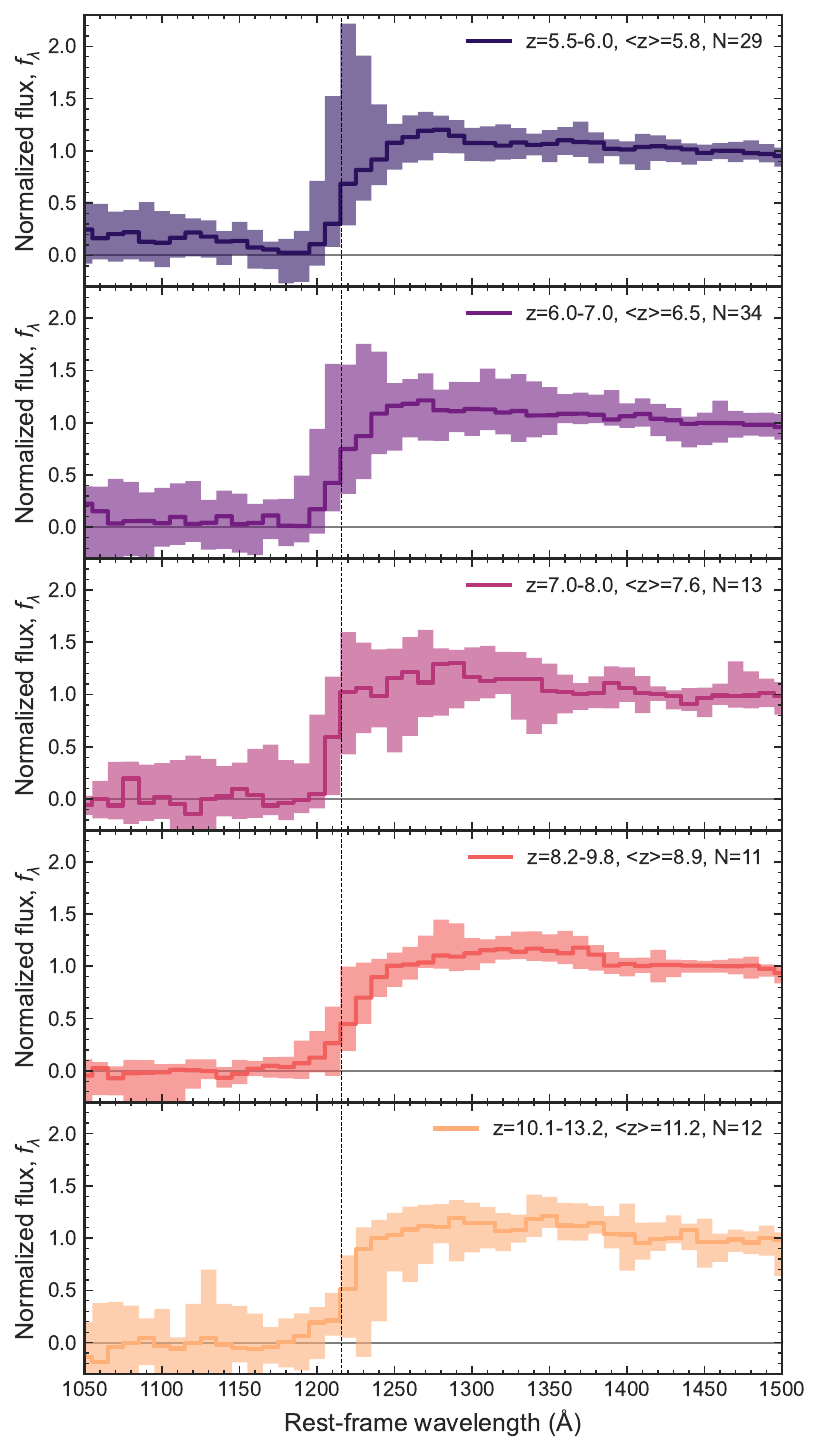}
    \caption{Median stacked spectra in our sample of \nsample\ sources in redshift bins. Shaded regions show the 16-84\% range of the spectra in each redshift bin. 
    We see a clear decrease in flux and the variance of the spectra around the \lya break with increasing redshift.
    }
    \label{fig:spec_stack_med}
\end{figure}

Studies of \lya emission in galaxies with JWST NIRSpec have found a decrease in the \lya EW distribution at $z\simgt7$ \citep{Napolitano2024,Nakane2024,Tang2024c,Jones2025,Kageura2025}, and that strong \lya emission becomes extremely rare at $z\simgt9$ -- with only one EW$\gtrsim15$\,\AA\ \lya detection identified \citep[at $z\approx13$,][]{Witstok2024b}. If this decline in the \lya EW distribution is due to damping wing absorption in an increasingly neutral IGM we should expect a corresponding decrease in the UV continuum redward of \lya.
To see how galaxy spectra evolve with redshift around the \lya break we first consider the evolution of stacked spectra. As we see considerable variance in the spectra, particularly at $z\sim5.5-7$, we then construct a `mean \lya transmission' for each galaxy, $\langle T \rangle$. We demonstrate the redshift evolution is most likely driven by an increasingly neutral IGM. 

We show stacked spectra for our sample in five redshift bins in Figure~\ref{fig:spec_stack_med}. We redshift the spectra to the rest-frame and normalise each spectrum by the median flux density between $1350-1550$\,\AA. We create 100 realisations of each spectrum, sampling from the noise. For the 7 galaxies at $z>10$ with redshift only from the break, in each realisation, we also sample a redshift based on the uncertainty from fitting the \lya break. For sources with optical line detections the typical spectroscopic redshift uncertainty ($\sigma_z\sim 0.002$) is sub-pixel for $z\sim5-14$ \lya breaks (where one prism wavelength pixel corresponds to $\Delta z \sim 0.05-0.15$) thus redshift uncertainties will not add significant uncertainty to the stacks.
We resample all spectra onto a common wavelength grid with pixel size $10$\,\AA\ and then stack in wavelength and redshift bins. 
Our stacks show the median and 16-84\% range of the normalised spectra in each wavelength pixel.

Figure~\ref{fig:spec_stack_med} shows a clear decrease in both the median and variance (the shaded 68\% range) of flux around \lya with increasing redshift. These stacks show: 1) at $z<6$ the 16th percentile range of the stack shows positive flux blueward of \lya \citep[$\sim 1050-1150$\,\AA, though with lower flux closest to line center as predicted due to gravitational infall][]{Laursen2011}, implying the majority of galaxies transmit some flux blueward of \lya, but at higher redshifts the median flux blueward of \lya is consistent with zero. We can also see this excess in individual spectra in Figure~\ref{fig:spec1}. This implies the IGM is not completely optically thick at the \lya resonance at $z<6$ ($\xHImean\simlt 10^{-4}$), as expected from quasar \lya forest observations \citep[e.g.,][]{Eilers2019,Bosman2022}. Recent JWST analyses by \citet{Meyer2025} and \citet{Umeda2025} have quantified this excess flux seen in prism spectra $z<6$ in more detail, and find it is consistent with measurements in quasars; 2) a rapid decrease in strong \lya emission at $z\simgt6$; 3) fully `damped' spectra at $z>8$, consistent with results in a smaller sample by \citet{Umeda2023}.

\begin{figure}
    \centering
    \includegraphics[width=\columnwidth]{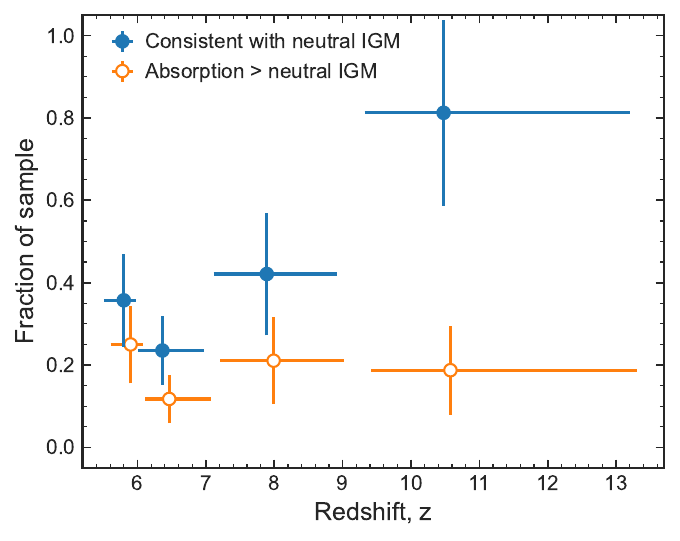}
    \caption{Fraction of sample showing spectra consistent with fully neutral IGM attenuation (blue points) and with absorption stronger than the neutral IGM, i.e. strong DLAs (orange points). While the fraction of strong DLA candidates drops slightly with redshift, the majority of $z>9$ spectra are consistent with a neutral IGM.
    }
    \label{fig:DLA_IGM}
\end{figure}

To assess the relative contribution of local absorption and IGM absorption to the decline in transmission, in Figure~\ref{fig:DLA_IGM} we plot the fraction of our sample with spectra consistent with a neutral IGM, and the fraction of strong DLA candidates (i.e. absorption stronger than the neutral IGM). 
We select sources as consistent with neutral IGM if the observed spectrum around the break is at least 1$\sigma$ lower, in at least 3 consecutive wavelength pixels, than the predicted continuum in an ionized IGM, convolved with the prism resolution, in a fully ionized IGM at the redshift of the source (Section~\ref{sec:fitting}, step 1). 
We select sources as strong DLA candidates if the observed spectrum around the break is $>1\sigma$ lower, in at least 3 consecutive wavelength pixels, than the predicted continuum in a fully neutral IGM at the redshift of the source (Section~\ref{sec:fitting}, step 1 + 5). 
This corresponds to DLAs with $\NHI \simgt 10^{21}$\,cm$^{-2}$ at $z\sim6$ and $\NHI \simgt 10^{21.5}$\,cm$^{-2}$ at $z\sim14$, irrespective of whether the source is an ionized region or not, as the DLA absorption becomes stronger than the IGM alone at these column densities (see Figure~\ref{fig:example_damped} and Figure~\ref{fig:example_spec_DLAs}).
Uncertainties on the fractions are calculated using Poisson statistics.

Figure~\ref{fig:DLA_IGM} shows the fraction of strong DLA candidates is $0.25\pm0.09$ at $z\sim5.5-6$ and $0.19\pm0.11$ at $z>9$, indicating minimal evolution in local absorption systems with increasing redshift which we will discuss further in Section~\ref{sec:disc_DLA}. By contrast, the fraction of spectra consistent with neutral IGM increases significantly from $0.36\pm0.11$ at $z\sim5.5-6$ to $0.81\pm0.23$ at $z>9$. Of the 12 $z>9$ spectra in our sample, only three (jades-1181-3991 (GNz11), ceers-2750-64, and jades-3215-20128771) have spectra showing emission in excess of the prediction for a neutral IGM. 
We further explore some simple physical models for the evolution of the stacked spectra in Appendix~\ref{app:stacks}, finding the evolution is most consistent with the majority of the redshift evolution being driven by the neutral IGM evolution. 

\begin{figure}
    \centering
    \includegraphics[width=\columnwidth]{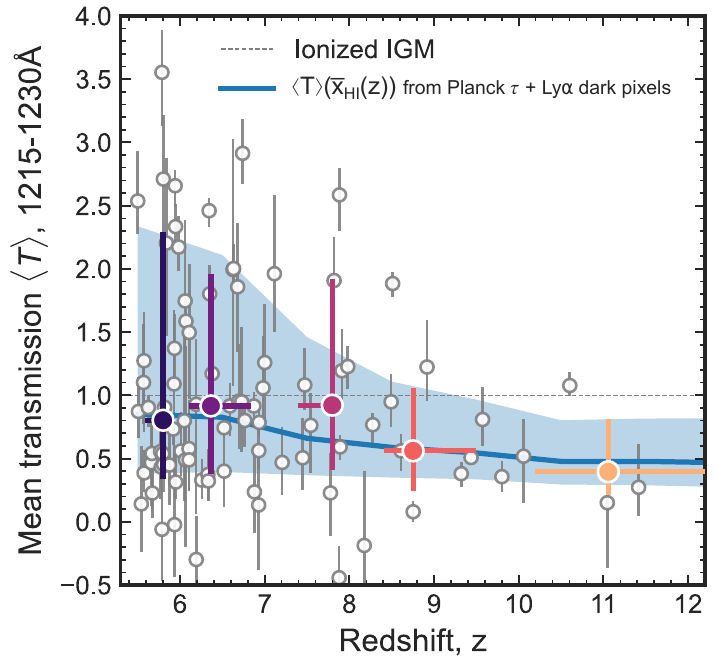}
    \caption{Mean transmission, $\langle T \rangle$ over $1215-1230$\,\AA\ rest-frame for each galaxy in our sample (grey circles), as a function of spectroscopic redshift, including only sources with redshifts from emission lines. $\langle T \rangle$ corresponds to the flux around the \lya break relative to the predicted galaxy continuum in an ionized IGM, with no contribution from \lya emission or absorption by HI. Thus $\langle T \rangle>1$ corresponds to \lya emission, and $\langle T \rangle<1$ to absorption. We show the median and 68\% range of $\langle T \rangle$ in four redshift bins (coloured circles). The blue line and shaded regions shows the predicted median and 68\% range of $\langle T \rangle$ assuming the $z\simlt6$ template spectra described in Section~\ref{sec:res_obs} and the reionization history inferred by \citet{Mason2019b}. The observed median and range are in close agreement with the IGM prediction.}
    \label{fig:T_z}
\end{figure}

To explore the variance we observe in the spectra (Figure~\ref{fig:spec_stack_med}) in more detail, in Figure~\ref{fig:T_z} we show the mean transmission $\langle T \rangle$ over the \lya-break ($1215-1230$\,\AA\ rest-frame, corresponding to a contribution from two wavelength pixels in the prism for all sources in our sample) for each galaxy as a function of redshift. The transmission is calculated as the ratio between the observed spectrum and the continuum model (see step 1, Section~\ref{sec:fitting}, excluding \lya emission, DLAs, or IGM absorption, corresponding to the thick blue lines in Figures~\ref{fig:spec1}-\ref{fig:spec4}) for each source. Using this definition $\langle T \rangle>1$ corresponds to \lya emission, and $\langle T \rangle<1$ is absorption. $\langle T \rangle <0$ corresponds to negative flux in the observed spectra due to noise fluctuations.
We show the median and 68\% range as error bars obtained from 1000 realisations of both the observed spectrum, resampling from the error spectrum, and the posterior for model continuum spectrum (see step 1, Section~\ref{sec:fitting} for more details on the uncertainty in the model continuum), convolved with the resolution of the prism. To calculate the transmission the spectra are rebinned on a common wavelength grid with wavelength pixel 5\,\AA. Because of the sensitivity of this to the precise spectroscopic redshift, we only include sources with redshifts measured from emission lines in Figure~\ref{fig:T_z}. We show the mean transmission for individual galaxies in grey as well as the median and 68\% range in 5 redshift bins.
We find both the median $\langle T \rangle$ and its 68\% range, as shown by the coloured points, decrease with redshift: $\langle T\rangle =0.80^{+1.49}_{-0.47}$ at $z<6$, falling to $\langle T\rangle =0.40^{+0.41}_{-0.19}$ at $z>10$. In particular, we see a strong decline in $\langle T \rangle>1$ for individual sources, which corresponds to a decline in strong \lya emission.

At $z\sim7-8$ the median stacked spectrum at $\approx 1216$\,\AA\ is higher than at all other redshifts (solid line in Figure~\ref{fig:spec_stack_med}) and we see the transmission is comparable to $z\sim6-7$ (Figure~\ref{fig:T_z}). We attribute this to cosmic variance in the IGM. This redshift bin is dominated by the large number of sources (6/11 sources) in the CEERS/EGS
field, which hosts the largest number of \lya-emitters known at $z>7$ and is likely a large ionized region \citep[e.g.,][]{Tilvi2020,Larson2022,Jung2024,Chen2024_fesc,Tang2023,Tang2024c,Napolitano2024}. We discuss the impacts of cosmic variance in Sections~\ref{sec:disc_EoR} and ~\ref{sec:disc_future}.

We compare our observed $\langle T \rangle$ with a prediction for the IGM transmission assuming the median reionization history $\xHImean(z)$ inferred by \citet{Mason2019b}, based on the \citet{Planck2018} CMB optical depth and the \lya forest dark pixel estimates of \xHImean at $z<6$ by \citet{McGreer2015} (blue line and shaded region showing median and 68\% range). To model $\langle T \rangle$ we create template spectra at $z<6$, using the fits to our $z<6$ sample to create high resolution model spectra which include local absorption (see more details in Appendix~\ref{app:stacks}), and apply \lya damping wings drawn from our IGM simulations (described in Section~\ref{sec:dw_sims}) given the IGM neutral fraction predicted as a function of redshift, assuming no evolution in local \NHI, as motivated by Figure~\ref{fig:DLA_IGM} and our analysis in Appendix~\ref{app:stacks}. 
For each template $z<6$ galaxy we draw damping wings to galaxies within 0.2\,mag of its UV magnitude, to account for brighter sources being more likely to be in bigger bubbles. Our prediction on $\langle T(z) \rangle$ is mostly determined by the overall IGM state given by \xHImean, as the dependence on \MUV is sub-dominant \citep{Mason2018}, especially given the median magnitude of the sample does not change significantly with redshift ($\MUV\approx-19.3$ at $z<6$ to $\MUV\approx-19.7$ at $z>8$).
We convolve these to the prism resolution and calculate $\langle T \rangle$ as described for the observed spectra.

We plot the median and 68\% range of the predicted $\langle T(z) \rangle$ as the blue line and shaded region on Figure~\ref{fig:T_z}, where the range is a direct consequence of the size distribution of ionized regions with increasing redshift (Figure~\ref{fig:T_igm_sightline}).
The median and 68\% range of the observed $\langle T \rangle$ closely tracks this prediction for an increasingly neutral IGM, excluding the $z\sim7-8$ bin. In particular, the decline in the variance of $\langle T \rangle$ with redshift is consistent with the expectations for an increasingly neutral IGM: in the late and mid-stages of reionization $\xHImean\simlt 0.5$ ($z\simlt 7$), most observable galaxies reside in ionized regions \citep{Lu2024}, meaning we can still expect to detect strong \lya emission.
In the earliest stages of reionization at $z\simgt9$, ionized regions become too rare and small to transmit significant flux, thus both the median and variance of $\langle T \rangle$ drops significantly \citep[e.g.,][]{Mason2018b}.

These results are consistent with the previous JWST analyses which have found a decrease in strong \lya emission with increasing redshift \citep{Nakane2024,Tang2024c,Jones2025,Kageura2025}, and increase in the strength of the \lya break with redshift \citep{Umeda2023}. Our results are qualitatively consistent with those of \citet{Heintz2024b} who explored the evolution of the \lya break in a larger sample, though with a lower S/N threshold, finding a decrease in \lya emission with increasing redshift and no strong evolution in the abundance of strong DLA candidates.
Our results are also consistent with an analysis of photometry by \citet{Asada2024}, who find an increase in an effective \NHI parameter (combining IGM and DLA damping) of $\sim$1\,dex from $z\sim6-10$, which can be produced by the transition to mostly neutral IGM (Figure~\ref{fig:example_damped}).

Our results demonstrate a clear reduction in the median and variance of flux around the \lya break with increasing redshift, dominated by a decline in strong \lya emission at $z>8$, with no significant evolution in the fraction of strong DLAs with redshift, implying the IGM drives the redshift evolution evolution. The majority of the spectra are consistent with a fully neutral IGM at $z\simgt9$.

\begin{figure*}
    \centering
    \includegraphics[width=\textwidth]{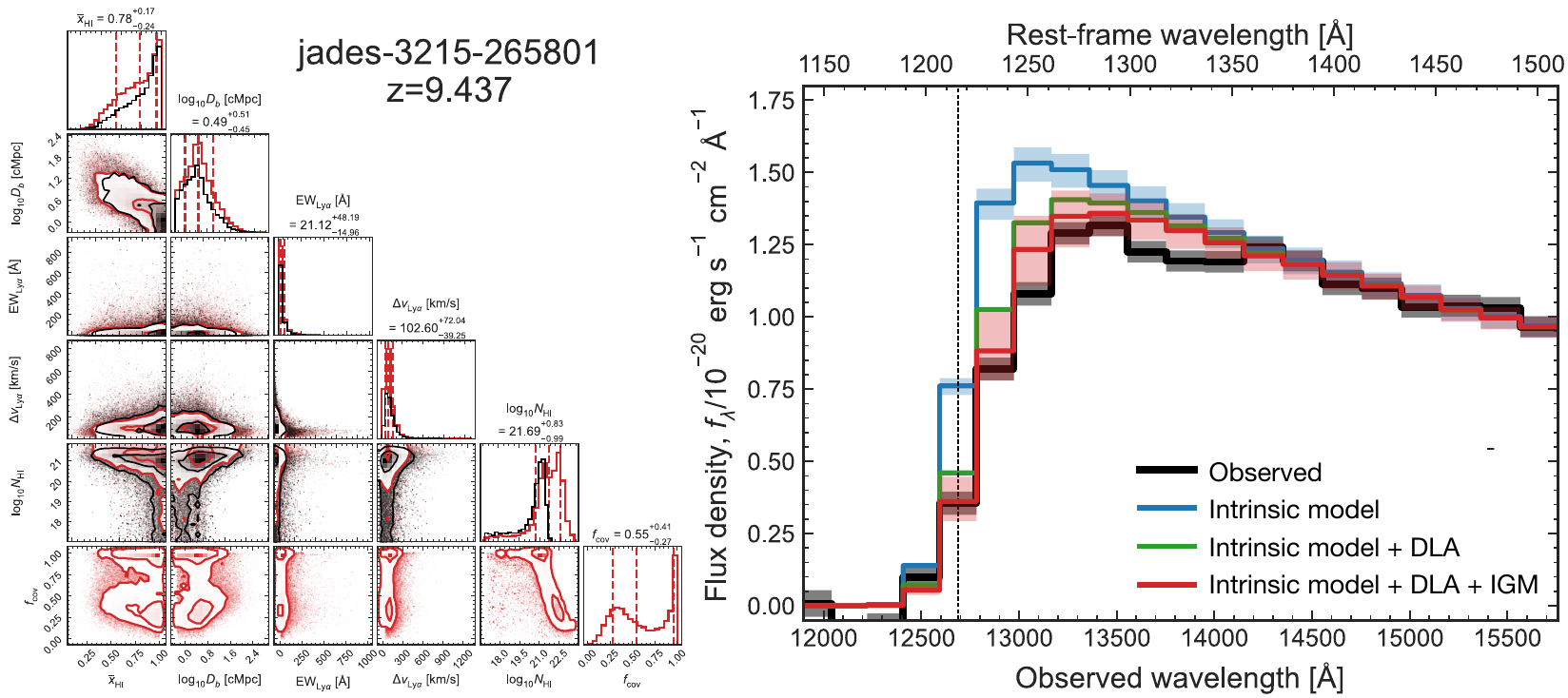}    \caption{Fitting the spectrum of the $z=9.44$ galaxy jades-3215-265801. 
    \textbf{Left:} 2D posteriors for $\xHI, D_b, EW_{\mathrm{Ly}\alpha,\mathrm{emit}}, EW_{\mathrm{Ly}\alpha,\mathrm{obs}}, \DV, \NHI, f_\mathrm{cov}$. 
    Red contours show the posteriors including $f_\mathrm{cov}$ and black contours show the posteriors fixing $f_\mathrm{cov}=1$. We see the IGM and \lya parameters are not sensitive to fixing $f_\mathrm{cov}$, though allowing $f_\mathrm{cov}<1$ will allow higher \NHI solutions.
    \textbf{Right:} Observed spectrum and uncertainty (black line and shaded region) compared with the best-fit `observed' (including absorption by IGM and DLAs), `intrinsic' models (emission only in an ionized IGM), `intrinsic+DLA' models (emission + DLA absorption in an ionized IGM) plotted in red, blue and green respectively. Lines show the median of the models, shaded regions show the 68\% range. 
    This spectrum is consistent with a highly neutral IGM ($\xHImean> 0.74 (1\sigma)$) at $z=9.44$.}
    \label{fig:jades_spectrum}
\end{figure*}

\subsection{IGM constraints from spectral fitting}  \label{sec:res_fit}

\begin{figure}
    \centering
    \includegraphics[width=\columnwidth]{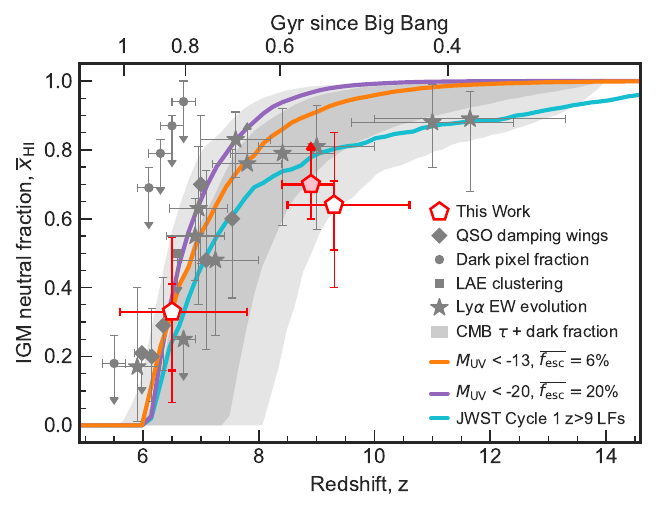}
    \caption{Timeline of reionization: volume-averaged mean hydrogen neutral fraction as a function of redshift. Our new constraints are plotted as red pentagons and error bars, light red error bars include the additional uncertainty due to IGM cosmic variance (Figure~\ref{fig:sightline_variance}). The filled red pentagon is the lower limit we obtain excluding GNz11 from our sample.
    We also plot, as grey points, constraints on \xHImean inferred from observations using inhomogeneous reionization simulations: pre-JWST measurements of the evolution of the \lya equivalent width distribution in Lyman-break galaxies (EW, \citealt{Mason2018,Mason2019c,Whitler2020,Bolan2022,Tang2024c,Kageura2025}), the clustering of \lya emitters (\citealt{Sobacchi2015}), and quasar damping wings \citep{Davies2018b,  Wang2020,Greig2019,Greig2024}; and model-independent constraints from the \lya forest dark pixel fraction \citep{Jin2023}.}
    \label{fig:timeline}
\end{figure}

We now select a sub-sample of our spectra with sufficient S/N to perform robust damping wing fits to obtain more quantitative constraints. Based on fits to mock spectra (see Section~\ref{sec:fitting} and Appendix~\ref{app:fitting}) we select sources with S/N$\geq5(15)$ per pixel where we can obtain robust \NHI($D_b$) estimates. This results in a subsample of \nsamplefitNHI(\nsamplefitxHI) sources with S/N$\geq$5(15), with \nhighzfit\ at $z>9$. We fit each galaxy's spectrum as described in Section~\ref{sec:fitting} and show the resulting fits to individual spectra in Appendix~\ref{app:spec}.

As an example, we show the fit for jades-3215-265801 at $z=9.44$ \citep{Bunker2023b,Curti2024} in Figure~\ref{fig:jades_spectrum}. This is one of the highest S/N ($\approx30$ per pixel) spectra in our sample and shows a clear attenuation around the \lya break relative to the expectation from the $>1500$\AA\ spectral fit (blue line and shaded region showing continuum model uncertainty). 
Our fit recovers a distance from neutral IGM, $\log_{10}D_b/\mathrm{cMpc} = 0.4^{+0.5}_{-0.4}$ resulting in a strong lower limit on $\xHImean> 0.74$ (1$\sigma$). Fixing $f_\mathrm{cov}=1$ we infer a local absorber HI column density $\log_{10} \NHI/\mathrm{cm}^{-2} = 20.8^{+0.5}_{-1.8}$, and obtain $\log_{10} \NHI/\mathrm{cm}^{-2} = 21.7_{-1.0}^{+0.8}$ allowing $f_\mathrm{cov}<1$. 
These results are consistent with the recent analysis by \citet{Curti2024} who did not consider $f_\mathrm{cov}<1$. 
The corner plot highlights there is a degeneracy between $D_b$ and \NHI (and thus also \xHImean), but the spectrum does constrain these parameters.
We show the fit using without including neutral IGM as a green line, showing the local absorber produces too much flux redward of \lya, demonstrating that a high $\xHImean$ (low $D_b$) is required to better explain this spectrum. A higher column density absorber would reduce the flux at line center and be inconsistent with the observed spectrum.
The shape of the 2D posterior for $D_b - \xHImean$ reflects our prior $p(D_b | \xHImean)$ (see Figure~\ref{fig:Db_dist}), whereby a low inferred $D_b$ implies high \xHImean, and vice versa. In this way, damping wing constraints on $D_b$ can be directly linked to \xHImean, propagating the uncertainty on $D_b$ self-consistently. As discussed in Section~\ref{sec:dw_igm} this mapping -- as for all \xHImean estimates using damping wing transmission -- depends on the reionization morphology. However, as demonstrated by \citet{Greig2017,Greig2019}, the inferred \xHImean is not expected to be significantly biased by the choice of reionization morphology. This is because, to first order, \xHImean is the most important factor in determining the reionization morphology, irrespective of the ionizing source model \citep{McQuinn2007b}.
We see the IGM and \lya parameters are not sensitive to our $f_\mathrm{cov}$ prior: the IGM damping impacts redder wavelengths than a DLA and the source shows no hint of \lya emission at the resolution of the prism so we recover our priors.

To estimate \xHImean as a function of redshift from our sample we combine the marginalised posteriors on \xHImean for each galaxy (Appendix~\ref{app:bayes}). We create two redshift bins at $z_\mathrm{bin}=5.5-8,>8$ (containing 8 and 6 galaxies respectively) to obtain $p(\xHImean | \langle z_\mathrm{bin} \rangle)$. We find $\xHImean=0.33^{+0.18}_{-0.27}, 0.64^{+0.17}_{-0.23}$ at $z \sim 5.5-8.0, 8.0-10.6$ ($\langle z \rangle = 6.5, 9.3$). We recover $\xHImean>0.70$ excluding GNz11. The uncertainties include a conservative addition of $\sigma(\xHImean) = 0.1$ to account for sightline variance given we have sampled only 3 fields (see Section~\ref{sec:disc_future}).
Figure~\ref{fig:timeline} shows our inferred timeline of reionization, including an additional uncertainty due to cosmic variance in the IGM (see Section~\ref{sec:disc_future}), along with other estimates from the literature which infer \xHImean using inhomogeneous reionization simulations, based on: the \lya equivalent width distribution in Lyman-break galaxies (EW, \citealt{Mason2018,Mason2019c,Whitler2020,Bolan2022,Tang2024c,Kageura2025}), the clustering of \lya emitters (\citealt{Sobacchi2015}), and quasar damping wings \citep{Davies2018b, Greig2019, Wang2020}; and the \lya forest dark pixel fraction \citep{Jin2023}.
Our results show a clear increase in the inferred neutral fraction with increasing redshift, albeit with large uncertainties.

For comparison, we also show the space of $\xHImean(z)$ allowed by the \citet{Planck2018} optical depth and \citet{McGreer2015} \lya forest dark pixel fraction constraints as inferred by \citet{Mason2019b}, along with three simple reionization history models \citep[following e.g.,][]{Madau1999} which all end around $z\sim6$: (1) integrating the \citet{Mason2023a} UV LF model down to $\MUV < -13$, assuming constant ionizing photon escape fraction of 6\%, (2) only including galaxies down to $\MUV < -20$, assuming constant ionizing photon escape fraction of 20\%, which produces the most rapid reionization; (3) the same as model (1) but fixing the UV luminosity density of the model at $z\geq9$ to approximate JWST UV LF results \citep[e.g.,][]{Donnan2024,Whitler2025}.
We will discuss our results in the context of our understanding of reionization in Section~\ref{sec:disc_EoR}.

\section{Discussion}  \label{sec:disc}

JWST has opened a unique new window on the earliest stages of reionization by providing deep rest-frame UV to optical spectroscopy of $z\sim6-14$ galaxies. In Section~\ref{sec:disc_EoR} we discuss our results in the context of our understanding of reionization.
In Section~\ref{sec:disc_DLA} we discuss the nature of local neutral hydrogen absorption systems and in Section~\ref{sec:disc_future} we discuss future prospects for improving IGM constraints from galaxy damping wings.

\subsection{The reionization process}  \label{sec:disc_EoR}

Our empirical constraints from the stacked spectra and mean transmission (Section~\ref{sec:res_obs}) imply the IGM is approaching almost fully neutral at $z\simgt9$. Our inferred constraints on \xHImean (Section~\ref{sec:res_fit}) also imply a mostly neutral IGM at $z\simgt8$. These results are independent confirmation of previous ground-based efforts to constrain the reionization history at $z>7$ via the damping wing attenuation in quasars \citep{Davies2018b,Wang2020,Greig2024}
 and decline of \lya emission in Lyman-break galaxies \citep{Stark2010,Schenker2014,Mason2019c,Bolan2022}. These results are also in agreement with recent independent analyses of JWST data based on the decline in the \lya EW distribution \citep{Tang2024c,Kageura2025} and damping wings \citep{Umeda2023,Park2024} which also point to a highly neutral IGM at $z\simgt8$. 
 Our approach builds on early JWST damping wing analyses by including additional sources of uncertainties and mapping spectra to \xHImean based on inhomogeneous IGM simulations.

 Mostly strikingly, in Section~\ref{sec:res_obs} we showed the spectra demonstrate a decrease in both the mean and variance of \lya transmission with increasing redshift. We interpret this as due to the decrease in size and variance of ionized regions with increasing redshift, as expected in the earliest stages of reionization \citep[e.g.,][]{Mesinger2007,Iliev2007}. \citet{Tang2024c} also find a decrease in the median \lya EW and variance of the EW distribution with redshift, which likely reflects the same signal. This can be seen as analogous to the decrease in the mean and variance of effective optical depths in the \lya forest at $z\simlt6$ \citep{Eilers2019,Bosman2022} which mark the end of inhomogeneous reionization as the mean and variance in the sizes of neutral regions decrease \citep[e.g.,][]{Keating2019}. Our results provide evidence we are now observing this process in reverse -- probing the earliest stages of reionization.

In Figure~\ref{fig:timeline} we show our \xHImean estimates along with previous constraints and simple theoretical models for the reionization timeline.
At $z\sim5.5-8$ our constraints are fully consistent with pre-JWST constraints from a number of independent probes (quasar damping wings, \lya forest dark pixel fraction, \lya EW distribution, \lya-emitter clustering). At $z>8$ our \xHImean constraint is slightly lower than, though consistent within error bars, the constraint by \citet[][]{Tang2024c} obtained from the \lya EW distribution in 48 $z>8$ galaxies the JWST public archive, the largest sample to-date used to constrain \xHImean at $z>8$.
If we exclude GNz11 our constraint on \xHImean at $z>8$ is a lower limit, $\xHImean > 0.70$ (68\% credible interval), driven mostly by the constraint from jades-3215-265801.
We attribute the difference between our result and that of \citet{Tang2024c} to several factors: our sample is significantly smaller due to our requirement of S/N$>15$ prism spectra at $z>8$ (just \nhighzfit\ sources), meaning we are more subject sample selection; most of our sources return low significance \xHImean constraints due to the moderate S/N; and finally, by fitting both IGM and local HI properties jointly we allow some of the decrease in transmission to be explained by local absorption (see e.g. Figure~\ref{fig:jades_spectrum}).

Upcoming Cycle 3 NIRSpec surveys \citep{Dickinson2024,Oesch2024} will significantly increase the sample of spectroscopically confirmed $z\simgt10$ galaxies to $\simgt 100$. High S/N spectra in these samples will hugely improve our ability to learn about the IGM at these redshifts, both via damping wing approaches as we have described, and the \lya EW distribution.
In the future, with larger samples, it could be most informative to infer distributions of $D_b$, rather than \xHImean, as a function of redshift to different simulations, as this should track the size evolution of ionized regions in a model-independent way and shed light on the morphology of reionization.

The $z\simgt10$ IGM contains key information about early star formation. In particular, models which end around the same time at $z\sim6$ can be driven by very different sources, but diverge at $z>9$, highlighting the importance of constraints on the IGM at these high redshifts. In Figure~\ref{fig:timeline} we show a reionization history corresponding to if the excess in the UV luminosity density observed with JWST holds down to low luminosities, as indicated by deep observations \citep{PerezGonzalez2024,Robertson2024,Whitler2025}. In this case, reionization could start early, and the IGM could be already $\sim20\%$ ionized at $z\sim10$ \citep[see also][]{Gelli2024}. This is interesting to note in relation to recent CMB analyses indicating the electron scattering optical depth may be higher than measured by \citet{Planck2018} \citep{Pagano2020,deBelsunce2021,Giare2024}. As discussed by \citet{Asthana2024}, an early start to reionization is not inconsistent with the requirement from the Lyman-$\alpha$ forest that reionization is complete by $z\sim5.3-6$. Current $z\simgt8$ constraints, the tightest coming from the evolution of the \lya EW distribution \citep{Nakane2024,Tang2024c,Jones2025,Kageura2025}, all imply a mostly neutral IGM at $z>8$ \citep[e.g., $\xHImean = 0.81^{+0.12}_{-0.24}$ at $z\sim8-10$][]{Tang2024c}, but do not yet reach the precision to rule out that the IGM may already be $\sim10\%$ ionized at $z\sim10$. Future observations with large samples of deeper spectra will improve our estimates of \xHImean (see Section~\ref{sec:disc_future}), providing important constraints on the onset of star formation.

\subsection{Nature and evolution of local absorbers}  \label{sec:disc_DLA}

In addition to neutral IGM, our sample demonstrates absorption due to HI gas within, or in close proximity to, the galaxies, as damped \lya absorbers (DLAs). As described in Section~\ref{sec:dw_DLA}, the existence of strong local absorption is not unexpected: HI dominates the volume of most galaxies, indeed the Milky Way disk is $\NHI \sim 10^{22}$\,cm$^{-2}$ \citep{Kalberla2009}, and the massive stars which dominate our spectra are likely to reside in the densest regions of the ISM and experience high HI columns. Evidence for neutral gas in the ISM and CGM of galaxies is observed both in absorption and emission over a wide range of redshifts at $z\sim0-6$ \citep[e.g.,][]{Shapley2003,Steidel2010,Wisotzki2016,Tanvir2019,Pahl2019,Krogager2024}. These observations of both absorption features and \lya emission \citep[whose lineshape is primarily set by \NHI e.g.,][]{Verhamme2015} imply high column densities of neutral gas in the ISM and/or CGM \citep[with median $\NHI \sim 10^{20.3-21.0}$\,cm$^{-2}$ in $z\sim3$ LBGs,][]{Reddy2016}, though likely with non-uniform covering fractions (and low dust sightlines) enabling high EW \lya escape close to systemic velocity in some galaxies \citep[e.g.,][]{Shapley2003,Heckman2001,Du2018b,Hu2023,Tang2024b}.

JWST has extended the detection of DLAs in galaxy spectra to $z>5$ \citep[e.g.,][]{Heintz2023b,Heintz2024,Chen2024_fesc,Hainline2024,DEugenio2023}, providing evidence for some systems with column densities $\log_{10} \NHI\simgt 22$. 
As described in Section~\ref{sec:dw_DLA}, only systems with column densities $\NHI \simgt 10^{22}$\,cm$^{-2}$ become challenging to distinguish from IGM absorption.
Such high column densities may be expected in the regions around young stars, before stellar feedback begins to disperse dense birth clouds. High resolution radiative hydrodynamic simulations predict $\NHI\sim10^{21-23}$\,cm$^{-2}$ in these regions, and that feedback should open low density channels (i.e. a non-uniform covering fraction) and finally disperse the cloud within $\simlt10-40$\,Myr \citep[][]{Kimm2019,Ma2020,Kakiichi2021}, though the feedback mechanisms are still debated. Additionally, considerable opacity may come from the dense filaments and/or clumps in the CGM and local environment of massive halos \citep[e.g.,][]{Rudie2012,Turner2017}.

In the context of reionization, it is important to understand to what extent the opacity due to local HI evolves with redshift and can impact IGM constraints.
We first explore the nature and evolution of \lya opacity due to local HI in our sample. We then discuss the impact of the decreased UV background during reionization on the \lya opacity within ionized regions. 

\begin{figure}
    \centering
    \includegraphics[width=\columnwidth]{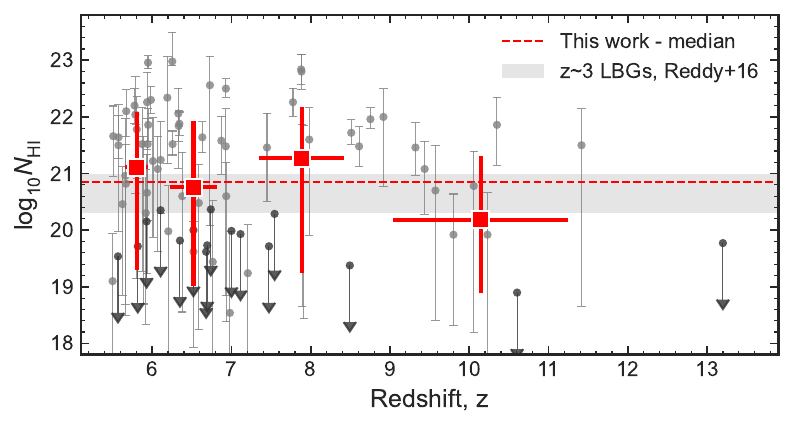}
    \caption{Inferred local HI column density as a function of redshift. Grey points show the median and 68\% range inferred from fits to individual spectra. Red markers show the median and 68\% range in four redshift bins. The grey shaded region shows the range of \NHI estimated in median stacked spectra of $\sim1000$ $z\sim3$ LBGs by \citet{Reddy2016}. 
    }
    \label{fig:N_HI}
\end{figure}

In Figure~\ref{fig:N_HI} we show the inferred HI column densities for our sample, obtained from fitting \nsamplefitNHI\ S/N$>$5 spectra as described in Section~\ref{sec:fitting}, after marginalising over the IGM attenuation. 
We also show the median and 68\% range, obtained from sampling the posteriors of our fits, of \NHI in 4 redshift bins. We find no significant redshift evolution over $z\sim5.5-8$, similar to constraints by \citet{Heintz2023b} and \citet{Umeda2023}, with a slight decrease at $z>8$. This is consistent with our empirical constraint in Figure~\ref{fig:DLA_IGM}.
We find a median $\NHI = 10^{20.8}$\,cm$^{-2}$, comparable to $z\sim3$ LBGs \citep{Reddy2016}, with a broad range consistent with observations from GRB sightlines \citep{Tanvir2019}, which probe the ISM around massive stars.
We find \NHI is somewhat sensitive to the \lya emission prior (Appendix~\ref{app:fitting}), finding median $\NHI = 10^{20.4}$\,cm$^{-2}$ if we use essentially a conditional prior on \lya emission given \NHI, but that the redshift trend is unchanged.

While the median \NHI we infer is similar to $z\sim3$ results, we do find a broad distribution. Consistent with our empirical findings that $\approx20\%$ of sources with breaks stronger than the neutral IGM alone, Figure~\ref{fig:DLA_IGM}), we find 18\% of sources with median $\NHI \simgt 10^{22}$\,cm$^{-2}$, though the uncertainties are large\footnote{We note \NHI can be sensitive to the continuum model, e.g. using a power-law fit to the $>1400$\,\AA\ continuum can result in $\sim1$\,dex higher \NHI than using the BEAGLE continuum model. As described in Section~\ref{sec:fitting}, photoionization models should provide better fits to the UV continuum compared to power-law fits as they include nebular continuum}.
Only 6 sources (8\% of the sample) have 68\% confidence intervals which do not extend below $\NHI < 10^{22}$\,cm$^{-2}$. These sources are: 
jades-1210-13176, which shows both \lya emission and the most extreme damped profile in our sample \citep[this has been previously discussed by][as potentially a nebular continuum dominated source, high \NHI proximate DLA, or AGN respectively. We find it can be fit well using a non-uniform covering fraction (Figure~\ref{fig:spec1})]{Cameron2023,Terp2024,Tacchella2024}; 
two sources in an extreme overdensity at $z=7.88$ in Abell 2744 \citep{Morishita2022b}, previously identified by \citet{Chen2024_fesc}, including one with \lya emission; 
two sources which also show \lya emission (jades-1210-9880 and uncover-4-36755) and `damped' profiles. However, potential absorption is present only in 3 pixels redward of \lya and the fits appear to overestimate the damping (Figure~\ref{fig:spec2}), thus we do not consider these 2 sources robust DLA candidates;
and finally, ceers-P7Pr-1023 ($z=7.78$) which shows a strong damped profile with $\log_{10} \NHI = 22.5\pm0.2$ and no \lya emission. \citet{Tang2023} noted this source is red ($\beta = -0.9$) suggesting significant dust, which is usually correlated with high \NHI at lower redshift \citep{Reddy2016}. The presence of \lya emission plus absorption in 4/6 of these candidates hints at non-uniform covering fractions caused by young stars starting to disperse their birth clouds, or alternative explanations such as strong nebular emission \citep{Katz2024}.

\begin{figure}
    \centering
    \includegraphics[width=\columnwidth]{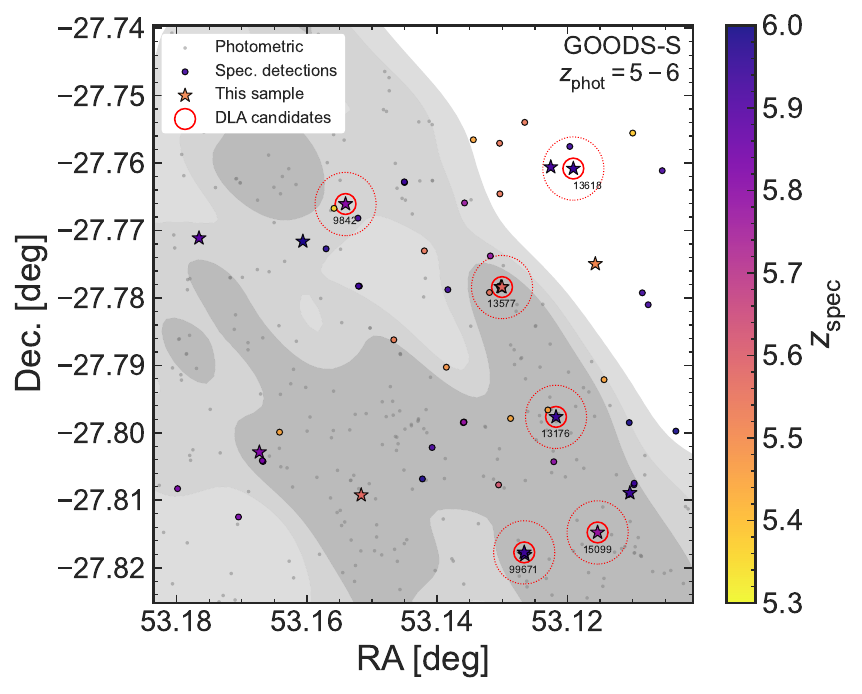}
    \caption{
    The $z=5.5-6$ subset of our sample in GOODS-S. We highlight our sample with stars and additional spectroscopically confirmed sources at $z=5-6$ from JADES and FRESCO as coloured points. Photometric candidates from JADES with $z_\mathrm{phot}=5-6$ are show as grey points, with grey shading marking their density distribution.
    DLA candidates (as described in Section~\ref{sec:disc_DLA}) are marked with red circles with projected radii of 30 and 100\,pkpc ($\simlt 16^{\prime\prime}$).
    The DLA candidates are more likely to have nearby neighbours (both in projection and physical distance) compared to sources which do not show strong damping wing signatures.}
    \label{fig:DLA_jades}
\end{figure}

We now consider the possible origins of moderate column density local DLAs ($\NHI \simgt 10^{21}$\,cm$^{-2}$) in our sample. We examine the $z<6$ sub-sample, where the IGM damping wing impact is expected to be minimal. We use a simple selection of DLA candidates following the approach described in Section~\ref{sec:res_obs}. We select sources where the observed spectrum around the break is at least 1$\sigma$ lower than the BEAGLE predicted intrinsic continuum (Section~\ref{sec:fitting}), including uncertainties in both the observed spectrum and predicted continuum, in a fully neutral IGM at the redshift of the source in at least 3 consecutive wavelength pixels, corresponding to DLAs with $\NHI \simgt 10^{21}$\,cm$^{-2}$ (Figure~\ref{fig:example_damped}). This selection is consistent with the results of Figure~\ref{fig:N_HI} but allows us to individually inspect every source, as some spectra show artifacts which could bias the damping wing fits.

We first examine whether there is a correlation between DLA candidates and dust attenuation, which could indicate absorption by dense gas in the source galaxy. 
\citet{Reddy2016} found that high neutral hydrogen column densities and covering fractions correlate with reddening by dust in $z\sim3$ Lyman-break galaxies, which would be expected if high dust fractions trace high gas fractions. Apart from ceers-P7Pr-1023, we do not find our DLA candidates have significantly higher dust attenuation, based on our BEAGLE fits, compared to our full sample. However, the majority of our sample have very low dust attenuation. If the absorbing gas is located in the ISM this suggests low metallicity or low dust-to-gas ratios \citep[consistent with the declining UV beta slopes at $z>7$ observed with JWST,][]{Topping2024,Cullen2024,Morales2024}, see also \citet{Tacchella2024}.
This is consistent with recent observations by \citet{Tang2024} of \lya line profiles at $z\sim5-6$, finding strong \lya-emitters (EW $>40$\,\AA) have high \lya velocity offsets from systemic (median $\DV \approx 230$\,km/s), implying scattering in $\NHI \sim 10^{19-20}$\,cm$^{-2}$ gas with a high covering fraction and low dust opacity \citep{Laursen2009,Verhamme2015}.

We also consider whether \lya absorption is enhanced in close associations ($\simlt 500$\,pkpc) of galaxies, which should trace the most massive halos ($M_h\sim10^{11-12}\Msun$, $R_\mathrm{vir}\sim25-50$\,pkpc) and protoclusters, where hydrodynamic simulations predict both an increased prevalence of filaments and dense neutral gas in the CGM, and higher gas mass in the ISM which could provide high opacities from star-forming regions \citep[e.g.,][]{Stern2021,Tortora2023,Gelli2025} and $z\sim2-3$ observations suggested enhanced absorption \citep[e.g.,][]{Turner2017}.
\citet{Chen2024_fesc} recently discovered three sources in an association of $>10$ galaxies within $r \simlt 60$\,pkpc at $z\approx7.9$ in the Abell 2744 field \citep[consistent with a protocluster forming in a $M_h\simgt4\times10^{11}\Msun$ halo,][]{Morishita2022b} show strong \lya absorption. 
We test this hypothesis more systematically in GOODS-S which has the highest density of spectroscopy of any field observed by JWST to-date. We focus on $z_\mathrm{spec}=5-6$ as the spectroscopic samples are largest here and the IGM damping wing should be minimal. In Figure~\ref{fig:DLA_jades} we show the positions of our sample, highlighting DLA candidates (defined as above) with red circles, along with spectroscopically confirmed galaxies from JADES and FRESCO \citep{Oesch2023,Tang2024,Meyer2024,Covelo-Paz2024}.

We see the majority of DLA candidates have close neighbours both in projection and 3D. We find all 6 DLA candidates in GOODS-S have close spectroscopically confirmed foreground neighbours ($\simlt 12^{\prime\prime}$, corresponding to impact parameter, $b \simlt 75$\,pkpc). In 5/6 cases these neighbours are offset in redshift by $\Delta z \simlt 0.01$. This corresponds to 3D separation $<500$\,pkpc, thus likely to be physically associated \citep[][]{Chiang2017} and/or could act as proximate absorbers.
Furthermore, several of these 5/6 DLA candidates have multiple close neighbours in 3D: 
jades-1210-13577 ($z=5.575$) has one neighbour within a 3D radius of 200\,pkpc ($z=5.573$) and sits directly behind (impact parameters $<10$\,pkpc) a close association of three sources at $z=5.567$; 
jades-3215-99671 has three neighbours within 500\,pkpc, including one with impact parameter $\approx 8$\,pkpc;
jades-1210-13176 also has three neighbours within 500\,pkpc, including one with impact parameter $\simlt 1$\,pkpc;
The only DLA candidate without close neighbours in 3D, jades-1210-15099 ($z=5.777$) has a foreground source with an impact parameter of 75\,pkpc, but the redshift of the foreground source ($z\approx5.1$) is too low to to be physically associated or to act as a proximate DLA (we find a best-fit proximate DLA would lie at $z\approx5.72$). There are no obvious spectral features which distinguish this source from the other DLA candidates.
Of the 12 sources in our sample at this redshift range with no strong DLA signature, only 6/12 have neighbours within $<500$\,pkpc.

This adds increasing evidence that strong DLA systems are associated with more massive halos.
However, the prism resolution (spectroscopic redshift uncertainty typically $\sigma_z\sim 0.002$) means the uncertainty in line-of-sight distance is $\simgt100$\,pkpc: better characterising these environments will require $R\simgt1000$ spectroscopy.
We should then expect the prevalence of strong DLAs to decrease at higher redshifts as halos assemble hierarchically. We see tentative evidence for this in Figures~\ref{fig:DLA_IGM} and \ref{fig:N_HI}, but larger samples will be required to confirm this.

Finally, we discuss the impact of increased opacity due to lower column density absorbers in the ionized IGM. Observations of the \lya forest have revealed the UV background photoionization rate drops by a factor $\sim10$ at $z\sim5-6$ \citep{Becker2013,Gaikwad2023,Davies2024}, as expected at the end stages of reionization before ionized regions fully merge. Hydrodynamical simulations predict a corresponding increase of Lyman-limit and sub-DLA absorption systems in ionized regions at $z\sim5-6$ \citep[$\NHI \sim 10^{17-19}$\,cm$^{-2}$,][]{Bolton2013,Nasir2021} as the lower UV background reduces the density threshold for self-shielding.
However, a substantial increase in DLAs ($\NHI > 10^{20.3}$\,cm$^{-2}$) is not predicted as those systems are already dense enough to self-shield. 
As we showed in Figure~\ref{fig:example_damped}, $\NHI < 10^{20.3}$\,cm$^{-2}$ absorbers are subdominant to the neutral IGM damping wing at $\simgt1220$\,\AA.
An increase in sub-DLAs and LLS can suppress \lya emission \citep{Bolton2013,Weinberger2019}, however \citet{Mesinger2015} demonstrated evolution in the neutral IGM dominates the opacity, assuming \lya is offset by $\DV\simgt200$\,km/s from systemic, where the damping wing from sub-DLAs is minimal.
At $z\simgt5$ such high offsets appear common, even in strong \lya emitters: with \citet{Tang2024} finding a median $\DV\approx230$\,km/s in strong \lya emitters (EW $> 40$\,\AA).
Thus we do not expect our results to be significantly impacted by an increase in opacity in the ionized IGM.

We conclude that current data suggest no strong redshift evolution of local HI column densities at $z\sim3-13$.
Future deep, high resolution NIRSpec spectra could provide more insights into the location and nature of absorbing gas, by measuring damped \lya troughs to determine the redshift of absorbing gas, detecting low ionization interstellar absorption lines, which trace high HI columns and covering fractions in the host galaxy \citep[e.g.,][]{Shapley2003}, and could also be used to determine the redshift of proximate absorbers \citep[e.g.,][]{Christensen2023,Davies2023}. High resolution NIRSpec spectra would also provide crucial tests for continuum models.

\subsection{Future prospects}  \label{sec:disc_future}

\begin{figure}
    \centering
    \includegraphics[width=\columnwidth]{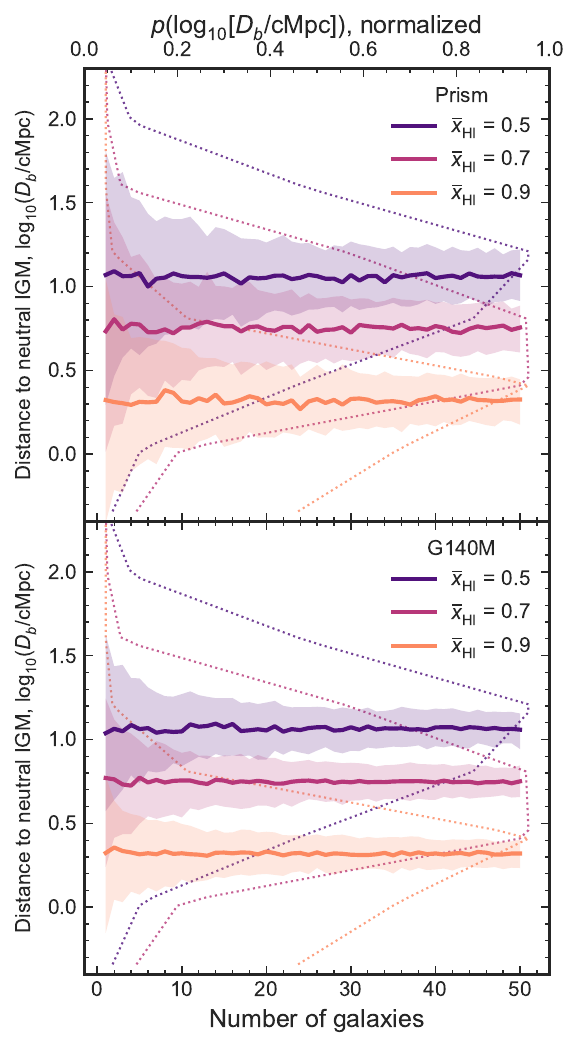}
    \caption{Forecasted number of galaxies required to constrain \xHImean to higher significance with the NIRSpec prism (top) and G140M grating (bottom) modes using our approach. Dotted lines show normalized $p(D_b | \xHI)$ distributions from our simulations at $\xHI=[0.5,0.7,0.9]$. Solid lines and shaded regions show the recovered median $D_b$ and 68\% credible interval from sampling $N$ galaxies each with $D_b$ drawn from these distributions, assuming a median uncertainty on $D_b$ of 0.7 and 0.3 dex for the prism and G140M observations respectively (corresponding to S/N$\approx 20$ and 5 per pixel respectively).}
    \label{fig:Ngals_xHI}
\end{figure}

\begin{figure}
    \centering
    \includegraphics[width=\columnwidth]{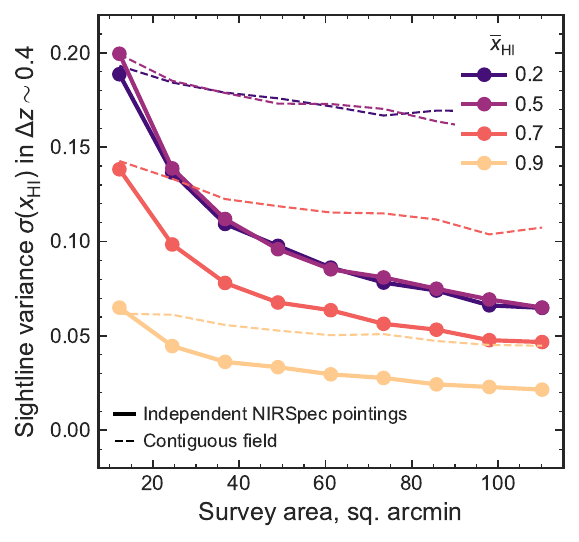}
    \caption{Cosmic variance in the volume-averaged neutral fraction \xHImean measured in sightlines of length 100\,cMpc ($\Delta z\sim0.4$) as a function of survey area and the true neutral fraction, \xHImean. We compare the sightline variance if the area is split between independent NIRSpec fields versus one contiguous field of the same area. Due to the large correlated ionized structures across $\simgt100\,$sq. arcmin, multiple independent fields are required to recover unbiased estimate of \xHImean.}
    \label{fig:sightline_variance}
\end{figure}

With the exquisite spectroscopic capabilities of JWST, the prospects for using galaxies to probe the earliest stages of reionization at $z>9$ are promising.
We first discuss prospects for improving our understanding of the impact of the IGM, \lya emission and local absorbers on prism spectra with higher S/N and higher spectral resolution data, and then discuss the prospects for both overcoming and utilising cosmic variance in the IGM.

Firstly, the most obvious improvements to our approach for fitting IGM damping wings will come from higher S/N and higher resolution spectra. While we see clear evolution in the spectral stacks and evolution of the mean transmission around the \lya break, fitting the damping wings is still challenging for most sources in the public archive given the low S/N of the spectra. In Appendix~\ref{app:fitting} we demonstrate we require S/N per pixel $\simgt15$ to gain unbiased constraints on $D_b$ from prism spectra, which can thus be mapped to constraints on \xHImean providing we sample a sufficient number of galaxies, which we will discuss below. However, even with very high S/N, the prism provides a rather blunt view of the IGM: we find a median uncertainty of $\approx0.5$\,dex on $D_b$ in our tests with S/N$=$100 per pixel. Our knowledge is limited by the $R\sim40$ resolution of the prism at $\sim1\,\mu$m, where the impact of the damping wing is compressed into $\sim5$ spectral pixels (see Figure~\ref{fig:example_damped_Db}). 
Therefore, while prism spectra provide a powerful initial view of the early stages of reionization \citep[e.g.,][]{Curtis-Lake2022,Umeda2023}, a full understanding requires higher resolution spectroscopy and precise constraints on the evolution of \lya emission with redshift \citep[e.g.,][]{Nakane2024,Tang2024c}.

At higher resolution, constraints on $D_b$ using our damping wing fitting approach become much more precise. We demonstrate this in Figure~\ref{fig:Ngals_xHI} where we show the number of galaxies required to constrain $\Delta \xHImean\simlt 0.2$ using either the prism or G140M gratings (see also Appendix~\ref{app:fitting}). 
As described in Section~\ref{sec:dw_igm}, constraints on \xHImean effectively come from measuring the distribution of $D_b$ at a given redshift and comparing it to predictions from simulations (e.g. Figure~\ref{fig:Db_dist}).
To investigate the number of galaxies required to robustly infer \xHImean, we sample $D_b$ from the distributions in our simulations at a given \xHImean (Section~\ref{sec:dw_sims}, shown as dotted lines in Figure~\ref{fig:Ngals_xHI}), and calculate the median $D_b$ we would recover from $N$ mock galaxies using our approach.
For this forecast, we assume S/N$\approx20$ per pixel for the prism and S/N$\approx5$ for the grating. We chose these values based on our tests to mock spectra (Appendix~\ref{app:fitting}), where we recover an average uncertainty on $D_b$ of $\approx0.7$ and $\approx0.3$ dex for the prism and G140M observations respectively. We sample $D_b$ from the simulated $p(D_b \,|\,\xHImean)$ distributions, additionally sampling from the Gaussian uncertainties on $D_b$ given above, for 200 realisations of $N$ galaxies. In this way, our forecasts account for both the measurement uncertainty on $D_b$ and the broad  $p(D_b \,|\,\xHImean)$  distribution predicted by the simulations.
We find G140M observations can constrain $\xHImean> 0.9$ with $\simgt 6$ galaxies, compared to $\simgt20$ galaxies with the prism.
Current grating spectra do not reach this S/N in the rest-frame UV
but a S/N$\approx5$ spectrum for $m_\mathrm{AB} = 26$ source would require $\sim30$\,hr integration in G140M, feasible for the brightest $z>8$ sources.
The increase in precision expected with grating spectra is due to the increased resolution around the \lya break, enabling a better estimate of the IGM damping wing (see Figure~\ref{fig:mock_SNR}).

Ultimately, higher resolution spectroscopy will provide the best constraints on early IGM properties as weak \lya emission can be resolved, which is most sensitive to the IGM opacity in the early stages of reionization (see Figure~\ref{fig:example_damped_Db}). Furthermore, grating spectroscopy will provide important validation of our approach for marginalising over \lya emission in prism spectra and better distinguish the impact of local absorbers.
In the prism the \lya line is spread over most of the pixels including the damping wing feature \citep[see Figure~\ref{fig:example_damped_Db} and][]{Keating2023a,Park2024}. \citet{Jones2024_Lya} and \citet{Chen2024_fesc} showed this limits the minimum detectable \lya EW of $\simgt 50$\,\AA\ for galaxies with our median $\MUV \sim -19$. 
Thus, accurate estimates of \xHImean and \NHI using our approach in prism spectra rely on accurate models, or direct measurements, of the emergent \lya emission (see Section~\ref{sec:fitting}). 
While significant progress has been made since the launch of JWST in linking \lya emission to other observables \citep[e.g.,][]{Prieto-Lyon2023b,Chen2024_fesc,Tang2024}, which we have utilised here, deep NIRSpec grating spectra can easily resolve weak \lya emission \citep[see Figure~\ref{fig:example_damped_Db}, e.g.,][]{Saxena2024}. In high S/N grating spectra ($\simgt5$ per pixel) we can make direct measurements of the UV continuum $\simgt 5$\,\AA\ redward of \lya, without any contamination from the line (and also NV$\lambda$1240 which may be present in sources dominated by young massive stars), and measure absorption troughs and metal absorption lines to more confidently establish the presence of DLAs.

Combining grating and prism spectra will therefore be an important next step to validating our approach \citep[see also][]{Curti2024}, as we can better recover \NHI and \xHImean from prism spectra if the \lya EW is known.
However, the current public sample of sources with robust \lya detections in grating data is still small (just 11 at $z>6.5$), and only 3 sources at $z>9$ have G140M spectra \citep{Tang2024c}.
In addition, high resolution spectra will provide critical tests of our ability to model the continuum in prism spectra. For example, several sources in our sample show relatively flat continua around the break, potentially due to unresolved interstellar absorption features \citep{Boyett2023}, which can be resolved with deep G140M spectra.
Future deep grating surveys will greatly improve our knowledge of the earliest stages of reionization.

Secondly, overcoming the large `cosmic variance' in the IGM will require more independent sightlines \citep[e.g.,][]{Taylor2014,Bruton2023}. This is because the typical sizes of ionized regions are comparable to or larger than both the field of view of JWST and the line-of-sight distance which contributes to the IGM damping wing \citep[$R\simlt10-100$\,cMpc, see Section~\ref{sec:dw_igm}, e.g.][]{Lu2024}. 
Thus, spatially correlated ionized regions impose an uncertainty floor in the neutral fraction which can be measured in a single field with JWST.
We demonstrate this in Figure~\ref{fig:sightline_variance} where we show the standard deviation of the volume-averaged IGM neutral fraction within mock survey volumes: we make 100\,cMpc ($\Delta z \approx 0.4$) skewers with different field areas for a range of \xHImean in our $z=9$ simulations. We compare surveys of multiple independent NIRSpec pointings to contiguous fields. It is clear that independent pointings reduce this sightline variance as $\sqrt{N_\mathrm{fields}}$, whereas the uncertainty remains fairly constant even for contiguous areas $>100\,$sq. arcmin (similar to the CEERS or JADES fields). Specifically, when the cosmic $\xHImean\simlt0.5$ the volume probed in a single 100 sq. arcmin field is expected to have $\sigma(\xHImean) \approx 0.2$ (see Figure~\ref{fig:T_igm_sightline}).

We can clearly see the impact of this cosmic variance in our sample, which covers three fields (GOODS-S, EGS and Abell 2744). For example, the majority of our $z\sim7-8$ spectra (8/13 sources) come from the EGS field observed by CEERS, which is known to be a large candidate ionized region \citep{Tilvi2020,Jung2022,Tang2023,Chen2024_fesc,Napolitano2024}. This is likely the cause of the high mean transmission at this redshift (Figure~\ref{fig:T_z}). Current results are also limited by our small sample size -- this is apparent in our $z\simgt8$ constraints on \xHImean, where the lack of strong damping in GNz11 lowers our \xHImean estimate \citep[see also,][]{Bruton2023b}: overcoming this sample variance will require tens of deep spectra.
Future surveys of more sightlines could exploit the sightline variance itself, as it is related to the typical sizes of ionized regions \citep[e.g.,][]{Lu2024}. 

\section{Conclusions}\label{sec:conclusions}

We have investigated the redshift evolution of the \lya break in \nsample\ $z\sim5.5-13$ galaxies with publicly available JWST/NIRSpec prism spectra in the context of reionization. 
We fit a sub-sample of high S/N spectra using an approach which takes into account \lya emission, local HI absorption and IGM HI absorption using sightlines drawn from realistic inhomogeneous reionization simulations. 
Our main conclusions are as follows:
\begin{enumerate}
    \item We observe a decline in both the mean and variance of flux around the \lya-break with increasing redshift in our sample, demonstrating strong \lya emission is disappearing at $z\simgt7$ and the spectra become increasingly `damped'. We find a median and 68\% range of transmission is $\langle T\rangle =0.80^{+1.49}_{-0.47}$ at $z<6$, falling to $\langle T\rangle =0.40^{+0.41}_{-0.19}$ at $z>10$. We attribute this to the decreasing mean and variance in the size of ionized regions as expected in the early stages of reionization. At $z>9$, $\approx80\%$ of spectra are consistent with a neutral IGM, compared to $<40\%$ at $z<9$.
    \item  We fit spectra to obtain posterior distributions for the distance of galaxies from neutral IGM, the volume-averaged IGM neutral fraction \xHImean, and the local absorber column density \NHI, for each galaxy. 
    We find IGM properties can be reliably recovered using our approach in prism spectra with S/N$\simgt$15 per pixel, though even with the highest S/N the low resolution of the prism limits recovered distance to the neutral IGM to $\simlt0.5$\,dex. We demonstrate this can be reduced substantially with high resolution grating data.
    \item Using \nsamplefitxHI\ sources with sufficient S/N we obtain posterior distributions for \xHImean in two redshift bins. 
    We find $\xHImean=0.33^{+0.18}_{-0.27}, 0.64^{+0.17}_{-0.23}$ (including sightline variance) at
    $z \approx 6.5, 9.3$ ($\xHImean>0.70$ excluding GNz11), providing additional evidence for a mostly neutral IGM at $z>8$, consistent with independent JWST analyses of the \lya EW distribution \citep{Nakane2024,Tang2024c,Jones2025,Kageura2025} and galaxy damping wings \citep{Umeda2023}.
    \item Exploring local HI absorption in our sample, we find a median $\NHI \approx 10^{20.8}$\,cm$^{-2}$, comparable to that observed in $z\sim3$ LBGs \citep{Reddy2016}, with no significant redshift evolution.
    At $z\sim5.5-6$ in GOODS-S, where our sample has high spectroscopic completeness, we find 5/6 sources which show strong DLA absorption signatures have at least one spectroscopically confirmed neighbour within $r_\mathrm{3D}<500$\,pkpc, compared to 6/12 for sources without DLA absorption signatures. This adds to the evidence that strong \lya absorption may be preferentially associated with galaxies in the most massive dark matter halos \citep{Chen2024_fesc}.
\end{enumerate}

The spectroscopic sensitivity and wavelength coverage of JWST/NIRSpec provide a unique opportunity to reveal the earliest stages of hydrogen reionization. Upcoming Cycle 3 surveys \citep{Dickinson2024,Oesch2024} are expected to obtain prism spectra of $\simgt100$ $z>10$ galaxies, providing an unprecedented dataset to constrain the properties of the IGM at $z\sim10-15$ and infer the properties of the faint first galaxies beyond even JWST's detection limits. Fully exploiting JWST observations to understand the early evolution of the IGM using the approach described here will require S/N$\simgt15$ prism spectra and deep, high resolution follow-up studies.

\begin{acknowledgements}
    We thank Sarah Bosman, Fred Davies, Peter Jakobsen, Koki Kakiichi, Kasper Heintz, James Muzerolle, Hyunbae Park, Anne Verhamme and participants of the NORDITA workshop programme “Cosmic Dawn at High Latitudes” for useful discussions.
    This work is based on observations made with the
NASA/ESA/CSA James Webb Space Telescope. The
data were obtained from the Mikulski Archive for Space
Telescopes at the Space Telescope Science Institute,
which is operated by the Association of Universities
for Research in Astronomy, Inc., under NASA contract
NAS 5-03127 for JWST. These observations are associated with program GTO 1180, 1181, 1210, and GO 3215
(JADES; doi:10.17909/8tdj-8n28), 
ERS 1345 and DDT 2750 (CEERS; doi:10.17909/z7p0-8481), and 
GO 2561 (UNCOVER). 
The authors acknowledge the JADES, CEERS, and UNCOVER teams led by Daniel
Eisenstein \& Nora Lüetzgendorf, Steven
L. Finkelstein, Pablo Arrabal Haro, and Ivo Labbé \& Rachel Bezanson for developing their observing programs.

CAM acknowledges support by the European Union ERC grant RISES (101163035), Carlsberg Foundation (CF22-1322), and VILLUM FONDEN (37459). Views and opinions expressed are those of the author(s) only and do not necessarily reflect those of the European Union or the European Research Council. Neither the European Union nor the granting authority can be held responsible for them.
    TYL acknowledges support by VILLUM FONDEN (37459). The Cosmic Dawn Center (DAWN) is funded by the Danish National Research Foundation under grant DNRF140.
    This work has been performed using the Danish National Life Science Supercomputing Center, Computerome.
\end{acknowledgements}

%
%
\bibliographystyle{aa}
\bibliography{library}{}

\begin{appendix}

\section{Comparison with analytic damping wings} \label{app:ME98}

As described in Section~\ref{sec:dw} there is significant sightline variance in the IGM during reionization, thus the assumption of a uniform IGM can bias the recovery of $\xHImean$, as the relationship between $\xHImean$, $D_b$ and the transmission is not deterministic. This has been previously discussed in detail by \citet{Mesinger2008} and we demonstrate this effect with our simulations here.

\begin{figure}[]
    \centering
    \includegraphics[width=0.8\columnwidth]{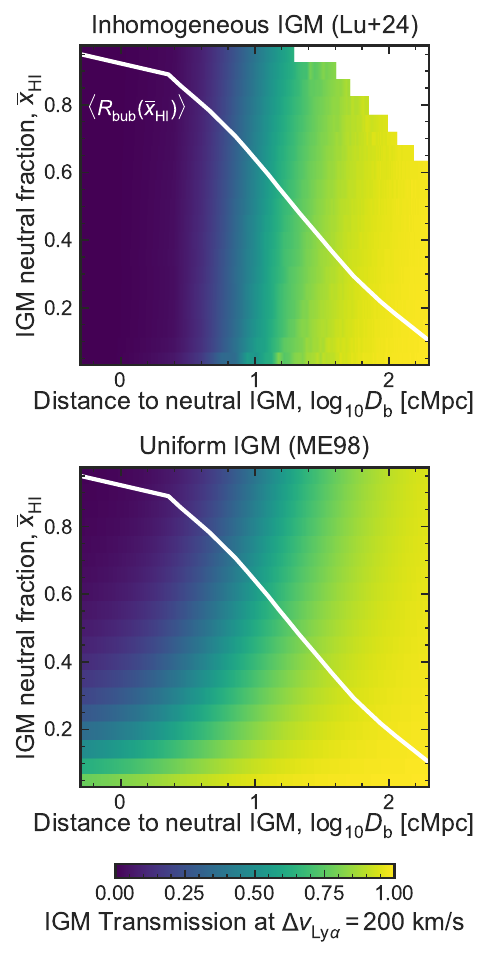}
    \caption{\lya damping wing transmission at $\DV = 200$\,km/s as a function of the distance of a galaxy from the nearest neutral patch and the IGM mean neutral fraction \xHImean.
    \textbf{Top:} The median transmission using inhomogeneous IGM simulations \citep{Lu2024} \textbf{Bottom:} Transmission calculated using the analytic calculation by \citet{Miralda-Escude1998b,Dijkstra2014}. White lines show the mean bubble size as a function of \xHImean predicted by \citet{Lu2024}. Using the realistic simulations, it is is clear the distance to the first neutral patch dominates the transmission compared to the average IGM neutral fraction, \xHImean.
    }
    \label{fig:Tcompare_ME98}
\end{figure}

For every galaxy in our grid of simulations spanning $\xHI\in[0,1]$ at $z=9$, we calculate the distance to the first neutral region, $D_b$, and the \lya transmission 200\,km/s redward of \lya line centre from the IGM damping wing (calculated over each galaxy's sightline as described in Section~\ref{sec:dw_igm}).
In the top panel of Figure~\ref{fig:Tcompare_ME98} we plot the median transmission at $+200$\,km/s redward of \lya, taking the median of galaxies in bins of $D_b$, as a function of \xHImean. In the lower panel we plot the \lya transmission 200\,km/s redward of \lya line centre predicted by the \citet{Miralda-Escude1998b} uniform IGM approximation, also as a function of $D_b$ and \xHImean. We also show the mean bubble size predicted in the simulations as a function of \xHImean.
The transmission obtained from the inhomogeneous IGM simulations demonstrates the damping wing optical depth depends most strongly on $D_b$, with little dependence on \xHImean \citep[see also,][]{Mesinger2008,Chen2024_DW,Keating2023b}.

We see the uniform IGM approximation works least well when both \xHImean and $D_b$ are low, $\xHImean\simlt0.5$, and $D_b<30$\,cMpc. 
Under the uniform IGM assumption, a galaxy a short distance from a neutral patch in a mostly ionized IGM is predicted to have very high transmission $\sim100\%$, as the approximation assumes the neutral patch is $\ll 100\%$ neutral, decreasing its optical depth, when in reality the optical depth should be higher.
The uniform IGM approximation will thus overpredict \lya transmission in this case, and thus lead to overestimates in \xHImean.
The sensitivity of the damping wing transmission to the distance of galaxies to neutral gas is clear motivation for using a realistic IGM simulation as a prior for $p(D_b \,|\, \xHImean)$. At a given redshift, the inferred distribution of $D_b$ provides information about \xHImean.

\section{Transmission profiles} \label{app:DLAs}

\begin{figure}[h]
    \centering
    \includegraphics[width=0.8\columnwidth]{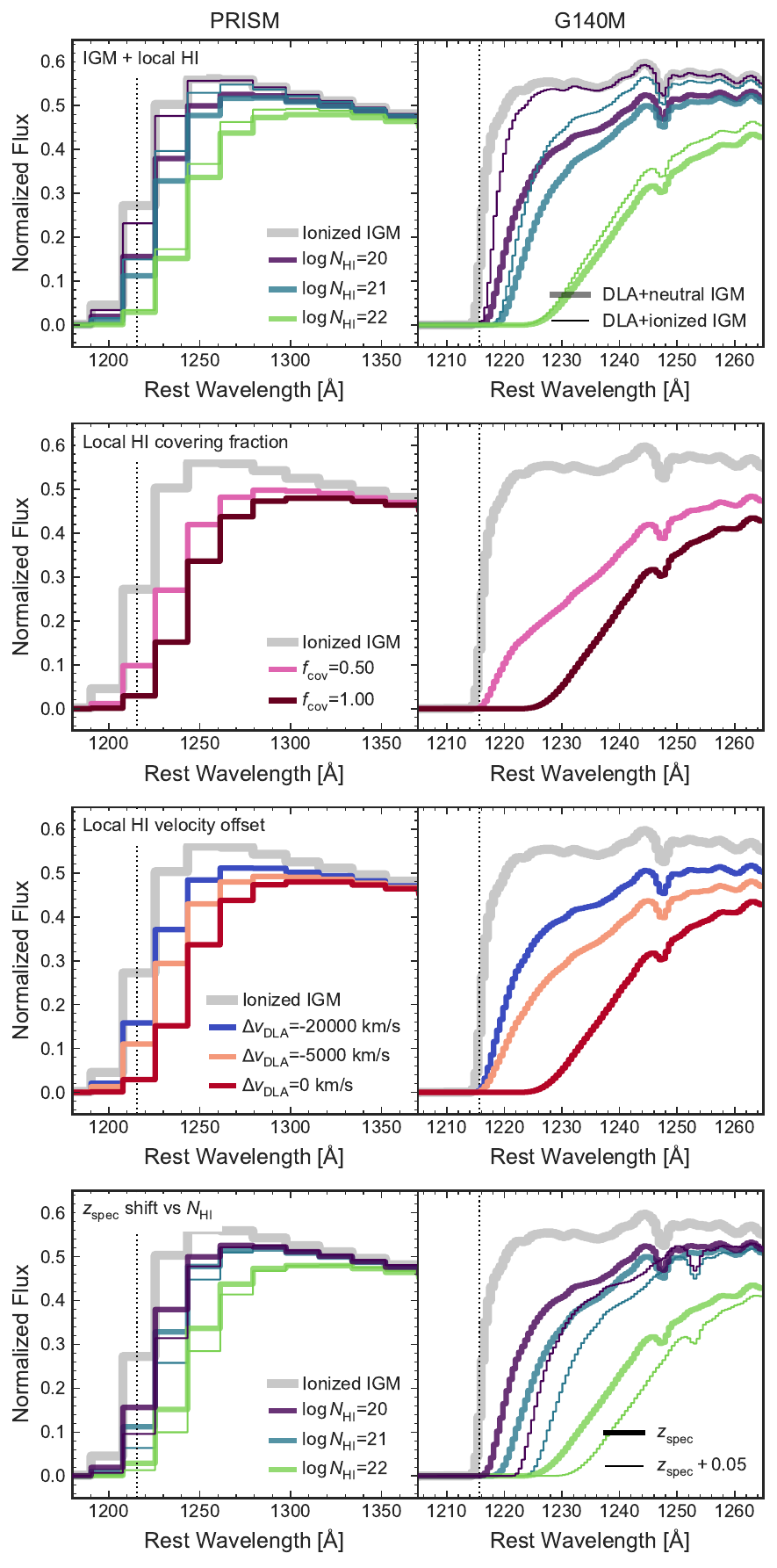}
    \caption{Example spectra ($f_\lambda$), convolved to the resolution of the NIRSpec prism (upper panels) and G140M grating (lower panels), showing the impact of a non-uniform covering fraction of local absorbers, $f_\mathrm{cov}$ (left panels);  if the absorber is located at a lower redshift than the emitting source, but with a peculiar velocity, $\Delta v_\textsc{dla}$ towards the source (central panels); and if the source is at a slightly higher redshift than the estimated $z_\mathrm{spec}$ (right panels; i.e. if the redshift can only be measured from the break, not emission lines), highlighting some of the degeneracies between these effects at the resolution of the prism.
    }
    \label{fig:example_spec_DLAs}
\end{figure}

Here we describe some additional effects which impact \lya transmission profiles, illustrated at the resolution of NIRSpec prism and G140M grating in Figure~\ref{fig:example_spec_DLAs}: (1) the HI column density \NHI; (2) the covering fraction of local HI $f_\mathrm{cov}$; (3) a proximate absorber along the line of sight; and (4) galaxies for which a precise spectroscopic redshift cannot be measured from emission lines. In all plots the grey solid line shows the input mock spectrum at $z=10$ without any attenuation from DLAs or the IGM.

The top panel of Figure~\ref{fig:example_spec_DLAs} shows the impact of local HI absorbers both with and without attenuation from the neutral IGM (thick vs thin lines) as a function of \NHI (coloured lines). At fixed \NHI, the neutral IGM produces more attenuation at redder wavelengths than local absorbers alone, meaning these can be distinguished with sufficient S/N and resolution.

The second panel shows the transmission due to local HI gas, including a non-uniform covering fraction $f_\mathrm{cov}$. In this case the transmission is given by \citep[e.g.,][]{Rivera-Thorsen2015}:
    \begin{equation} \label{eqn:T_DLA}
        \mathcal{T}_\mathrm{DLA}(\Delta \lambda) = 1- f_\mathrm{cov}\left(1-e^{-\tau_\mathrm{DLA}(\Delta \lambda)}\right).
    \end{equation}
Where $\tau_\mathrm{DLA}$ is given by Equation~\ref{eqn:tau_DLA}. Here we show an example with $\NHI = 10^{22}$\,cm$^{-2}$, applying a fully neutral IGM to the damped cases. Reducing the covering fraction increases the transmitted flux redward of \lya line centre.

The third panel also shows a $\NHI = 10^{22}$\,cm$^{-2}$ local absorber, but where the absorbing gas is not located in the emitting source but has a peculiar velocity, $\Delta v_\textsc{dla}$, which also increases transmission redward of \lya line centre. 
At the resolution of the prism, cases with non-zero covering fraction, $f_\mathrm{cov}\sim0.5$ and large peculiar velocities, $\Delta v_\textsc{dla} \sim -5000$\,km/s, can produce very similar spectra, introducing a degeneracy. As discussed in Section~\ref{sec:disc_DLA} we consider the local non-uniform covering fraction a likely more physical picture of the local absorption.

In the bottom row of Figure~\ref{fig:example_spec_DLAs} we show that even without a precise spectroscopic redshift from emission lines it should still be possible to get information about the IGM, but that there is a degeneracy between \NHI and redshift. We show that at the resolution of the NIRSpec prism, in a fully neutral IGM, a source at redshift $z_\mathrm{spec}$ with a local absorber column density of $\log_{10}\NHI \simlt 21$ has an almost identical transmission profile to a source with $\log_{10}\NHI \simlt 20$ but at $z\approx z_\mathrm{spec}+0.05$. This is because the DLA removes flux very close to line center. However, as the neutral IGM reduces the flux at $>1240$\,\AA\ more strongly than the DLA, we should still be able to recover some information about the IGM in either case (i.e. the coloured lines are significantly different from the grey line). This means we are still able to use sources at $z>10$, even without a spectroscopic redshift determination from e.g. [OIII] emission lines.

\section{Evolution of stacked spectra} \label{app:stacks}

\begin{figure*}
    \centering
    \includegraphics[width=0.9\textwidth]{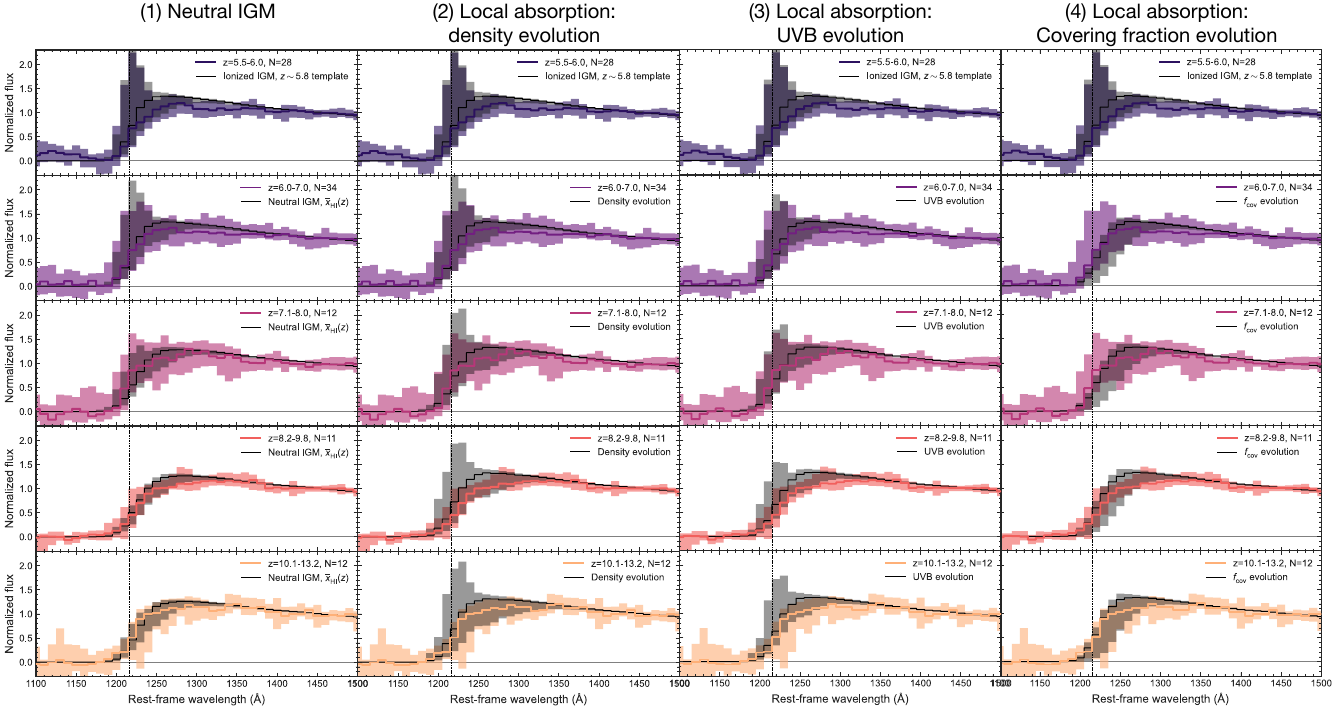}
    \caption{Median stacked spectra in redshift bins as in Figure~\ref{fig:spec_stack_med}. At $z<6$ (top panel) we show the median and 68\% range of our template spectra in a fully ionized IGM, convolved to the prism resolution (grey line and shaded region). At $z\geq6$ we show the predicted median and 68\% range of spectra assuming four IGM/CGM evolution models (grey line and shaded region, described in Appendix~\ref{app:stacks}): 
    (1) an increasingly neutral IGM following the \citet{Mason2019b} reionization history; 
    (2) DLA column densities increase with increase cosmic density, $\NHI \propto (1+z)^2$;
    (3) the same density evolution and at $z>6$ a minimum $\log_{10}\NHI > 20.3$ due to the lower UV background;
    (4) assuming the same density evolution and that $f_\mathrm{cov}=1$ at $z>6$.
    The spectra are most consistent with the majority of the evolution being driven by the IGM evolution.}
    \label{fig:spec_stack_models}
\end{figure*}

To gain intuition into the redshift evolution of stacked spectra (Figure~\ref{fig:spec_stack_med}) we explore four simple physically motivated models.
Because of the low resolution of the prism, this requires using high-resolution $R\simgt1000$ spectral templates, applying the models, and then convolving with the prism resolution. As no $R>1000$ templates for the faint $z\sim5-6$ galaxies in our sample exist yet, we construct them from fits to our $z<6$ sample as described in Section~\ref{sec:fitting}, naturally including any local absorption in the ISM and CGM of galaxies at $z\sim5.5-6$. We take samples from the posteriors for each $z<6$ galaxy (29 galaxies) to generate template spectra with a resolution of $R\sim4000$ to which we can apply transmission models.

We show the median and 68\% range of these templates, normalised and convolved to the resolution of the prism, as the grey line and shaded region in the top panels of Figure~\ref{fig:spec_stack_models}. 
We then apply four simple models and describe their predictions for the evolution of the spectra:
\begin{enumerate}
    \item \textbf{Neutral IGM:} We apply damping wings drawn from our IGM simulations (Section~\ref{sec:dw_sims}) to the templates, assuming $\xHImean(z)$ predicted by \citet{Mason2019b}.
    This model predicts a decrease in both the mean flux and variance around \lya as ionized regions become too rare and small to transmit significant flux \citep[e.g.,][]{Mason2018b}. In Figure~\ref{fig:spec_stack_models}, we see our stacks agree qualitatively very well with this model.

    \item \textbf{Local absorber density evolution:} Assuming DLA column density increases with increasing cosmic density, $\NHI \propto (1+z)^2$. This assumes that most of the evolution is due to an increase in density in the CGM, and that the CGM is mostly neutral at $z\sim6$ \citep[as predicted by some hydrodynamic simulations, e.g.,][]{Stern2021}.
    This model predicts it would still be possible to observe strong \lya emission at $z>8$, producing spectra which are inconsistent with the observations at $z>8$ where we do not detect strong \lya emission. This is because at $z<6$ many spectra require low column densities ($\NHI <10^{19}$\,cm$^{-2}$) to explain the strong \lya emission in this bin, and the $(1+z)^2$ evolution will not produce enough opacity to significantly damp \lya in all sources at $z>8$. Thus we do not consider pure density evolution in the CGM a primary driver of the observed evolution in the prism spectra.

    \item \textbf{UV background evolution:} Assuming an increase in LLS and sub-DLAs ($17.2 < \log_{10}\NHI < 20.3$),
    within ionized regions at $z\simgt6$ as the UV background drops before ionized regions merge \citep[as predicted by hydrodynamical simulations][]{Bolton2013,Rahmati2015,Nasir2021}.
    To explore this we impose a minimum $\log_{10}\NHI > 20.3$ in the local absorber model, but note this likely overestimates the importance of absorbers. 
    This model is also not a good match to the observations for two reasons: firstly, the evolution in self-shielded systems is expected to occur rapidly at the end of reionization \citep[e.g.][]{Nasir2021}, so we would expect a sharp increase in absorption systems with redshift at $z\simgt6$. In our data we see strong \lya emission at $z\sim5-7$, implying the spectra cannot be fully explained by a rapid evolution in self-shielding systems at the end of reionization. Secondly, the predicted continuum in this model at $z>8$ is higher than the observed spectra, as not all sources have high enough column densities to significantly damp the continuum, implying additional neutral IGM attenuation is still needed (see Figure~\ref{fig:example_damped}).

    \item \textbf{Covering fraction evolution:} 
    We assume an extreme model where the covering fraction of local HI $f_\mathrm{cov}=1$ at $z>6$. This is motivated by some hydrodynamical simulations showing an increase in covering fractions in the CGM with increasing redshift \citep{Rahmati2015,Tortora2023}. 
    However, we note $f_\mathrm{cov}$ depends strongly on feedback prescriptions and resolution in simulations \citep{Faucher-Giguere2016,vandeVoort2019}. 
    This model underpredicts the observed spectra at $z\sim6-7$ (because it predicts strong \lya is all absorbed), but overpredicts the observed spectra at $z>8$ (for the same reason as in the UVB evolution case). A gradual increase in $f_\mathrm{cov}$ with redshift could contribute to some of the observed evolution of the stacks. We will discuss this further in Section~\ref{sec:disc_DLA}.
    
\end{enumerate}

Thus we conclude that, while local absorption is present in the observed spectra over all redshifts, the observations are most consistent with the majority of the redshift evolution being driven by the neutral IGM evolution. 

\section{Bayesian inference setup and priors} \label{app:bayes}

We use Bayesian inference to infer the parameters $\xHImean$, $D_b$ and $\theta_\mathrm{gal}$ ($=\theta_\mathrm{\lya}, \theta_\mathrm{DLA}$) for each galaxy. For $z>10$ sources without spectroscopic redshifts from emission lines we also fit for $z_\mathrm{spec}$, using a Gaussian prior for the redshift based on an initial fit to the \lya break.
The posterior for each galaxy, with observed spectrum $f_\mathrm{i,obs}$ and properties $\phi_{\mathrm{gal},i}$ (\MUV, OIII+H$\beta$ EW) is:
\BEA \label{eqn:like_fi}
p( \xHImean, D_b, \theta_\mathrm{gal} \,|\, f_\mathrm{i,obs}, \phi_\mathrm{gal}) = ( f_\mathrm{i,obs} \,|\, \theta_\mathrm{gal}, D_b, \xHImean) \nonumber \\
\times p(\theta_\mathrm{gal} \,|\, \phi_\mathrm{gal}) p(D_b \,|\, \xHImean, \phi_\mathrm{gal}) p(\xHImean)
\EEA
Here $( f_\mathrm{i,obs} \,|\, \theta_\mathrm{gal}, D_b, \xHImean)$ is the likelihood (Equation~\ref{eqn:likelihood}) as described in Section~\ref{sec:fitting}.

We use a conditional prior $p(D_b \,|\, \xHImean, \phi_\mathrm{gal})$ from our simulations (Section~\ref{sec:dw_sims}), selecting sightlines based on \MUV to account for brighter galaxies being more likely to be in larger bubbles \citep[step 5 above,][though this does not have a large impact on our results]{Mason2018b}. 
Using this prior makes the inference of \xHImean for each galaxy explicitly conditional on $D_b$, thereby propagating uncertainties in the inferred $D_b$ values into the uncertainties on \xHImean.
As the damping wings are relatively independent of \xHImean at fixed $D_b$ (see Appendix~\ref{app:ME98}), the redshift evolution of $D_b$ should provide the most empirical evidence for IGM evolution, regardless of the mapping to \xHI.
At $z\leq6.3$ we use half-Gaussian priors on $\xHImean$ based on the \lya+$\beta$ forest dark pixel fraction constraints by \citet{Jin2023}. At higher redshifts we assume a uniform prior on $\xHImean=[0,1]$.

We use the empirical distributions of \lya EW by \citet{Tang2024} as a prior on the \textit{emergent} \lya EW (i.e. after transmission through the ISM and CGM, calculated after step 4 above), such that at $z\sim5-6$ we should recover the observed \lya EW distribution, and at $z>6$ the observed EW distribution does not exceed that at $z\sim5-6$ \citep[which is reasonable based on NIRSpec grating and ground-based spectra, e.g.][]{Pentericci2014,Mason2019c,jung_texas_2020,Tang2024c}. These distributions are derived from $>700$ $z\sim5-6$ Lyman-break galaxies with ground-based \lya spectroscopy from Keck and JWST photometry. Due to the high resolution ($R\sim4000$) of the ground-based spectroscopy, these EW measurements will not be impacted by local absorption, unlike in the prism where \lya emission and local absorption are blended. Thus, this prior should be informative for recovering the \NHI distribution from the prism.
For sources at $z\simlt 10$, where [OIII]+H$\beta$ is detectable in NIRCam and NIRSpec, we use the EW model by \citet{Tang2024} conditional on [OIII]+H$\beta$ EW (whereby sources with strong [OIII]+H$\beta$ EW are more likely to have strong \lya). For sources without [OIII]+H$\beta$ measurements we use the EW distribution conditional on \MUV. Following \citet{Tang2024} we apply a slit-loss correction of 0.8 to map predicted \lya fluxes based on VLT/MUSE measurements to the NIRSpec slits.
Similarly, we use the \lya velocity offsets model by \citet{Mason2018} as a prior on the \textit{emergent} velocity offset.
We use a uniform prior on $\log_{10}\NHI \in [17,23.5]$, noting that for $\log_{10}\NHI/\mathrm{cm}^{-2} \simlt 20$, unless there is strong \lya emission, at the resolution of the prism we can only obtain an upper limit on \NHI (see Figure~\ref{fig:example_damped}). Motivated by $z\sim3$ results by \citet{Reddy2016} which imply HI covering fractions in the ISM are high ($\simgt90\%)$,we use a prior which is uniform in $\log (1-f_\mathrm{cov})$: $p(f_\mathrm{cov}) = 1/(1-f_\mathrm{cov}+\epsilon$). This avoids non-physical scenarios with very high \NHI and low $f_\mathrm{cov}$.

\begin{figure}[]
    \centering
    \includegraphics[width=\columnwidth]{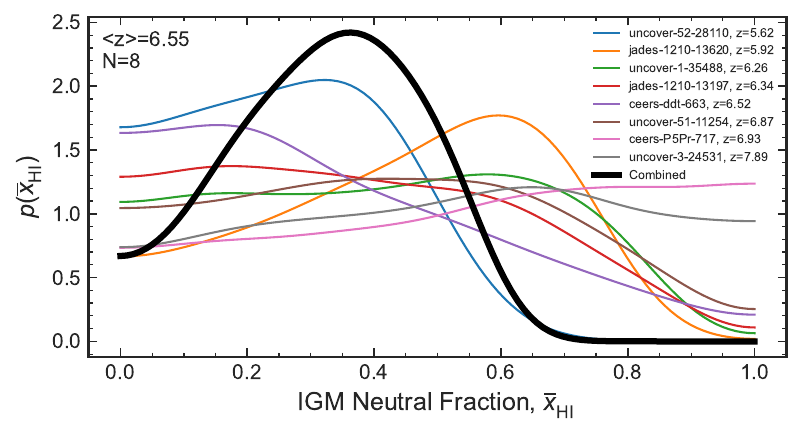}
    \includegraphics[width=\columnwidth]{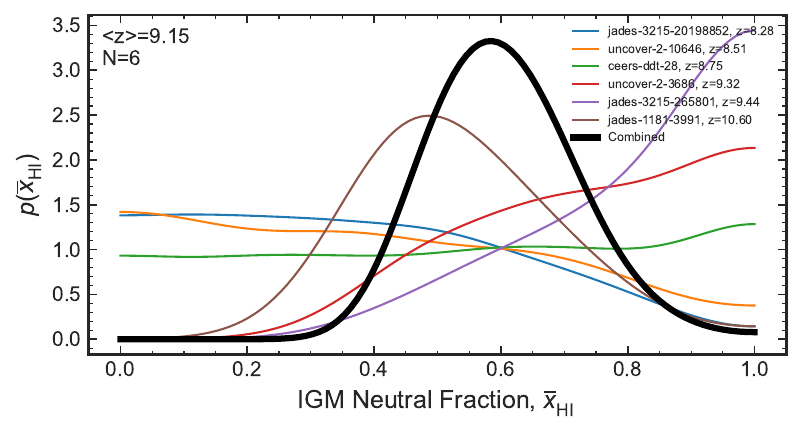}
    \caption{\xHImean posteriors for each galaxy in our high SNR sample (thin coloured lines) and the combined posterior in each redshift bin (thick black line). We note this posterior does not include the additional IGM cosmic variance (Section~\ref{sec:disc_future}) we add to the constraints shown in Figure~\ref{fig:timeline}, see Section~\ref{sec:res_fit} for discussion.}
    \label{fig:pxHI}
\end{figure}

We obtain the posterior for each galaxy, $p( \xHImean, D_b, \theta_{\mathrm{gal},i} \,|\, f_\mathrm{i,obs})$ using Markov Chain Monte Carlo with the \texttt{emcee} sampler \citep{Foreman-Mackey2013}.
We fit over the rest-frame wavelength range $1100-1400$\,\AA, which we find provides the most robust recovery of parameters in mock spectra, and contains no other UV emission lines except NV which is expected to be very weak for stellar populations $\simgt5$\,Myr \citep{Chisholm2019}. 
We use 50 walkers and $\sim10^{4-5}$ steps, such that the chain is $>50\times$ the integrated autocorrelation time for the number of fitted parameters, and discard the first 50\% of the chain. 
To obtain the marginalised $p(\xHImean \,|\, \{ f_\mathrm{obs} \})$ we take the resulting \xHImean samples from each galaxy and fit a smooth function with a Gaussian Kernel Density Estimation.
The final posterior of $p(\xHImean \,|\, \{ f_\mathrm{obs} \})$ at a given redshift is then the product of the individual \xHImean posteriors in each bin (Section~\ref{sec:res_fit}). In Figure~\ref{fig:pxHI} we show the individual \xHImean posteriors for the 14 SNR$>15$ galaxies in our sample, in two redshift bins (see Section~\ref{sec:res_fit}), and their combined posterior.

\section{Validation of fitting} \label{app:fitting}

\begin{figure*}[]
    \centering
    \includegraphics[width=0.9\textwidth]{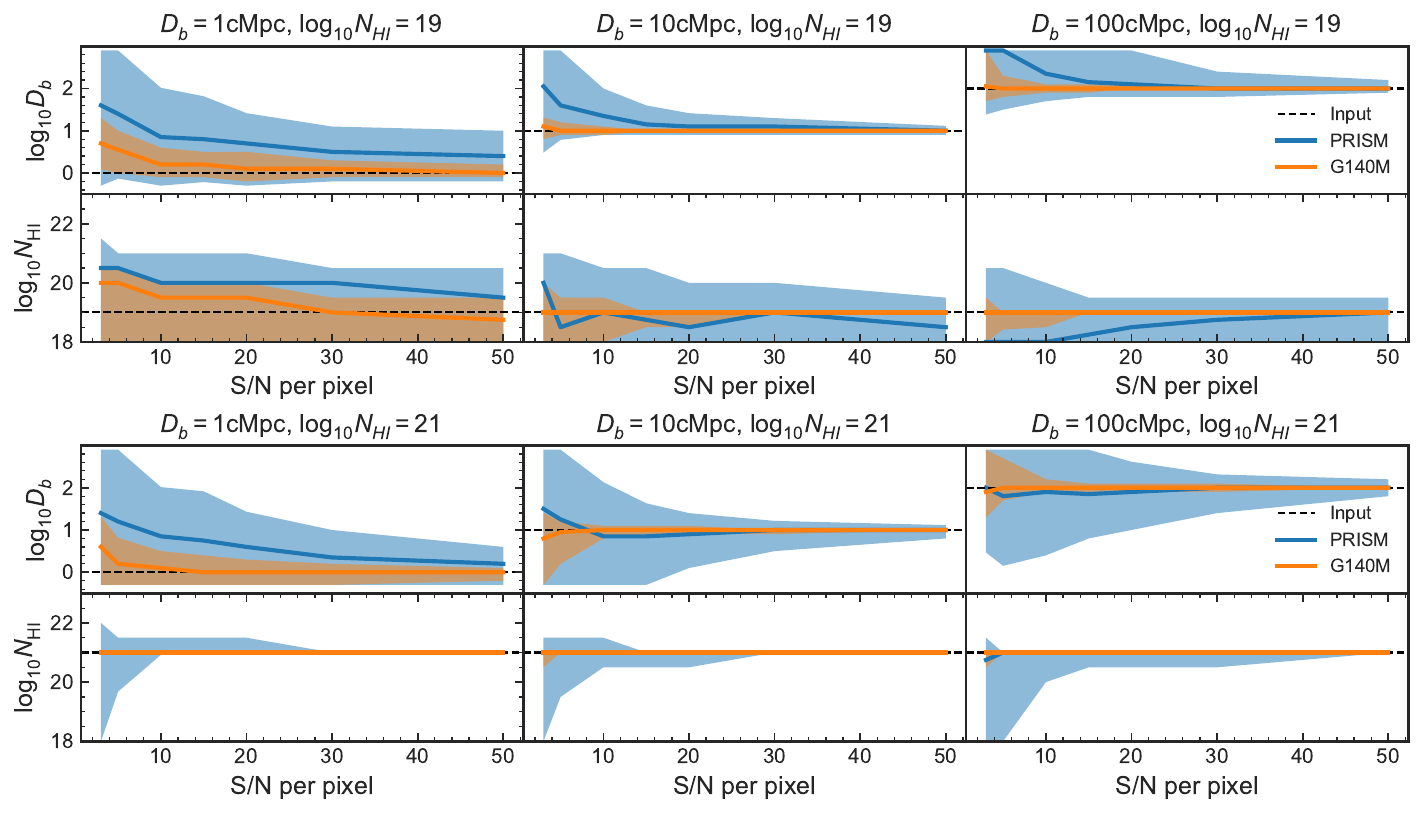}
    \caption{Accuracy of recovered distance to neutral IGM, $D_b$, and local column density, \NHI as a function of S/N per pixel in mock NIRSpec prism and G140M observations. We show the median and 68\% range of the maximum likelihood recovered values from 100 realisations. The low resolution of the prism limits the accuracy of constraints, while robust constraints could be obtained even for low S/N ($\sim5$ per pixel) with G140M.}
    \label{fig:mock_SNR}
\end{figure*}

\begin{figure}[]
    \centering
    \includegraphics[width=\columnwidth]{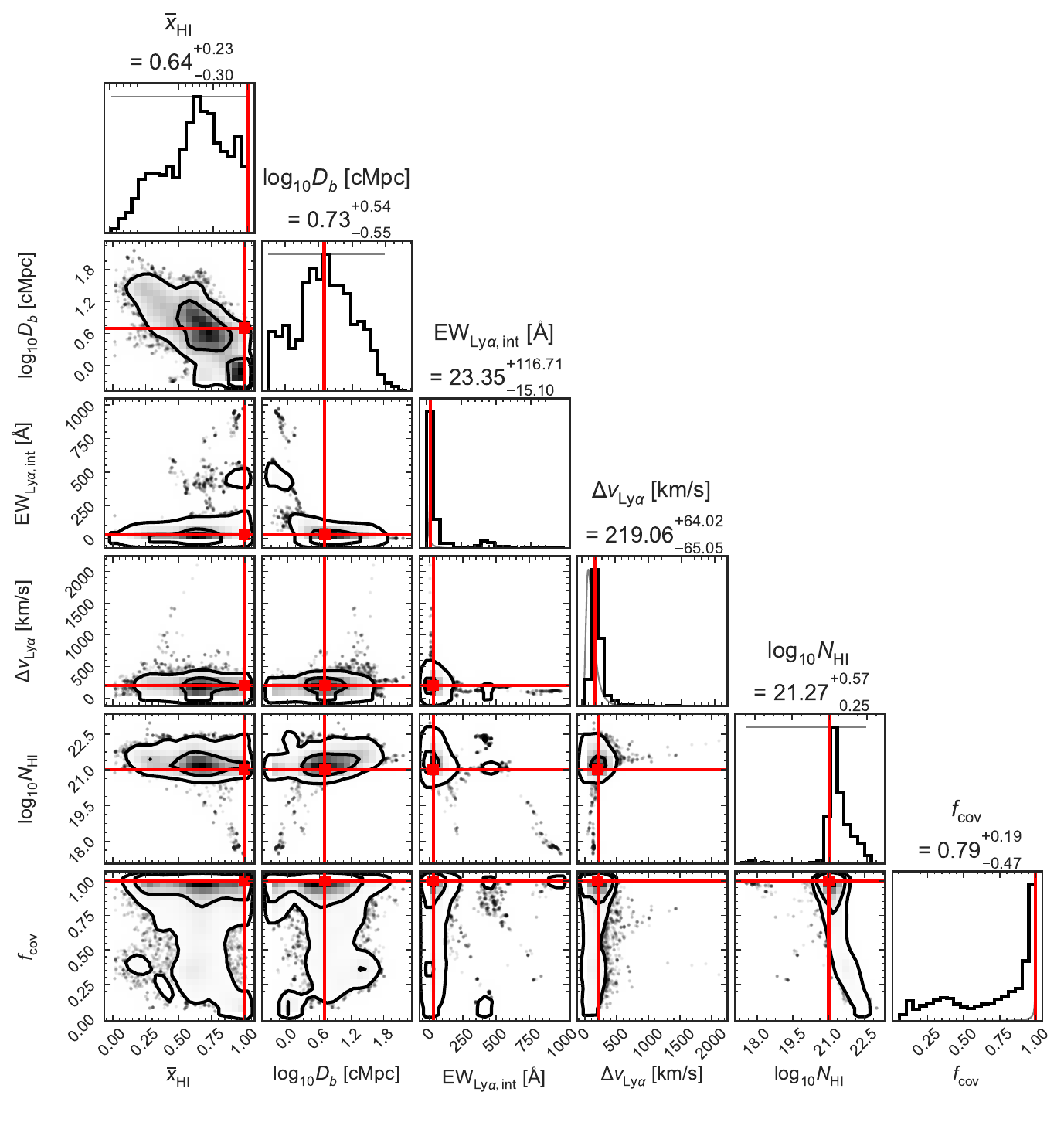}
    \caption{Recovered posteriors for an example mock prism spectrum with S/N$=50$ per pixel. All input parameters (shown as red crosshairs) are recovered well within 2$\sigma$. Our constraint on \xHImean is driven by the prior $p(D_b | \xHImean)$ (Figure~\ref{fig:Db_dist}). As expected from the prior, given our recovered $D_b\approx5$\,cMpc, the most likely $\xHImean\approx0.7$, which drives our constraint here. This demonstrates how accurately recovering \xHImean from prism spectra requires $\simgt10$ galaxies per redshift bin to estimate distribution of $D_b$ (see discussion in Section~\ref{sec:disc_future}.}
    \label{fig:valid_corner}
\end{figure}

We validate our fits using mock data and describe our key results here. As described in Sections~\ref{sec:dw_igm} and ~\ref{sec:disc_future}, and Appendix~\ref{app:ME98}, the IGM damping wing optical depth depends most strongly on $D_b$, and so by essentially estimating the distribution of $D_b$ at a given redshift, we can map to \xHImean \citep[Figure~\ref{fig:Db_dist}, see also e.g.][]{Kist2025}. Thus, we first demonstrate what S/N per pixel is required to robustly recover the most important parameters -- $D_b$ and \NHI. We then show an example of fitting a high S/N mock prism spectrum showing the recovery of the other parameters. In Section~\ref{sec:disc_future} we investigate how many sources are required to accurately recover \xHImean to $\simlt 0.2$.

We generate mock spectra from the BEAGLE fits to our sample as templates, and apply IGM damping wings and DLA optical depths on a grid of $D_b$ and $\NHI$ values. We then add noise, accounting for the covariance between adjacent pixels (Equation~\ref{eqn:cov}), to the model spectra and fit the spectra using the approach described in Section~\ref{sec:fitting}, verifying that input parameters can be recovered well for high S/N spectra.

We demonstrate the impact of S/N on our parameter recovery in Figure~\ref{fig:mock_SNR}. This shows the median and 68\% range of the maximum likelihood recovered values of distance to neutral IGM, $D_b$, and local column density, \NHI as a function of S/N per pixel, from mock spectra. For each S/N value we generate 100 realisations of the flux given the covariance matrix (obtained by rescaling the observed error spectrum for our template source).
We show the recovered parameters for 6 input combinations of $D_b=[1,10,100]$\,cMpc and $\log_{10}\NHI=[19,21]$ for both mock prism observations and G140M observations.

Figure~\ref{fig:mock_SNR} demonstrates that IGM and DLA properties can be robustly recovered from prism spectra, but that there can be large uncertainties in parameters recovered from prism spectra, even for very high S/N spectra ($\sim0.5$\,dex for S/N$\sim50$ spectra). This is due to the low resolution, limiting our ability to distinguish small changes in the shape of the continuum. By contrast, G140M observations promise to provide precise constraints on both $D_b$ and $\NHI$ for spectra with S/N$\simgt5$.

We see S/N$\simgt15$ is required to robustly recover the input $D_b$ from prism spectra, and that for small input bubble sizes, there can be a bias to larger bubble sizes and higher \NHI with low S/N spectra. This is because the shape of the damping wing is mostly insensitive to $D_b \simlt 5$\,cMpc (see Figure~\ref{fig:example_damped_Db}), so negative noise fluctuations will not significantly shift the inferred $D_b$ lower, while positive noise fluctuations will always result in a higher inferred $D_b$. Future work could potentially mitigate this bias by e.g. modifying the likelihood form or using machine learning approaches \citep{Chen2023a,Park2024}. This bias is not present in grating observations due to the higher resolution around the break.
We find \NHI can be recovered well, to within $\simlt 0.2$\,dex, for S/N$\simgt5$ per pixel in both prism and grating spectra. We note that the shape of the damping wing in the continuum as seen in prism is mostly insensitive to $\NHI \simlt 10^{20}$\,cm$^{-2}$, so in those cases we only return upper limits.

In Figure~\ref{fig:valid_corner} we show an example of the recovered posteriors from fitting a mock spectrum. For this example we apply IGM damping wings (using a simulated sightline at $\xHImean\approx1$ with $D_b=5$\,cMpc), a DLA with $\log_{10}\NHI=21.0$ and $f_\mathrm{cov}=1$, and an emergent \lya emission line (i.e. pre-IGM absorption) with EW=$30$\,\AA, $\DV=FWHM=200$\,km/s. We add uncertainties to the spectrum assuming S/N$=50$ per pixel. We see all input parameters can be recovered well within $2\sigma$, though the uncertainties are large, as expected from Figure~\ref{fig:mock_SNR}. This figure demonstrates there is some degeneracy between \NHI and $D_b$, and between \NHI and $f_\mathrm{cov}$, but that a low $D_b \simlt 20$\,cMpc can be clearly recovered even when \NHI is high ($>10^{21}$\,cm$^{-2}$). We see \xHImean is the least well-recovered parameter. As described in Section~\ref{sec:dw_igm}, our recovery of \xHImean depends on estimating the distribution of $D_b$ at a given redshift, so our recovered \xHImean depends on our prior $p(D_b | \xHImean)$. In this example, we see we recover $\xHImean = 0.64^{+0.23}_{-0.30}$. This is consistent with the expectation from Figure~\ref{fig:Db_dist} that 5\,cMpc bubbles are most common when $\xHImean \approx 0.7$. This highlights how observations of multiple galaxies are required to overcome the sightline variance in the IGM and accurately recover \xHImean at a given redshift. We explore this more quantitively in Section~\ref{sec:disc_future}.

\section{Sample and spectra} \label{app:spec}

\begin{figure*}
    \centering
    \includegraphics[width=\textwidth]{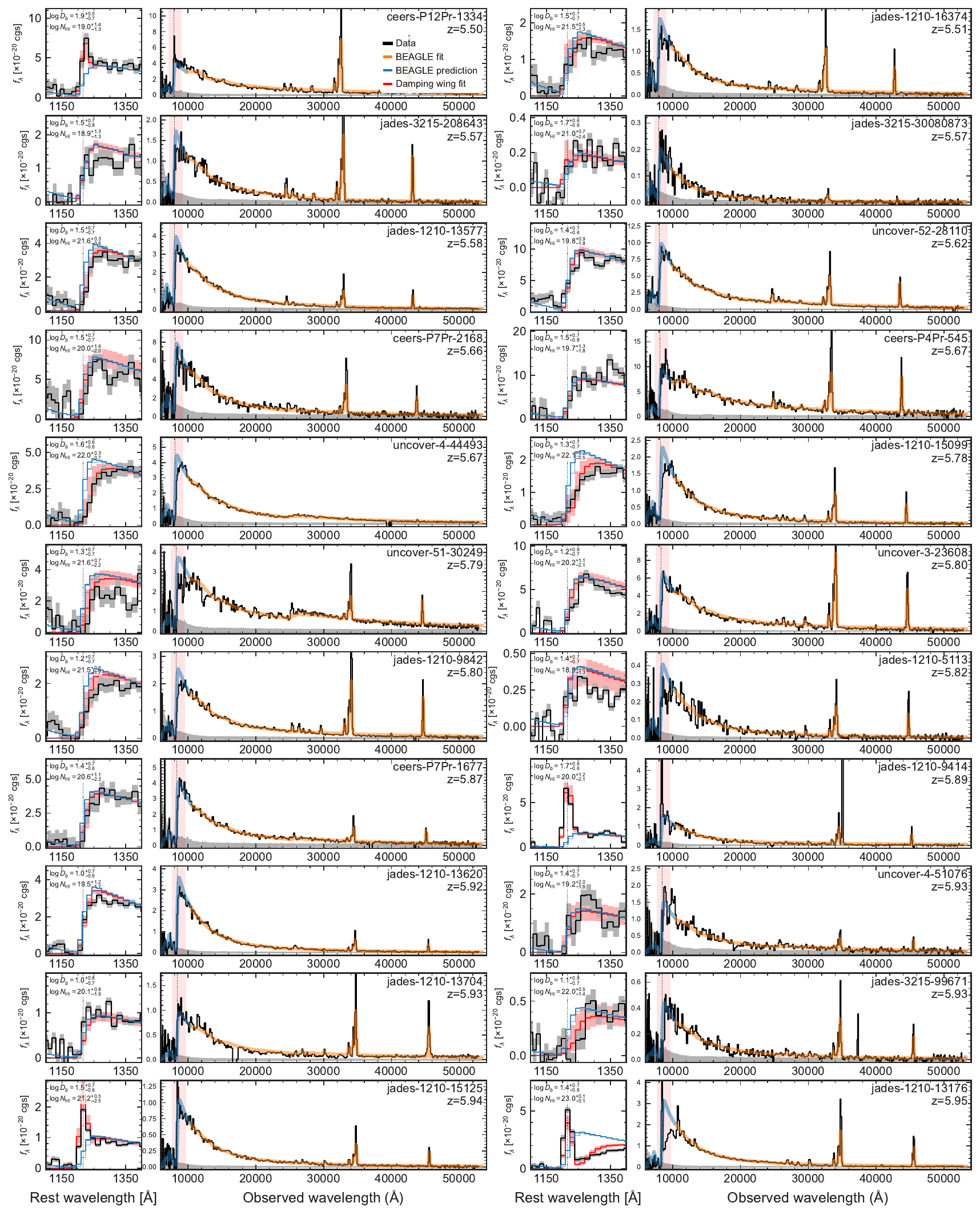}
    \caption{Spectra of our sample in ascending redshift order. We show the NIRSpec prism spectra in black with the error spectrum in shaded grey. The best-fit BEAGLE model is shown in orange, convolved to the prism resolution. The blue lines are the extrapolation of the BEAGLE model to the rest-frame $<1500$\,\AA, which is the predicted continuum spectrum we use for our damping wing fits (Section~\ref{sec:fitting}). Small panels show the damping wing fit zoomed in around 1100-1400\,\AA\ in the rest-frame (marked as the red shaded region on the full spectra). The thick blue lines show the BEAGLE continuum model extrapolation in an ionized IGM, the thin blue lines show the continuum model assuming a fully neutral IGM with no ionized bubble around the source. The red line and shaded region showing median and 68\% range of samples from posteriors.}
    \label{fig:spec1}
\end{figure*}

\begin{figure*}
    \centering
    \includegraphics[width=\textwidth]{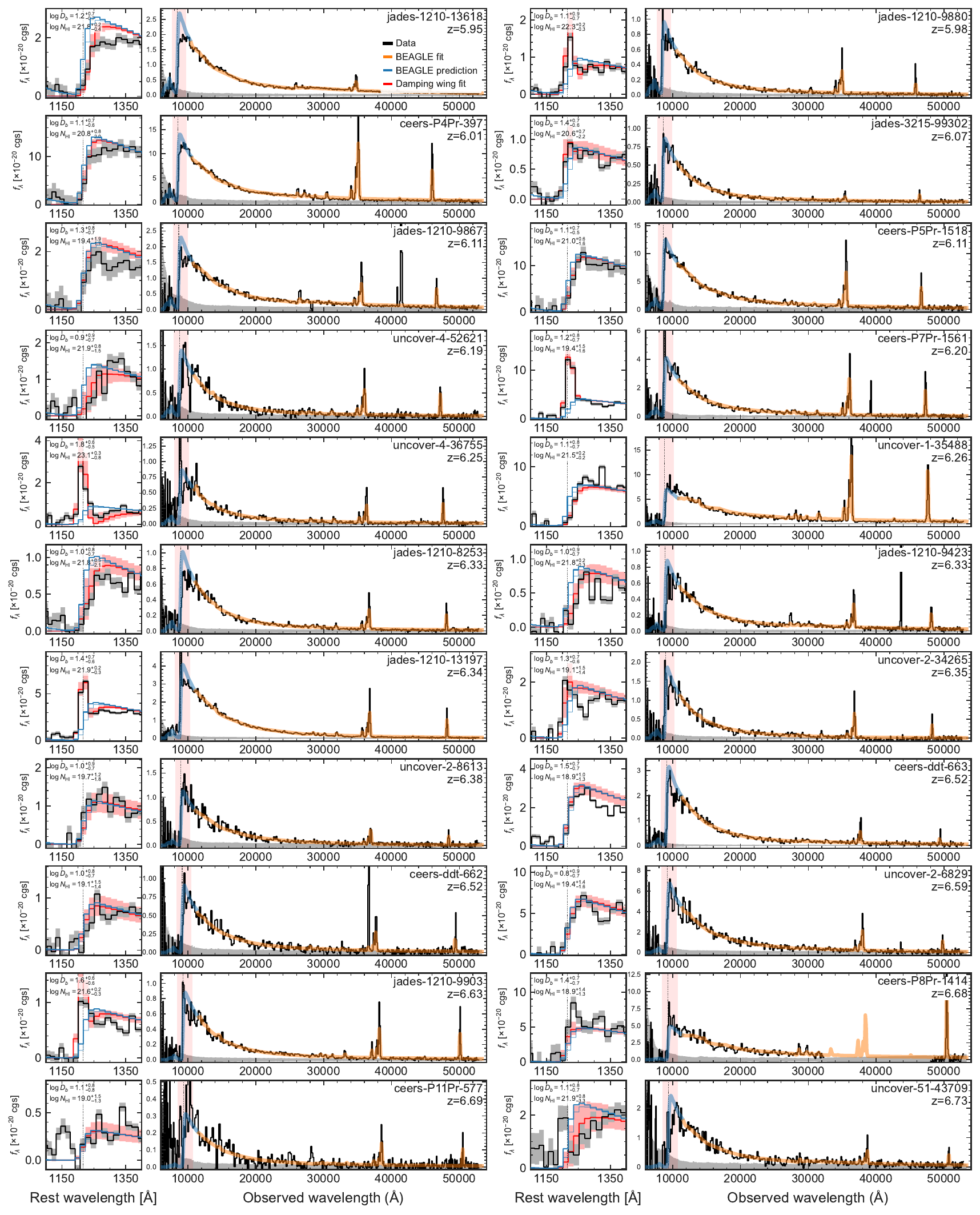}
    \caption{Same as Fig.~\ref{fig:spec1}}
    \label{fig:spec2}
\end{figure*}

\begin{figure*}
    \centering
    \includegraphics[width=\textwidth]{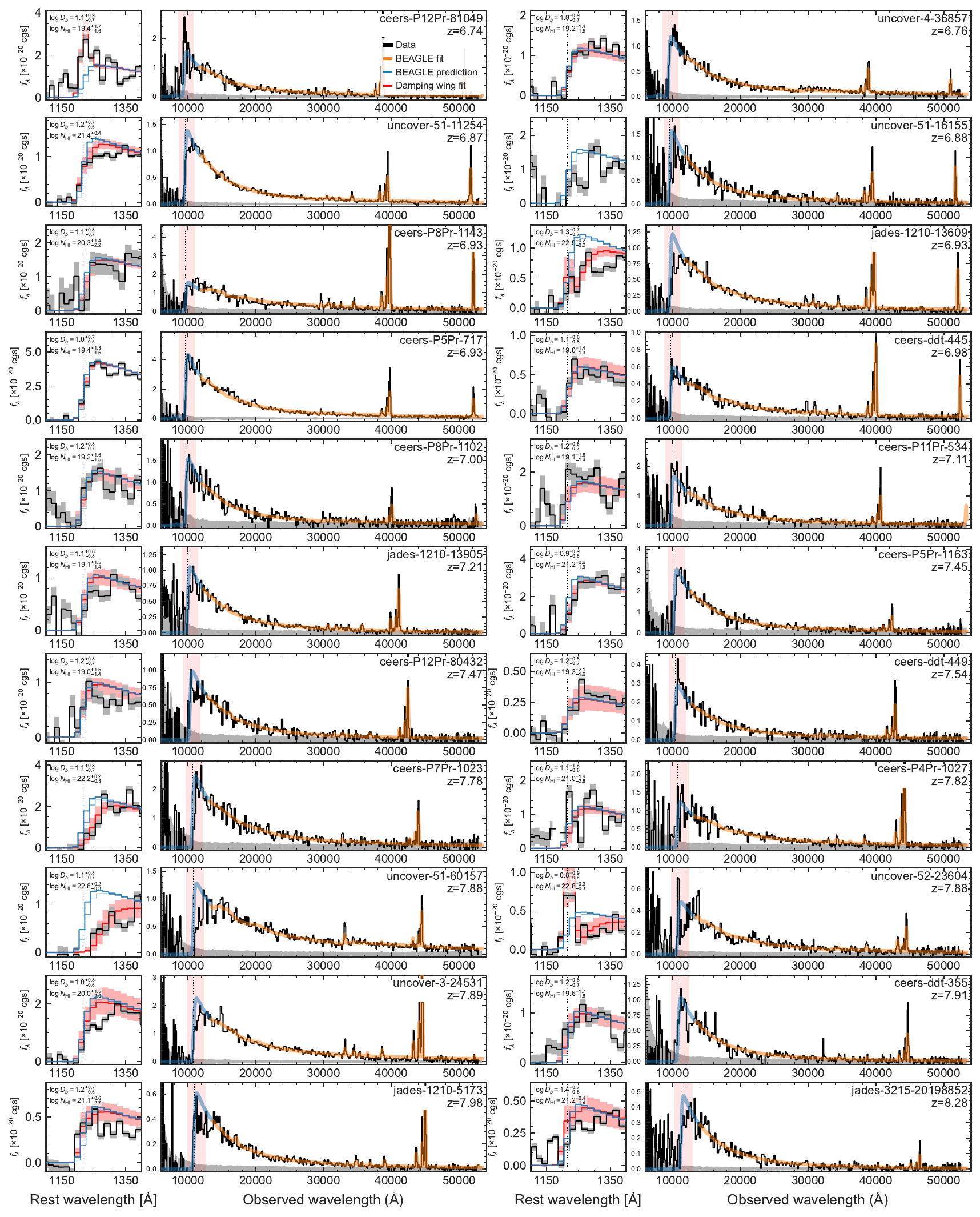}
    \caption{Same as Fig.~\ref{fig:spec1}}
    \label{fig:spec3}
\end{figure*}

\begin{figure*}
    \centering
    \includegraphics[width=\textwidth]{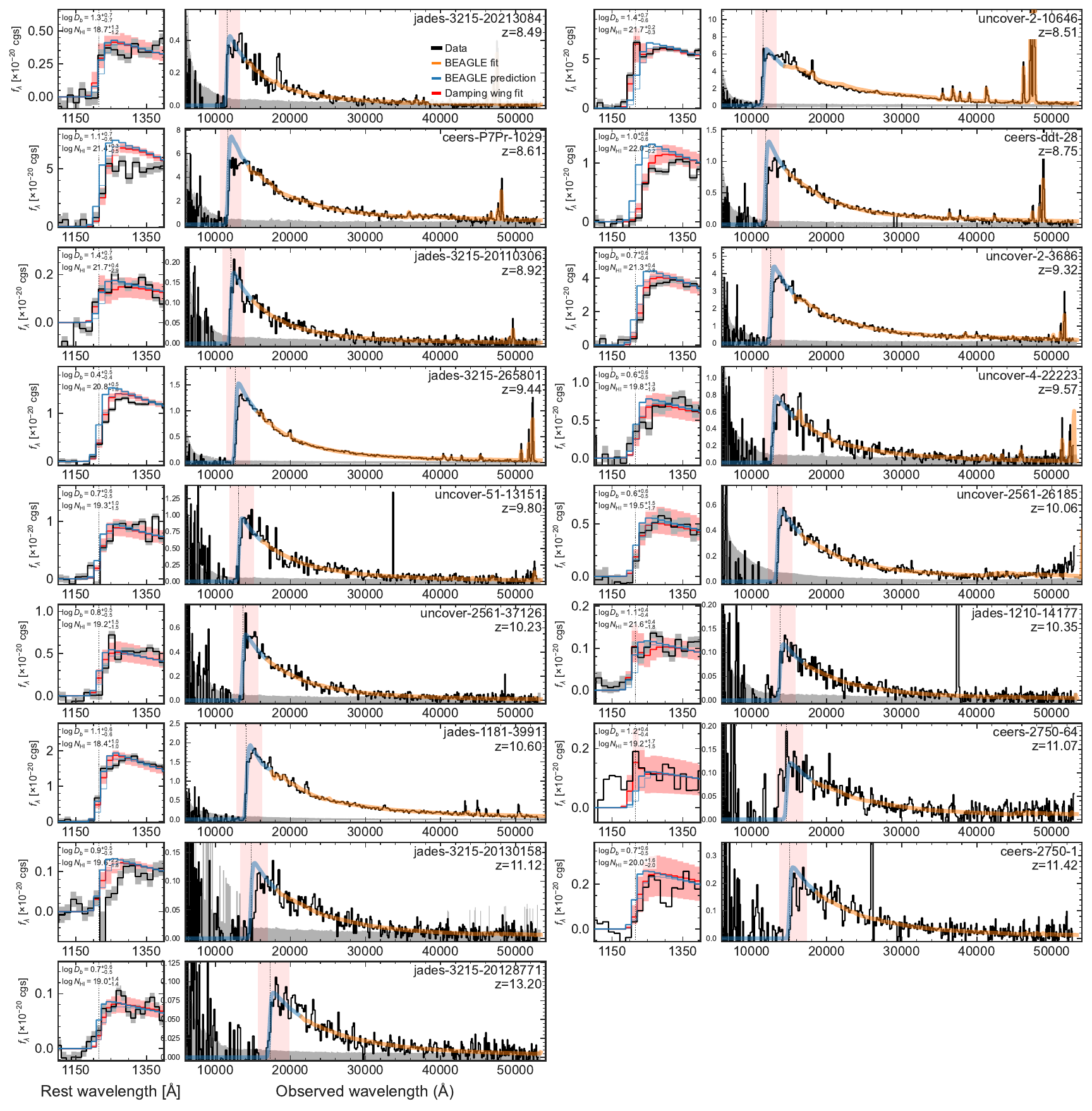}
    \caption{Same as Fig.~\ref{fig:spec1}}
    \label{fig:spec4}
\end{figure*}

In Table~\ref{tab:sample} we list IDs, coordinates, spectroscopic redshifts and \MUV for our sample. In Figures~\ref{fig:spec1}-\ref{fig:spec4} we show the individual prism spectra (black) and error spectrum (grey shaded region). The median BEAGLE fit to the spectrum is shown in orange over the region we fit to ($>1500$\,\AA\ in the rest-frame) and in blue in the region where we do not fit -- which represents the predicted unattenuated continuum which we use to fit the damping wings (Section~\ref{sec:fitting}). Smaller panels show a zoomed region around the \lya break, showing the observed spectra, median BEAGLE continuum prediction assuming ionized IGM (thick blue line) and fully neutral IGM (thin blue line), and the best-fit damping wing model (red line and shaded region marking median and 68\% range of samples of the posterior).


\end{appendix}

\onecolumn
\setlength{\tabcolsep}{4pt} 
\renewcommand{\arraystretch}{0.8} 
\setlength{\LTpre}{0pt} 
\setlength{\LTpost}{0pt} 
\setlength{\LTpre}{-6pt}
\setlength{\LTpost}{-6pt}

\begin{longtable}{lrrrr} 
\caption{\label{tab:sample}Sample used in our analysis. \textdagger\ denotes sources with S/N$>$15 which are included in our IGM inference. $^*$ denotes sources with redshift measured only from the \lya break, where we also fit for redshift in our damping wing fits.}\\
    \hline\hline
    ID & R.A. (deg) & Dec. (deg) & $z_\mathrm{spec}$ & \MUV \\
    \hline
\endfirsthead

    \caption{Sample (continued).}\\
    \hline\hline
    ID & R.A. (deg) & Dec. (deg) & $z_\mathrm{spec}$ & \MUV \\
    \hline
\endhead

    \hline \multicolumn{5}{c}{Continued on next page} \\
\endfoot

    \hline
\endlastfoot

ceers-P12Pr-1334 & 214.7683562 & 52.7176417 & 5.500 & $-20.07_{-0.26}^{+0.21}$ \\
jades-1210-16374 & 53.1157262 & -27.7749567 & 5.508 & $-18.47_{-0.27}^{+0.21}$ \\
uncover-3-12065 & 3.5700593 & -30.4036886 & 5.542 & $-19.23_{-0.05}^{+0.05}$ \\
jades-3215-208643 & 53.1302108 & -27.7783582 & 5.568 & $-18.56_{-0.19}^{+0.16}$ \\
jades-3215-30080873 & 53.1516681 & -27.8092365 & 5.574 & $-18.21_{-0.27}^{+0.22}$ \\
jades-1210-13577 & 53.1300486 & -27.7783888 & 5.575 & $-19.89_{-0.06}^{+0.05}$ \\
uncover-52-28110$^{\dag}$ & 3.6206593 & -30.3742656 & 5.625 & $-20.71_{-0.02}^{+0.02}$ \\
ceers-P7Pr-2168 & 215.1526021 & 53.0570611 & 5.661 & $-20.29_{-0.16}^{+0.14}$ \\
ceers-P4Pr-545 & 214.8644108 & 52.8536583 & 5.668 & $-19.27_{-0.04}^{+0.04}$ \\
uncover-4-44493 & 3.5931618 & -30.3464757 & 5.672 & $-20.92_{-0.03}^{+0.02}$ \\
ceers-P4Pr-403 & 214.8289679 & 52.8757000 & 5.770 & $-20.38_{-0.04}^{+0.04}$ \\
jades-1210-15099 & 53.1153793 & -27.8147737 & 5.777 & $-19.33_{-0.11}^{+0.10}$ \\
jades-3215-210003 & 53.1318414 & -27.7737748 & 5.786 & $-18.64_{-0.19}^{+0.16}$ \\
uncover-51-30249 & 3.5961401 & -30.3713764 & 5.787 & $-19.28_{-0.10}^{+0.10}$ \\
uncover-3-23608 & 3.5428145 & -30.3806460 & 5.796 & $-19.77_{-0.04}^{+0.04}$ \\
jades-3215-109389 & 53.1221037 & -27.8042913 & 5.803 & $-18.39_{-0.15}^{+0.13}$ \\
jades-1210-9842 & 53.1540712 & -27.7660718 & 5.805 & $-19.16_{-0.17}^{+0.14}$ \\
jades-1210-5113 & 53.1673026 & -27.8028741 & 5.818 & $-17.97_{-0.17}^{+0.15}$ \\
jades-3215-201127 & 53.1668515 & -27.8041256 & 5.838 & $-19.28_{-0.14}^{+0.12}$ \\
ceers-P7Pr-1677 & 215.1887383 & 53.0643778 & 5.874 & $-20.88_{-0.05}^{+0.05}$ \\
jades-1210-9414 & 53.1765679 & -27.7711311 & 5.892 & $-18.42_{-0.12}^{+0.10}$ \\
jades-1210-13620$^{\dag}$ & 53.1225640 & -27.7605900 & 5.919 & $-19.69_{-0.13}^{+0.12}$ \\
uncover-4-51076 & 3.5536890 & -30.3300569 & 5.929 & $-18.79_{-0.08}^{+0.08}$ \\
jades-1210-13704 & 53.1265384 & -27.8180900 & 5.934 & $-19.02_{-0.13}^{+0.12}$ \\
jades-3215-99671 & 53.1266416 & -27.8177312 & 5.935 & $-18.24_{-0.17}^{+0.14}$ \\
jades-1210-15125 & 53.1104167 & -27.8089236 & 5.942 & $-18.79_{-0.15}^{+0.14}$ \\
jades-1210-13176 & 53.1217573 & -27.7976379 & 5.947 & $-19.84_{-0.04}^{+0.03}$ \\
jades-1210-13618 & 53.1191033 & -27.7608124 & 5.949 & $-19.87_{-0.13}^{+0.12}$ \\
jades-1210-9880 & 53.1606215 & -27.7716100 & 5.984 & $-18.87_{-0.11}^{+0.10}$ \\
ceers-P4Pr-397 & 214.8361971 & 52.8826917 & 6.010 & $-20.99_{-0.02}^{+0.02}$ \\
ceers-P4Pr-362 & 214.8126892 & 52.8815361 & 6.050 & $-18.52_{-0.10}^{+0.09}$ \\
ceers-P4Pr-603 & 214.8672471 & 52.8367361 & 6.060 & $-20.91_{-0.07}^{+0.07}$ \\
ceers-P4Pr-618 & 214.8764692 & 52.8394111 & 6.064 & $-19.74_{-0.08}^{+0.07}$ \\
jades-3215-99302 & 53.1258179 & -27.8182275 & 6.070 & $-18.46_{-0.13}^{+0.12}$ \\
jades-1210-9867 & 53.1561000 & -27.7758826 & 6.106 & $-20.14_{-0.07}^{+0.06}$ \\
ceers-P5Pr-1518 & 215.0068021 & 52.9650417 & 6.107 & $-21.18_{-0.09}^{+0.09}$ \\
ceers-P4Pr-355 & 214.8064821 & 52.8788278 & 6.108 & $-19.81_{-0.04}^{+0.04}$ \\
ceers-P8Pr-1065 & 215.1168542 & 53.0010806 & 6.190 & $-20.03_{-0.42}^{+0.30}$ \\
uncover-4-52621 & 3.5606470 & -30.3261073 & 6.191 & $-18.71_{-0.17}^{+0.14}$ \\
ceers-P7Pr-1561 & 215.1660971 & 53.0707556 & 6.203 & $-20.22_{-0.10}^{+0.09}$ \\
uncover-4-36755 & 3.6003302 & -30.3606852 & 6.253 & $-18.18_{-0.16}^{+0.14}$ \\
uncover-1-35488$^{\dag}$ & 3.5789839 & -30.3625979 & 6.257 & $-19.74_{-0.04}^{+0.04}$ \\
jades-1210-8253 & 53.1666018 & -27.7724021 & 6.329 & $-19.06_{-0.08}^{+0.08}$ \\
jades-1210-9423 & 53.1758189 & -27.7744750 & 6.334 & $-19.12_{-0.10}^{+0.09}$ \\
jades-1210-13197$^{\dag}$ & 53.1349181 & -27.7727107 & 6.343 & $-19.97_{-0.04}^{+0.04}$ \\
uncover-2-34265 & 3.6071804 & -30.3648155 & 6.350 & $-18.82_{-0.08}^{+0.07}$ \\
uncover-2-8613 & 3.6006012 & -30.4102722 & 6.379 & $-17.25_{-0.07}^{+0.06}$ \\
ceers-ddt-663$^{\dag}$ & 214.8789692 & 52.8967472 & 6.520 & $-20.14_{-0.04}^{+0.04}$ \\
ceers-ddt-662 & 214.8778829 & 52.8976750 & 6.520 & $-19.02_{-0.08}^{+0.07}$ \\
uncover-2-6829 & 3.5937927 & -30.4154212 & 6.588 & $-19.23_{-0.04}^{+0.04}$ \\
jades-1210-5447 & 53.1628763 & -27.7692935 & 6.626 & $-17.70_{-0.16}^{+0.14}$ \\
jades-1210-9903 & 53.1690468 & -27.7788335 & 6.632 & $-18.68_{-0.11}^{+0.10}$ \\
ceers-P8Pr-1414 & 215.1280287 & 52.9849361 & 6.680 & $-20.92_{-0.03}^{+0.03}$ \\
ceers-P11Pr-577 & 214.8928608 & 52.8651583 & 6.694 & $-18.46_{-0.10}^{+0.09}$ \\
uncover-51-43709 & 3.5759403 & -30.3480270 & 6.728 & $-17.51_{-0.19}^{+0.16}$ \\
ceers-P12Pr-81049 & 214.7898221 & 52.7307889 & 6.739 & $-19.77_{-0.05}^{+0.04}$ \\
uncover-4-36857 & 3.5828283 & -30.3602961 & 6.764 & $-18.53_{-0.15}^{+0.14}$ \\
uncover-51-11254$^{\dag}$ & 3.5804464 & -30.4050217 & 6.872 & $-18.75_{-0.05}^{+0.05}$ \\
uncover-51-16155 & 3.5829561 & -30.3952308 & 6.878 & $-16.90_{-0.09}^{+0.08}$ \\
ceers-P8Pr-1143 & 215.0770063 & 52.9695056 & 6.928 & $-20.24_{-0.30}^{+0.23}$ \\
jades-1210-13609 & 53.1173008 & -27.7640888 & 6.930 & $-19.23_{-0.26}^{+0.21}$ \\
ceers-P5Pr-717$^{\dag}$ & 215.0814058 & 52.9721806 & 6.934 & $-21.54_{-0.09}^{+0.08}$ \\
ceers-ddt-445 & 214.9416108 & 52.9291306 & 6.980 & $-19.34_{-0.15}^{+0.13}$ \\
ceers-P8Pr-1102 & 215.0910475 & 52.9542861 & 7.000 & $-20.04_{-0.00}^{+0.00}$ \\
ceers-P11Pr-534 & 214.8591171 & 52.8536389 & 7.114 & $-20.57_{-0.08}^{+0.07}$ \\
jades-1210-13905 & 53.1183411 & -27.7690127 & 7.206 & $-18.70_{-0.33}^{+0.26}$ \\
ceers-P5Pr-1163 & 214.9904679 & 52.9719889 & 7.448 & $-20.24_{-0.00}^{+0.00}$ \\
ceers-P12Pr-80432 & 214.8120558 & 52.7467472 & 7.473 & $-19.99_{-0.07}^{+0.06}$ \\
ceers-ddt-449 & 214.9404892 & 52.9325556 & 7.544 & $-18.95_{-0.16}^{+0.14}$ \\
ceers-P7Pr-1023 & 215.1884129 & 53.0336472 & 7.776 & $-20.87_{-0.00}^{+0.00}$ \\
ceers-P4Pr-1027 & 214.8829958 & 52.8404167 & 7.819 & $-20.61_{-0.03}^{+0.03}$ \\
uncover-51-60157 & 3.6038888 & -30.3822626 & 7.879 & $-20.09_{-0.08}^{+0.07}$ \\
uncover-52-23604 & 3.6052466 & -30.3805843 & 7.883 & $-18.07_{-0.12}^{+0.11}$ \\
uncover-3-24531$^{\dag}$ & 3.6013404 & -30.3792037 & 7.891 & $-20.22_{-0.04}^{+0.04}$ \\
ceers-ddt-355 & 214.9447642 & 52.9314500 & 7.912 & $-19.36_{-0.11}^{+0.10}$ \\
jades-1210-5173 & 53.1568262 & -27.7671606 & 7.981 & $-18.87_{-0.16}^{+0.14}$ \\
ceers-P8Pr-1149 & 215.0897142 & 52.9661833 & 8.175 & $-20.60_{-0.00}^{+0.00}$ \\
jades-3215-20198852$^{\dag}$ & 53.1077596 & -27.8129310 & 8.276 & $-18.37_{-0.18}^{+0.16}$ \\
jades-3215-20213084 & 53.1589078 & -27.7650743 & 8.493 & $-18.94_{-0.16}^{+0.14}$ \\
uncover-2-10646$^{\dag}$ & 3.6369614 & -30.4063615 & 8.511 & $-21.57_{-0.05}^{+0.05}$ \\
ceers-P7Pr-1029 & 215.2187625 & 53.0698611 & 8.610 & $-21.63_{-0.00}^{+0.00}$ \\
ceers-ddt-28$^{\dag}$ & 214.9386421 & 52.9117500 & 8.753 & $-20.72_{-0.04}^{+0.04}$ \\
jades-3215-20110306 & 53.1691312 & -27.8029220 & 8.917 & $-18.09_{-0.23}^{+0.19}$ \\
uncover-2-3686$^{\dag}$ & 3.6171995 & -30.4255353 & 9.321 & $-21.72_{-0.05}^{+0.04}$ \\
jades-3215-265801$^{\dag}$ & 53.1124268 & -27.7746194 & 9.437 & $-20.23_{-0.15}^{+0.13}$ \\
uncover-4-22223 & 3.5681144 & -30.3830525 & 9.573 & $-17.27_{-0.23}^{+0.19}$ \\
uncover-51-13151 & 3.5924999 & -30.4014640 & 9.801 & $-17.62_{-0.04}^{+0.04}$ \\
uncover-2561-26185 & 3.5670710 & -30.3778610 & 10.057 & $-18.90_{-0.06}^{+0.06}$ \\
ceers-1345-80041$^*$ & 214.7325250 & 52.7580900 & $10.070_{-0.190}^{+0.140}$ & $-20.10_{-0.10}^{+0.10}$ \\
uncover-2561-37126$^*$ & 3.5901110 & -30.3597420 & $10.230_{-0.020}^{+0.020}$ & $-20.02_{-0.04}^{+0.04}$ \\
jades-1210-14177$^*$ & 53.1588400 & -27.7734920 & $10.350_{-0.060}^{+0.070}$ & $-18.82_{-0.08}^{+0.10}$ \\
jades-1181-3991$^{\dag}$ & 189.1060540 & 62.2420490 & 10.603 & $-22.00_{-0.04}^{+0.05}$ \\
ceers-2750-10 & 214.9066330 & 52.9455040 & 11.046 & $-20.30_{-0.20}^{+0.10}$ \\
ceers-2750-64$^*$ & 214.9227830 & 52.9115280 & $11.070_{-0.260}^{+0.130}$ & $-19.30_{-0.20}^{+0.20}$ \\
jades-3215-20130158 & 53.1647620 & -27.7746260 & 11.122 & $-19.69_{-0.10}^{+0.12}$ \\
ceers-2750-1 & 214.9431480 & 52.9424420 & 11.416 & $-20.22_{-0.10}^{+0.11}$ \\
jades-3215-20096216 & 53.1663460 & -27.8215570 & 12.512 & $-19.08_{-0.12}^{+0.11}$ \\
uncover-2561-13077$^*$ & 3.5708690 & -30.4015850 & $13.079_{-0.020}^{+0.020}$ & $-19.04_{-0.10}^{+0.11}$ \\
jades-3215-20128771$^*$ & 53.1498810 & -27.7765020 & $13.200_{-0.070}^{+0.040}$ & $-18.82_{-0.07}^{+0.10}$
\end{longtable}

\end{document}